\documentclass[draft=false,twocolumn,showpacs,preprintnumbers]{aa}
\usepackage{natbib,twoopt}
\usepackage{graphicx,units,multirow,subfigure,color,amsmath,amssymb,epsfig,makecell}
\usepackage{courier}
\usepackage[varg]{txfonts}
\usepackage[usenames,dvipsnames,svgnames,table]{xcolor}
\usepackage[absolute,overlay]{textpos}

\newcommand{\beq}{\begin{equation}}
\newcommand{\eeq}{\end{equation}}
\newcommand{\ben}{\begin{eqnarray}}
\newcommand{\een}{\end{eqnarray}}
\newcommand{\besub}{\begin{subequations}}
\newcommand{\eesub}{\end{subequations}}
\newcommand{\bi}{\begin{itemize}}
\newcommand{\ei}{\end{itemize}}

\newcommand{\bea}{\begin{align}}
\newcommand{\eea}{\end{align}}

\newcommand{\BIG}{{\sf BIG}}
\newcommand{\SLIM}{{\sf SLIM}}
\newcommand{\QUAINT}{{\sf QUAINT}}
\newcommand{\galprop}{\texttt{Galprop}}
\newcommand{\usine}{{\sc usine~v3.5}}
\newcommand{\minuit}{{\sc minuit}}
\newcommand{\minos}{{\sc minos}}
\newcommand{\usinebis}{{\sc usine}}
\newcommand{\hesse}{{\sc hesse}}

\newcommand{\deut}{\ensuremath{^2}\textrm{H}}

\newcommand{\het}{\ensuremath{^3}\textrm{He}}
\newcommand{\hef}{\ensuremath{^4}\textrm{He}}

\newcommand{\hettohef}{\het{}/\hef{}}

\newcommand{\chimindof}{\ensuremath{\chi^2_{\rm min}}/{\rm dof}}
\newcommand{\chinui}{\ensuremath{\chi^2_{\rm nui}}}
\newcommand{\chipernui}{\ensuremath{\chi^2_{\rm nui}/{n_{\rm nui}}}}
\newcommand{\veryshortarrow}[1][3pt]{\,\mathrel{%
   \vcenter{\hbox{\rule[-.5\fontdimen8\textfont3]{#1}{\fontdimen8\textfont3}}}%
   \mkern-4mu\hbox{\usefont{U}{lasy}{m}{n}\symbol{41}}}}

\def\aap{A\&A}%
\def\apjl{ApJ}%
\def\apjs{ApJS}%
\def\ao{Appl.~Opt.}%
\bibpunct{(}{)}{;}{a}{}{,} 

\definecolor{light-gray}{gray}{0.95}
\definecolor{dark-gray}{gray}{0.4}

\def\pbar{\ensuremath{\overline{p}}}

\usepackage[pdftex,             
            breaklinks=true,%
            colorlinks=true,%
            pdfauthor={Weinrich et al.},%
            pdftitle={Combined analysis of AMS-02 (Li,Be,B)/C, N/O, $^3$He, and $^4$He data}%
           ]{hyperref}
\makeatletter
\newcommandtwoopt{\citeads}[3][][]{\href{http://adsabs.harvard.edu/abs/#3}{\def\hyper@linkstart##1##2{}\let\hyper@linkend\@empty\citealp[#1][#2]{#3}}}
\newcommandtwoopt{\citepads}[3][][]{\href{http://adsabs.harvard.edu/abs/#3}{\def\hyper@linkstart##1##2{}\let\hyper@linkend\@empty\citep[#1][#2]{#3}}}
\newcommandtwoopt{\citetads}[3][][]{\href{http://adsabs.harvard.edu/abs/#3}{\def\hyper@linkstart##1##2{}\let\hyper@linkend\@empty\citet[#1][#2]{#3}}}
\newcommandtwoopt{\citealpads}[3][][]{\href{http://adsabs.harvard.edu/abs/#3}{\def\hyper@linkstart##1##2{}\let\hyper@linkend\@empty\citealp[#1][#2]{#3}}}
\newcommandtwoopt{\citealtads}[3][][]{\href{http://adsabs.harvard.edu/abs/#3}{\def\hyper@linkstart##1##2{}\let\hyper@linkend\@empty\citealt[#1][#2]{#3}}}
\newcommandtwoopt{\citeyearads}[3][][]{\href{http://adsabs.harvard.edu/abs/#3}{\def\hyper@linkstart##1##2{}\let\hyper@linkend\@empty\citeyear[#1][#2]{#3}}}
\newcommandtwoopt{\citeadsstar}[3][][]{\href{http://adsabs.harvard.edu/abs/#3}{\def\hyper@linkstart##1##2{}\let\hyper@linkend\@empty\citealp*[#1][#2]{#3}}}
\newcommandtwoopt{\citepadsstar}[3][][]{\href{http://adsabs.harvard.edu/abs/#3}{\def\hyper@linkstart##1##2{}\let\hyper@linkend\@empty\citep*[#1][#2]{#3}}}
\newcommandtwoopt{\citetadsstar}[3][][]{\href{http://adsabs.harvard.edu/abs/#3}{\def\hyper@linkstart##1##2{}\let\hyper@linkend\@empty\citet*[#1][#2]{#3}}}
\newcommandtwoopt{\citeyearadsstar}[3][][]{\href{http://adsabs.harvard.edu/abs/#3}{\def\hyper@linkstart##1##2{}\let\hyper@linkend\@empty\citeyear*[#1][#2]{#3}}}
\newcommandtwoopt{\citeauthoradsstar}[3][][]{\href{http://adsabs.harvard.edu/abs/#3}{\def\hyper@linkstart##1##2{}\let\hyper@linkend\@empty\citeauthor*[#1][#2]{#3}}}
\newcommandtwoopt{\citepthesis}[3][][]{\href{http://tel.archives-ouvertes.fr/docs/#3}{\def\hyper@linkstart##1##2{}\let\hyper@linkend\@empty\citep[#1][#2]{#3}}}
\newcommandtwoopt{\citetthesis}[3][][]{\href{http://tel.archives-ouvertes.fr/docs/#3}{\def\hyper@linkstart##1##2{}\let\hyper@linkend\@empty\citet[#1][#2]{#3}}}
\makeatother

\begin{document}

\input epsf
\title{Combined analysis of AMS-02 (Li,Be,B)/C, N/O, $^3$He, and $^4$He data}

\author{N. Weinrich\inst{1}
  \and Y. G\'enolini\inst{2}\thanks{\url{yoann.genolini@nbi.ku.dk}}
  \and M. Boudaud\inst{3}\thanks{Deceased}
  \and L. Derome\inst{1}\thanks{\url{laurent.derome@lpsc.in2p3.fr}}
  \and D. Maurin\inst{1}\thanks{\url{david.maurin@lpsc.in2p3.fr}}
}

\authorrunning{N. Weinrich, Y. Génolini, M. Boudaud et al.}

\institute{
LPSC, Universit\'e Grenoble Alpes, CNRS/IN2P3, 53 avenue des Martyrs, 38026 Grenoble, France
\and Niels Bohr International Academy \& Discovery Center, Niels Bohr Institute, University of Copenhagen, Blegdamsvej 17, DK-2100 Copenhagen, Denmark
\and Instituto de F\'isica Te\'orica UAM/CSIC, Calle Nicol\'as Cabrera 13-15, Cantoblanco E-28049 Madrid, Spain
}

\date{Received / Accepted}

\abstract
{AMS-02 measured several secondary-to-primary ratios enabling a detailed study of Galactic cosmic-ray transport.}
{We constrain previously derived benchmark scenarios (based on AMS-02 B/C data only) using other secondary-to-primary ratios, to test the universality of transport and the presence of a low-rigidity diffusion break.}
{We use the 1D thin disc/thick halo propagation model of \usine{} and a $\chi^2$ minimisation accounting for a covariance matrix of errors (AMS-02 systematics) and nuisance parameters (cross-sections and solar modulation uncertainties).}
{The combined analysis of AMS-02 Li/C, Be/C, and B/C strengthens the case for a diffusion slope of $\delta=0.50\pm 0.03$ with a low-rigidity break or upturn of the diffusion coefficient at GV rigidities. Our simple model can successfully reproduce all considered data (Li/C, Be/C, B/C, N/O, and \hettohef{}), although several issues remain: (i) the quantitative agreement depends on the assumptions made on the not well constrained correlation lengths of AMS-02 data systematics; (ii) combined analyses are very sensitive to production cross sections, and we find post-fit values differing by $\sim5-15\%$ from their most likely values (roughly within currently estimated nuclear uncertainties); (iii) two very distinct regions of the parameter space remain viable, either with reacceleration and convection, or with purely diffusive transport.
 }
{To take full benefit of combined analyses of AMS-02 data, better nuclear data and a better handle on energy correlations in the data systematic are required. AMS-02 data on heavier species are eagerly awaited to further explore cosmic-ray propagation scenarios.}

\keywords{Astroparticle physics -- Cosmic rays}

\maketitle
\setcounter{tocdepth}{2}

\section{Introduction}

The high-precision cosmic-ray (CR) data released in the last years, in particular by the AMS-02 experiment, confirmed or unveiled anomalies \citepads{2015ICRC...34....9S,2018JApA...39...41S}, e.g., spectral breaks in both primary and secondary species \citepads{2015PhRvL.114q1103A,2015PhRvL.115u1101A,2018PhRvL.120b1101A}. The latter reinvigorated the discipline, with a flurry of activities around CR transport, interpreting these existing or apparent anomalies as many plausible scenarios: secondary production at source \citepads[e.g.,][]{2003A&A...410..189B,2009PhRvL.103h1103B,2012A&A...544A..16T,2014PhRvD..90f1301M,2017ApJ...844L..26T,2017PhRvD..95l3007C,2019PhRvD.100f3020Y}; spatially-dependent diffusion \citepads[e.g.,][]{2012PhRvL.109f1101B,2012ApJ...752L..13T,2015A&A...583A..95A,2016PhRvD..94l3007F,2018PhRvD..97f3008G,2018PhRvL.121b1102E}; space-time granularity effects \citepads[e.g.,][]{2012MNRAS.421.1209T,2013A&A...555A..48B,2015PhRvL.115r1103K,Genolini:2016hte,2018JCAP...11..045M,2019JCAP...01..046B}; etc---for recent reviews on GCRs from MeV to PeV energies, we refer the reader \citetads{2015ARA&A..53..199G,2017PrPNP..94..184A,2019PrPNP.10903710K,2019IJMPD..2830022G}.

In this work, we follow on our previous efforts to interpret AMS-02 data \citepads{2019PhRvD..99l3028G}. We rely on steady-state semi-analytical propagation models \citepads[e.g.,][]{1964ocr..book.....G,2001ApJ...547..264J,2001ApJ...555..585M} available in the public code \usinebis{} \citepads{2018arXiv180702968M}. To avoid mixing uncertainties of different natures when studying simultaneously source and transport parameters, we focus on flux ratios of so-called {\em secondary} species (absent from the sources, but produced by nuclear interaction in the interstellar medium, ISM) to {\em primary} species (dominantly from injection at source). These ratios are extremely sensitive to propagation parameters, while mostly insensitive to the source spectrum of primary species, provided that they share a common power-law in rigidity \citepads{2002A&A...394.1039M,2011A&A...526A.101P,2015A&A...580A...9G}. This approach has already been successfully used in several studies with simple cross checks on primary fluxes, in order to: (i) find evidence for a break in the diffusion coefficient at $\sim 250$~GV \citepads{2017PhRvL.119x1101G}; (ii) provide a refined methodology accounting for correlations in systematic errors and cross section uncertainties in the context of high-precision data \citepads{2019A&A...627A.158D}; (iii) provide new benchmark propagation models hinting at a new break in the low-rigidity regime at $\lesssim 5$~GV \citepads{2019PhRvD..99l3028G}; and (iv) perform a new calculation of the \pbar{} flux, showing that current AMS-02 data are fully consistent with a purely secondary origin \citepads{Boudaud:2019efq}.

We continue here our step-by-step approach to interpret more species measured by AMS-02. A companion paper (Weinrich et al., in prep.) will focuses on the determination of the halo size of the Galaxy, which is a crucial input to assess the significance of a possible dark matter component in the \pbar{} and positron data \citepads[e.g.,][]{2012CRPhy..13..740L}. However, the answers that can be provided only make sense if the robustness of the model and the consistency of transport parameters is demonstrated with all secondary-to-primary AMS-02 data---ideally from the lightest to the heaviest nuclei. In a previous work \citepads{2019PhRvD..99l3028G}, only AMS-02 B/C data were used, but Li/C and Be/C (or similarly Li/O and Be/O) with similar precision are now available \citepads{2018PhRvL.120b1101A}. More recently, on a smaller rigidity range and with slightly larger uncertainties, \hettohef{} data were also released \citepads{2019PhRvL.123r1102A}. One can also use the N/O ratio \citepads{2018PhRvL.121e1103A}, although N is a mixed species (both primary and secondary contributions) making it sensitive to source parameters. So far, these are the currently published high-precision secondary-to-primary AMS-02 ratios. The minimal requirement to advocate the validity of a model from \pbar{} to O elements is to find consistent transport parameters for all considered species. This {\em universality} was recently challenged in \citetads{2016ApJ...824...16J}, where different transport parameters were found for $Z=1-2$ elements and heavier species---AMS-02 data were however not available at the time of their analysis. Alternatively, assuming universality of the transport parameters and  analysing AMS-02 Li, Be, B, and N data, \citetads{2020ApJ...889..167B} concluded on the presence of a primary source of Li to reproduce existing data.

As already underlined, an important issue is that of spectral breaks in CR spectra and their interpretation. At high rigidity ($\sim 300$~GV), spectral breaks are seen in primary species \citepads{2015PhRvL.114q1103A,2015PhRvL.115u1101A} and in secondary-to-primary ratios \citepads{Aguilar2016}. A quantitative analysis of B/C data strongly favours a scenario with a diffusion break \citepads{2017PhRvL.119x1101G,2018JCAP...01..055R,2019PhRvD..99l3028G}. This conclusion should be strengthened by a combined analysis of several species. Qualitatively this is backed-up by the different spectral breaks observed in several primary and secondary species \citepads{2018PhRvL.120b1101A} and from their interpretation in a propagation model \citepads{2020ApJ...889..167B}. In this paper, we do not inspect further this finding---we rely on the results of \citetads{2019PhRvD..99l3028G} for the diffusion break parameters---and rather prefer to focus on a possible low-rigidity break. We stress that very precise low-energy Li/C, Be/C, and B/C data from ACE-CRIS \citepads{2009ApJ...698.1666G} exist. Using various secondary-to-primary ratios and extending the analysis to lower energy may strengthen or weaken the case for a diffusion break at a few GV as observed from B/C AMS-02 data only \citetads{2019PhRvD..99l3028G}. This break was also hinted by the interpretation of AMS-02 electrons and positrons in a pure diffusion propagation model \citepads{2019PhRvD.100d3007V}.

The paper is organised as follows: in Sect.~\ref{sec:setup}, we recall the propagation model and the configurations used for the analysis. In Sect.~\ref{sec:LiBeB}, we perform separate or simultaneous analyses of Li/C, Be/C, and B/C (or LiBeB for short). These analyses allow us to refine our benchmark transport models and strengthen the case for a departure from a universal power-law diffusion at low rigidity. In Sect.~\ref{sec:NO_3He4He}, we investigate whether \hettohef{} and N/O data can be accommodated by the model, and highlight the crucial role of the correlation length in the data systematic uncertainties. In Sect.~\ref{sec:aboutxs}, we take a deeper look at how cross-section nuisance parameters behave in the performed fits, further highlighting the needs for better measurements of cross sections \citepads{2018PhRvC..98c4611G}. We summarise these findings and conclude in Sect.~\ref{sec:conclusions}.

For readability, we postpone several technical details and checks in the appendices: App.~\ref{app:cov_mat} details the covariance matrix of systematic errors used for AMS-02 data; App.~\ref{app:nuis_xs} details the cross-section reactions used as nuisance parameters; App.~\ref{app:min_conv} illustrates the difficulties to achieve and ensure a good convergence of minimisation in the presence of many nuisance parameters; App.~\ref{app:LE-data} discusses the model consistency and possible constraints low-energy data may bring on the diffusion coefficient at low energy.

\section{Model and configurations (\BIG{}, \SLIM{}, \QUAINT{})}
\label{sec:setup}

The treatment of CR propagation and the fitting strategy mainly follow the choices detailed in \citetads{2019A&A...627A.158D} and \citetads{2019PhRvD..99l3028G}. Hereafter, we summarise the main steps and features of our modelling and approach.

\subsection{CR propagation framework}

We assume the CR density to obey a steady-state diffusion-advection equation---Eq.~(1) in \citetads{2019PhRvD..99l3028G}---, which also includes all relevant interactions between CRs and the interstellar matter. Fixing the geometry of the diffusion halo allows us to derive semi-analytically solutions that are computed with the code \usine\footnote{\url{https://lpsc.in2p3.fr/usine}} \citepads{2018arXiv180702968M}. We assume CRs propagate within an infinite slab of half-thickness $L$, with a null density at the borders to mimic magnetic confinement, so we disregard radial boundaries (sent to infinity). This configuration defines a simple 1D geometry with a single vertical coordinate $z$. CR sources and the gas are pinched in a thin plan of half-thickness $h=100$~pc at $z=0$, to which spallations and energy losses are restricted. This 1D geometry is sufficient to capture the full GCR propagation phenomenology \citepads[e.g.,][]{2001ApJ...547..264J,2010A&A...516A..67M,2015A&A...580A...9G}.

CR transport is driven by diffusion and convection. The diffusion tensor is assumed to be isotropic and homogeneous, boiling down to a scalar function of the CR rigidity $R=p/Ze$. Quasi-linear theory (QLT) predicts that the $R$-dependence should follow a simple power law $\propto R^\delta$, with $\delta=2-\nu$ related to the power-law index of the magnetic turbulence spectrum, $(\delta B/B)^2\propto k^\nu$ \citepads{1990acr..book.....B,2002cra..book.....S,2009ASSL..362.....S}. However, this behaviour strictly applies to the inertial regime. The actual diffusion coefficient should be seen as an \textit{effective} coefficient that could deviate from a pure power law (see \citealtads{2019PhRvD..99l3028G} for an extended discussion) and we use:
\begin{equation}
  \label{eq:def_K}
  K(R) = {\beta^\eta} K_{0} \;
  {\left\{ 1 \!+ \left( \frac{R}{R_{\rm l}} \right)^{\frac{\delta_{\rm l}-\delta}{s_{\rm l}}} \right\}^{s_{\rm l}}}
  {\left\{  \frac{R}{R_0\!=\!1\,{\rm GV}} \right\}^\delta}\,
  {\left\{  1 \!+ \left( \frac{R}{R_{\rm h}} \right)^{\frac{\delta-\delta_{\rm h}}{s_{\rm h}}}
    \right\}^{-s_{\rm h}}}\!\!\!\!\!\!.
\end{equation}
This coefficient is broken down in several limiting cases in Sect.~\ref{sec:benchmarks}. It enables two different softening of the diffusion coefficient at \textit{low} and at \textit{large} rigidity \citepads{2017PhRvL.119x1101G,2019PhRvD..99l3028G}. These deviations are parameterised by rigidity scales $R_l$ and $R_h$, respectively, and by power-law indices $\delta_l$ and $\delta_h$. We stress that, in the above equation, the normalisation $K_0$ is defined at $R_0=1$~GV. For a meaningful inter-comparison of the different propagation setups, we also sometimes refer to $K_{10}$ defined at $R_{10}=10$~GV (inertial regime), with $K_{10}=K_0 \times 10^\delta$ or equivalently $\log_{10}(K_{10})=\log_{10}(K_0) + \delta$.

The CR scattering on plasma waves leading to spatial diffusion also induces diffusion in momentum space (a.k.a. reacceleration). Following the treatment of \citetads{1988SvAL...14..132O,1994ApJ...431..705S,2001ApJ...547..264J}, the diffusion coefficient in momentum space $K_{pp}$ can directly be related to $K$:
\[
  K_{pp}(R,\vec{x})= \frac{4}{3} \frac{1}{\delta (4-\delta^2) (4-\delta)} \frac{V_a^2 p^2}{K(R)}\,,
\]
where we have introduced the speed of plasma waves $V_a$ (the Alfv\'enic speed). In our treatment, the reacceleration is pinched in the Galactic plane\footnote{The phenomenology of a more extended reacceleration zone is obtained rescaling $V_a^2$ by $h/z_{\rm A}$ \citepads{2002A&A...394.1039M}; $z_{\rm A}$ is the half-height over which reacceleration spreads in the magnetic slab \citepads{2001ApJ...547..264J}. For $h/z_{\rm A}\simeq {\cal O}(h/L)$, fitted values of $V_a$ should be scaled by a factor $\sqrt{L/h}$ before any comparison against theoretical or observational constraints~\citepads{2014MNRAS.442.3010T,2017A&A...597A.117D}.}.
Our modelling also includes convection which naturally arises from the global motion of the plasma; it is characterised by the convective speed $V_{\rm c}$ taken to be constant and positive above the galactic plan, and negative below.

For each run, the fluxes of the elements from Lithium (Li) to Silicon (Si) are computed assuming \het{} and the isotopes of Li, Be and B to be pure secondary species. The primaries are injected following a universal power law in rigidity with index $\alpha$ for He and heavier species. The normalisation of the primary components of all elements is fixed by the 10.6 GeV/n data point of HEAO-3, except for H, C, N, and O elements which are normalised to the more precise AMS-02 data at 50 GV.

\subsection{Benchmark models}
\label{sec:benchmarks}
The above-defined propagation scenario involves numerous physical processes in a 12-dimensional parameter space. There are eight parameters for spatial diffusion ($K_0,\,\delta,\,\eta,\,R_l,\,\delta_l,\,s_l,\,R_h,\,\delta_h,\,s_h$), one for reacceleration ($V_a$), one for convection ($V_{\rm c}$), and one for the halo size ($L$).

To speed up the analyses and convergence, several simplifications can be made. First, owing to the $K_0/L$ degeneracy for secondary-to-primary stable species \citepads[e.g.,][]{2001ApJ...555..585M}, we need to fix $L$, and we choose $L=5$~kpc to be consistent with values derived from the analysis of radioactive species (Weinrich et al., in prep.)\footnote{One of the studied secondary-to-primary ratios, Be/C, involves the $\beta$-unstable $^{10}$Be species, and thus depends on the exact halo size value.}. Second, we fix the three high-rigidity break parameters ($R_h,\delta_h,s_h$) to the values of \citetads{2019PhRvD..99l3028G}. That analysis concluded that uncertainties on the high-rigidity parameters neither impact the best values nor the uncertainties of the remaining propagation parameters (that we study here). Finally, as in \citetads{2019PhRvD..99l3028G}, we fix the smoothness of the low-rigidity break parameter $s_l$ = 0.04, which amounts to consider a fast transition.

With the above simplifications, the parameter space is reduced to eight dimensions. Following \citetads{2019PhRvD..99l3028G}, we define three benchmark configurations \BIG{}, \SLIM{}, and \QUAINT{} that account for different subsets of the parameter space. The salient features and free parameters of these configurations, possibly pointing to different underlying microphysical processes, are:
\begin{itemize}
  \item \BIG{} (double-break diffusion coefficient, convection, and reacceleration): the non-relativistic parameter $\eta$ is fixed to 1, since its effect is degenerated with that of $\delta_l$. It is the most general configuration, maximising the flexibility at low rigidity, with 6 free parameters $(K_0,\,\delta,\,R_l,\,\delta_l,\, V_{\rm c},\,V_a)$.

  \item \SLIM{} (subpart of \BIG{} with $V_a\!=\!V_{\rm c}\!=\!0$ and $\eta\!=\!1$): possible damping of small-scale magnetic turbulences may directly reflect on the low-rigidity change of the diffusion slope without convection neither reacceleration. An excellent fit on B/C data was obtained \citepads{2019PhRvD..99l3028G} from only 4 free parameters ($K_0,\,\delta,\,R_l,\,\delta_l)$.

  \item \QUAINT{} (subpart of \BIG{} with non-relativistic break): instead of a power-law break in spatial diffusion, deviation from QLT at low rigidity is accounted for by letting $\eta$ free, in addition to convection and reacceleration effects. This parametrisation is similar to that used in \citetads{2010A&A...516A..67M} and \citetads{2010APh....34..274D}, but with the extra high-rigidity break, leading to 5 free parameters $(K_0,\,\delta,\,\eta,\, V_{\rm c},\, V_a)$.
\end{itemize}

\subsection{Minimisation strategy}
\label{sec:strategy}

To extract the propagation parameters, we fit \BIG{}, \SLIM{}, and \QUAINT{} against different datasets. We extensively use the \minuit\ package \citepads{1975CoPhC..10..343J} interfaced with the \usinebis\ code \citepads{2018arXiv180702968M}, with the \minos{} method used to retrieve asymmetric error bars. App.~\ref{app:min_conv} shows that convergence with many (nuisance) parameters can be difficult to achieve, regardless of the algorithm used. To ensure true minima are found, ${\cal O}(100)$ minimisations from different starting points are carried out for all our analyses.

\subsubsection{$\chi^2$ with covariance and nuisance}
The quantity we minimise is a $\chi^2$ accounting for systematics in the data uncertainties and in the modelling:
\begin{equation}
  \chi^2 = \sum_{t}\left( \sum_{q_t} \left( {\cal D}_{\rm cov}^{t,q_t} + {\cal N}^{t,q_t}_{\rm Sol.Mod.}\right) \right) + \sum_r{\cal N}^{r}_{\rm XS},
  \label{eq:chi2}
\end{equation}
where $t$ and $q_t$ respectively run over data taking periods and quantities (e.g., Li/C, Be/C, B/C in a combined fit) measured at $t$, whereas $r$ runs over cross-section reactions.
We stress that modification of cross-section values impacts CR calculations (model uncertainties), but they are independent of any specific data taking period and quantity included in the fit (data-related uncertainties), hence $r$ sitting outside the $t$ and $q$ loops.

The ${\cal D}_{\rm cov}^{t,q_t}$ term includes ${ij}$ energy bins correlations ($n_E$ bins in total) from a covariance matrix of data uncertainties,
\begin{equation}
 {\cal D}_{\rm cov} = \sum_{i,j=1}^{n_E,n_E}({\rm data}_i-{\rm model}_i) \; \left({\cal C}^{-1}\right)_{ij} \; ({\rm data}_j-{\rm model}_j),
     \label{eq:cov}
\end{equation}
which reduces to $\sum_{k=1}^{n_E} \left({\rm data}_k-{\rm model}_k\right)^2/\sigma_k^2$ for data with uncorrelated systematics $\sigma_k$.

The ${\cal N}^{t,q}_{\rm Sol.Mod.}$ and ${\cal N}_{\rm XS}$ terms account for Gaussian-distributed nuisance parameters (solar modulation and cross sections) of the form
\begin{equation}
   {\cal N} = \frac{(y-\langle{y}\rangle)^2}{\sigma_y^2},
   \label{eq:chi2_nuis}
\end{equation}
where $\langle{y}\rangle$ and $\sigma_y^2$ are the mean and variance of the parameter, and $y$ the tested value in the fit. For more details and justifications, we refer the reader to \citetads{2019A&A...627A.158D} and App. B therein.

\subsubsection{Data uncertainties}
It is common practice to estimate the \textit{total} errors by summing systematics and statistics in quadrature. However, AMS-02 data are dominated by systematics below $\sim100$~GV and energy correlations can be important.
As shown in \citetads{2019A&A...627A.158D} for the B/C ratio, accounting for these correlations is crucial to obtain unbiased fits and parameters. Following the approach detailed in \citetads{2019A&A...627A.158D}---also used in \citetads{2019PhRvD..99l3028G} and \citetads{Boudaud:2019efq}---, we build, from the information given in the AMS-02 publications, a covariance matrix of systematic uncertainties for Li/C, Be/C, Be/C data, and also for ratios up to O (including N/O). We recall the method and show the resulting matrices in App.~\ref{app:cov_mat}; \het{} and \hef{} require extra care, as detailed in App.~\ref{app:covHe}.

For the low-energy datasets (ACE-CRIS, Ulysses, Voyager, etc.) analysed in App.~\ref{app:LE-data}, we stick to the \textit{total} errors since these data are dominated by statistical uncertainties and only cover a narrow energy range.

\subsubsection{Cross-section nuisance parameters}
One of the most important ingredients to compute secondary fluxes are the nuclear fragmentation (or spallation) cross sections. For the latter we use the \galprop{}\footnote{\url{https://galprop.stanford.edu/}} reference parametrisation.
As shown in \citetads{2019A&A...627A.158D} on simulated data, starting from the wrong cross sections can significantly bias the fit. To account for the large ($\sim 10-15\%$) uncertainties in the cross sections, \citetads{2019A&A...627A.158D} proposed a ``normalisation, slope and shape'' (NSS) strategy, in which a few selected reactions vary via a combination of global normalisation, power-law modification (slope) at low energy, and shift of the energy scale. This strategy was successfully used for the B/C analysis in \citetads{2019PhRvD..99l3028G}, applying NSS nuisance parameters to a few dominant reactions---selected from the ranking of \citetads{2018PhRvC..98c4611G}.

The NSS strategy is recalled in App.~\ref{app:nuis_xs}, where our selected reactions and NSS nuisance parameters for \het{}, Li, Be, B, and N are gathered. As discussed in Sect.~\ref{sec:LiBeB}, whereas the impact of cross-sections uncertainties is mostly absorbed by the diffusion coefficient normalisation and also mitigated by the impact of data systematic uncertainties in single-species fits \citepads{2019PhRvD..99l3028G}, this is no longer the case for multi-species fits.

\subsubsection{Solar modulation nuisance parameters}
\label{sec:phi_nui}
To describe Solar modulation, we rely on the force-field approximation \citepads{1967ApJ...149L.115G,1968ApJ...154.1011G,1969JGR....74.4973F,1987A&A...184..119P,1998APh.....9..261B,2004JGRA..109.1101C} with the Fisk potential $\phi_{\rm FF}$ as the only free parameter.

We use priors on the $\phi_{\rm FF}$ values of all datasets considered, consistent with neutron monitor (NM) data \citepads{2015AdSpR..55..363M} as follows: IS H and He fluxes have been determined in \citetads{2016A&A...591A..94G,2017A&A...605C...2G} and were used in \citetads{2017AdSpR..60..833G} to obtain $\phi_{\rm FF}$ time series from NM data. Averaging these time series over the appropriate data taking periods provide for each dataset the central value to be used as its prior nuisance parameter, $\langle{\phi}_{\rm FF}\rangle$, and we take $\sigma_{\phi}=100$~MV \citepads{2017AdSpR..60..833G}. This procedure applies to all data, in particular to the low-energy ones, coming from several experiments, as discussed in App.~\ref{app:LE-data}.

We assign the same nuisance parameter to the AMS-02 data on Li, Be, and B \citepads{2018PhRvL.120b1101A}, and also N \citepads{2018PhRvL.121e1103A}---they were analysed from the data taking period\footnote{This is at variance with \citetads{2019ApJ...873...77N}, who chose to allow for different force-field modulation values. As there is no clear motivation to allow for these differences for similar species (same $A/Z$), we believe that their conclusions are misleading.}.

 We stress that choosing consistent values for AMS-02 data, \citetads{2019ApJ...873...77N}, i.e. from May~2011 to May~2016. At variance, \het{} data were taken on a slightly longer period \citepads{2019PhRvL.123r1102A}, from May~2011 to November~2017, leading to an estimated Solar modulation difference of $\sim20$~MV. As \hettohef{} is only used for validation and in order not to add a new degree of freedom for this species, we enforce a single nuisance parameter $\phi_{\rm prior} = 676$~MV for all AMS-02 data used in this analysis.

\subsubsection{Post-fit check of nuisance parameters, \chipernui{}}
It is useful to define the specific contribution, \chinui{}, of the nuisance parameters to the overall $\chi^2$ given in Eq.~(\ref{eq:chi2}). Considering $n_{nui}= n_s+n_x$ nuisance parameters, $n_s$ for Solar modulation and $n_x$ for cross-section, we define
\begin{equation}
  \chipernui{} \equiv \left(\sum_{s=0}^{n_s}{\cal N}^s_{\rm Sol.Mod.} + \sum_{x=0}^{n_x}{\cal N}^{x}_{\rm XS}\right)/(n_s+n_x),
  \label{eq:chi2nuis}
\end{equation}
which is used as an {\em a posteriori} validation of the fit.

Because the number of degrees of freedom---number of data minus the number of free parameters---is quite large ($\sim 200)$ and the number of nuisance parameters smaller ($\lesssim 10-20$), situations in which $\chipernui{}>1$ for a very good fit $\chimindof{}\sim 1$ can arise. The value of \chipernui{} directly tells us how many $\sigma$ away (from their expected value) nuisance parameter post-fit values are. We control for each fit that these values are $\lesssim 1$.

\section{Combined analysis of Li/C, Be/C, and B/C (LiBeB)}
\label{sec:LiBeB}
The AMS-02 data considered in this section are ratios of Li, Be, and B to C (or O), coming from the same data taking period (2011-2016) and publication \citepads{2018PhRvL.120b1101A}.

\subsection{Preliminary remarks}
\label{sec:remarks}
\begin{figure}[t]
  \includegraphics[width=\columnwidth]{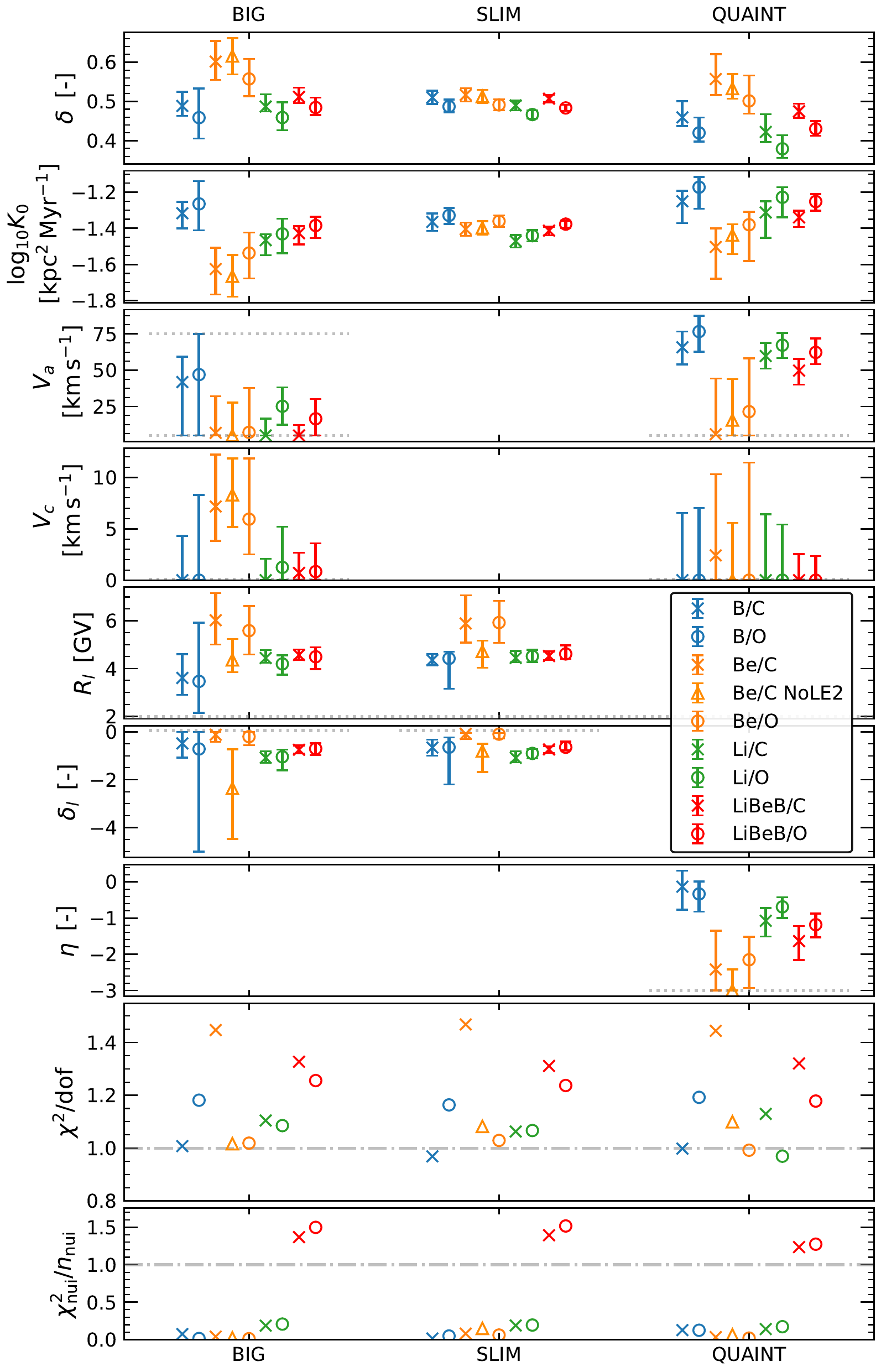}
  \caption{Best-fit transport parameters with their asymmetric uncertainties (from \minos{}) and corresponding \chimindof{} and \chipernui{} (next-to-last and last panels) for \BIG{}, \SLIM{}, and \QUAINT{} (first, second, and third column). Different secondary species are colour-coded, with different symbols for ratios to C (crosses) and to O (circles): `Be/C NoLE2' stands for Be/C AMS02 data without the two lowest rigidity points (orange triangles), and LiBeB/C (or /O) stands for the combined analysis of Li/C, Be/C, and B/C (or /O). See text for details.}
  \label{fig:fit_LiBeB}
\end{figure}
A first issue to consider is whether to analyse $x/$C or $x/$O ratios (with $x=$~Li, Be, or B). For the analysis of transport parameters, both should encode roughly the same information and provide the same results. In principle, using O would be a better choice, because it is a `pure' primary, but standard practice so far has been to use C, which has at most a $\sim 20\%$ secondary contribution at a few GeV/n \citepads[e.g.,][]{2018PhRvC..98c4611G}.

As a first consistency check, we show in Fig.~\ref{fig:fit_LiBeB} the best-fit transport parameters (and their $1\sigma$ asymmetric error bars) from the fit of Li$/y$ (green), Be$/y$ (orange), B$/y$ (blue) with $y=$~C (crosses) or $y=$~O (circles). As expected, consistent results at the $\sim 1\sigma$ level are obtained for all models, though some small deviations exist. The difference could be related to a mis-modelling of the production cross section of carbon isotopes (not taken as nuisance parameters here): a $\sim 15\%$ cross-section uncertainty would translate into a peak uncertainty on C of a few percent at GV rigidities, at the level of the $\sim3\%$ data uncertainty of AMS-02 data.

A bit more puzzling is the large \chimindof{} difference ($\sim 0.4$) between Be/C and Be/O, which is significant in all models (orange crosses vs circles in the bottom panel of Fig.~\ref{fig:fit_LiBeB}). The origin of this difference is attributed to the presence of an upturn in the low-rigidity Be/C data (orange symbols in the top panel of Fig.~\ref{fig:LiBeB_model_vs_data}). As a consistency check, we fitted Be/C without the two lowest-rigidity data points and find an excellent \chimindof{} (orange triangles, dubbed `Be/C NoLE2' in Fig.~\ref{fig:fit_LiBeB}). This procedure also brings the high-rigidity transport parameters in slightly better agreement (compare the orange crosses with their circle and triangle counterparts). Whether this is physically meaningful is difficult to say. With a better modelling of AMS-02 systematics, the situation could probably be clarified. If it remains, the presence of an upturn could hint at some subtle but important physics effect not considered yet.

We conclude that there is no significant difference using C or O (in the ratio), and in the following, we stick to the standard practice of using $x/$C ratios. However, in the combined analyses below, the reader should keep in mind that whenever Be/C data are included, the \chimindof{} is slightly `degraded' compared to the results we would have obtained using Be/O. It should not straightforwardly be interpreted as a poorer quality of the model.

\subsection{Separate vs combined analysis}
\begin{figure}[t]
  \includegraphics[width=\columnwidth]{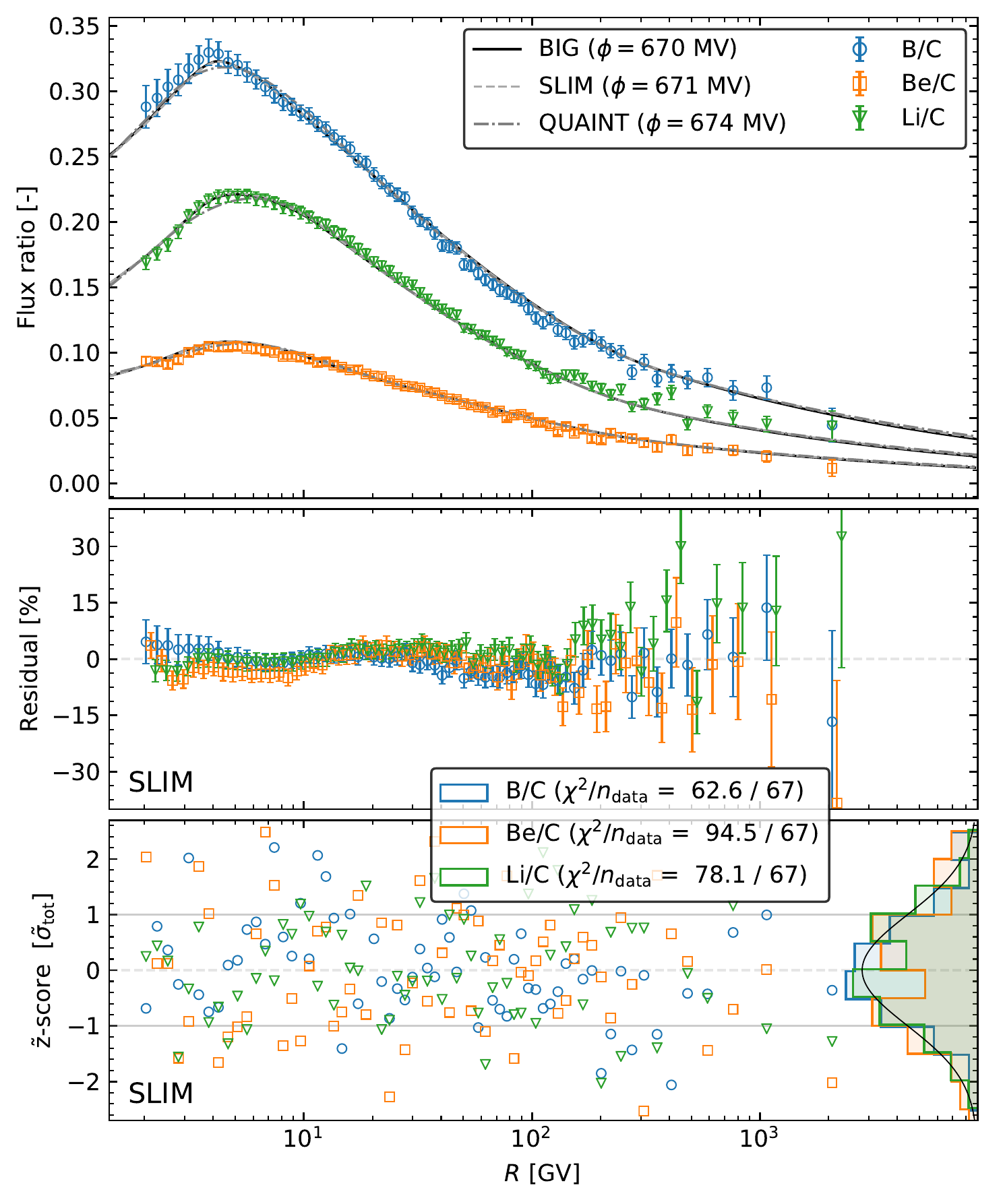}
  \caption{Flux ratios (top), residuals (centre) and $\tilde{z}$-scores (bottom) for B/C (blue circles), Be/C (orange squares), and Li/C (green triangles). The models (top panel) are calculated for \BIG{} (solid line), \QUAINT{}, and \SLIM{} (dashed-dotted line), from the best-fit transport parameters of the combined analysis of all three species. The residuals and $\tilde{z}$-score are shown for \SLIM{} only---the $\tilde{z}$-score is related to the usual $z$-score by a rotation in a base where the covariance matrix of data systematics is diagonal (see text for details).}
  \label{fig:LiBeB_model_vs_data}
\end{figure}
The cross symbols in Fig.~\ref{fig:fit_LiBeB} show the best-fit transport parameters from the analysis of AMS-02 Li/C (green), Be/C (orange), and B/C (blue) separately. The parameters are compatible at the $1\sigma$ level for all models for Li/C and B/C. For Be/C, the agreement is also very good for \SLIM{}, but with small deviations for \BIG{}, and \QUAINT{}. The difference is not significant enough to conclude on a preference for \SLIM{}.

As the data uncertainties have the same status (see App.~\ref{app:nuis_xs}), we can perform a combined analysis of the three species (red symbols) without further complication. The combined fit leads to slightly tighter constraints on the transport parameters (red symbols in Fig.~\ref{fig:fit_LiBeB}), except for \BIG{}---probably because of degeneracies between its too many parameters. The \chimindof{} are acceptable\footnote{We remind that we would obtain much better \chimindof{} without including the first two data points of Be/C.} (see also Table~\ref{tab:LiBeBC_parameter_results}), though we observe a jump of \chipernui{} from $\sim0$ (in single-species fits) to $\sim 1.5$ (in the combined fit). This is attributed to cross-section nuisance parameters wandering away from their input values. Indeed, the partial degeneracy between the diffusion coefficient normalisation $K_0$ and the production cross section is lifted in the simultaneous fit: $K_0$ is enforced to be the same for all species, so that cross-section degrees of freedom are now used, adding a penalty in \chipernui{}. We come back to this very important issue in Sect.~\ref{sec:aboutxs}.

The resulting best-fit ratios are shown against the data in Fig.~\ref{fig:LiBeB_model_vs_data}. The top and middle panels show the total uncertainties (statistical plus systematics in quadrature), and as such, they overestimate the real uncertainties used in the $\chi^2$ analysis. As introduced in \citetads{Boudaud:2019efq}, a graphical representation of the `rotated' score (denoted $\tilde{z}$-score) provides an unbiased view accounting for the role of correlations in data systematics. The rotated base is defined so that the covariance matrix of uncertainties is diagonal,
\begin{equation}
\tilde{\cal C} = U \, {\cal C} \, U^{\rm T}\,,
\label{eq:rotmat}
\end{equation}
with $U$ an orthogonal rotation matrix. In this new base, we define the rotated residual
\begin{equation}
\tilde{z}_{i} = \tilde{x}_{i}/\tilde{\sigma}_{i},
\end{equation}
with the rotated difference and rotated diagonal systematics respectively defined to be
\begin{equation}
\tilde{x}_{i} \equiv \sum_j U_{ij} ({\rm model}_j-{\rm data}_j) \quad {\rm and} \quad \tilde{\sigma}_{i}\equiv \sqrt{\cal C}_{ii}\,.
\end{equation}
In the rotated base, rigidities are replaced by pseudo (or rotated) rigidities, defined to be
\begin{equation}
\tilde{R}_{i} = \sum_{j} U_{ij}^{2} \, R_{j}\,.
\end{equation}
Because the rotation is small, with $U$ close to unity, the pseudo rigidity $\tilde{R}_{i}$, for the case of AMS-02 data, is not very different from the physical value $R_{i}$.

The $\tilde{z}$-score is shown in the bottom panel of Fig.~\ref{fig:LiBeB_model_vs_data} as a function of $\tilde{R}$ for Li/C, Be/C, and B/C. By construction
\begin{equation}
\chi^2 = \sum_{i} \tilde{z}_{i}^{2}\,,
\label{eq:chi2rot}
\end{equation}
and we can also build an histogram of $\tilde{z}_{i}$ values, as shown on the right-hand side of the panel. The latter should follow a centred Gaussian of width unity if the model matches the data. This is indeed mostly the case, except for Be/C (orange line) which has too large tails (we recall that the deviation is partly driven by the two Be/C low-energy data points). More quantitatively, the distance between the model and data from the global fit is $\chi^2_{\rm B/C}=62.8$, $\chi^2_{\rm Li/C}=78.5$, and $\chi^2_{\rm Be/C}=95.7$ for 68 data points each.

\subsection{Updated benchmark models and low-rigidity break}

\begin{table}[t]
\caption{Values of best-fit transport parameters (and $1\sigma$ uncertainties) from the combined analysis of Li/C, Be/B, and B/C AMS-02 data.}
{
\label{tab:LiBeBC_parameter_results}
\begin{tabular}{crrr}
\hline\hline
  Parameter [unit]\!\!\!\! & {BIG }  & {SLIM }  & {QUAINT } \\  \hline \\[-1em]
 \multicolumn{4}{c}{Intermediate-rigidity parameters}\\[5pt]
  $\log_{10} \left(\frac{{K}_0}{1~{\rm GV}}\right)^\dagger$ \,\,[-] & $ -1.43^{+  0.04}_{-  0.06}$ & $ -1.41^{+  0.02}_{-  0.03}$ & $ -1.34^{+  0.04}_{-  0.05}$\\ [5pt]
  $\delta$ \, [-]  & $ 0.51^{+ 0.02}_{- 0.02}$ & $ 0.51^{+ 0.01}_{- 0.01}$ & $ 0.47^{+ 0.02}_{- 0.02}$\\ [10pt]
 \multicolumn{4}{c}{Low-rigidity parameters}\\[5pt]
  $V_c\,\,\mathrm{[km\,s^{-1}]}$ & $ 0.7^{+ 2.0}_{- 0.9}$ & n/a    & $ 0.0^{+ 2.5}$\\ [5pt]
  $V_a\,\,\mathrm{[km\, s^{-1}]}$  & $ 0.0^{+ 12.2}$ & n/a    & $49.7^{+ 8.1}_{- 9.7}$\\ [5pt]
  $\eta$ \, [-] & 1 (fixed)  & 1 (fixed)  & $-1.64^{+ 0.42}_{- 0.52}$\\[5pt]
  $\delta_l$ \, [-] & $ -0.75_{- 0.16}^{+ 0.18}$ & $ -0.74_{- 0.14}^{+ 0.13}$ & n/a   \\ [5pt]
  $R_l$ \, [GV] & $  4.57^{+  0.23}_{-  0.20}$ & $  4.53^{+  0.19}_{-  0.18}$ & n/a   \\ [10pt]
 \multicolumn{4}{c}{$\chi^2$ indicators$^\ddagger$}\\[3pt]
  \chimindof{} & $258.8^{\star}/195$ & $258.3^{\star}/197$ & $258.8/196$\\[0pt]
  \chipernui{} & $21.9/16~~$ & $22.3/16~~$ & $19.8/16~~$ \\[2pt]
  \hline
\end{tabular}
\tablefoottext{$\dagger$}{$K_0$ is in $\mathrm{[kpc^2\, Myr^{-1}]}$.}\\
\tablefoottext{$\ddagger$}{$\chi^2_{\rm min}$ and \chinui{} are defined in Eqs.~(\ref{eq:chi2}) and (\ref{eq:chi2nuis})}.\\
\tablefoottext{$\star$}{$\chi^2_{\rm min}$ for \BIG{} and \QUAINT{} are not strictly equal because of their slightly different (fixed) high-rigidity parameters.}
}
\end{table}

We gather in Table~\ref{tab:LiBeBC_parameter_results} the best-fit transport parameters obtained from the combined analysis of LiBeB AMS-02 data. Compared to our previous analysis of B/C ratio only \citepads{2019PhRvD..99l3028G}, several differences are worth noting.

\paragraph{Intermediate-rigidity parameters ($K_0,\delta$)}
For \SLIM{}, the combined analysis leads to a similar value of the diffusion slope, though with smaller uncertainties ($\delta=0.51\pm0.02$). For the two other models with more free parameters (partially degenerated), the present analysis gives slightly larger values, so that all models now converge to the same $\delta$ value. In the combined analysis, the diffusion coefficient normalisation is $K_{10}/L\approx (0.024,0.025,0.027)\pm0.004$ for (\BIG{},\,\SLIM{},\,\QUAINT{}) to compare to $K_{10}/L\approx(0.030,0.028,0.033)\pm0.003$ in the B/C-only analysis \citepads{2019PhRvD..99l3028G}\footnote{Beware that $L=5$~kpc here, whereas $L=10$~kpc in \citetads{2019PhRvD..99l3028G}. We used $K_{10}=10^{\log_{10}(K_0) + \delta}$ to convert values of Table~\ref{tab:LiBeBC_parameter_results}.}. There is a significant trend towards lower values in all models ($\sim10\%$). This is another illustration of the fact that production cross-section degrees of freedom (nuisance parameters) are now used in the combined fit. Somehow, a smaller production cross section was required for B, leading to a smaller value for $K_0$ (actually $K_0/L$). We discuss the meaning of the derived cross-section values in Sect.~\ref{sec:aboutxs}.

\begin{figure}[t]
  \includegraphics[width=\columnwidth]{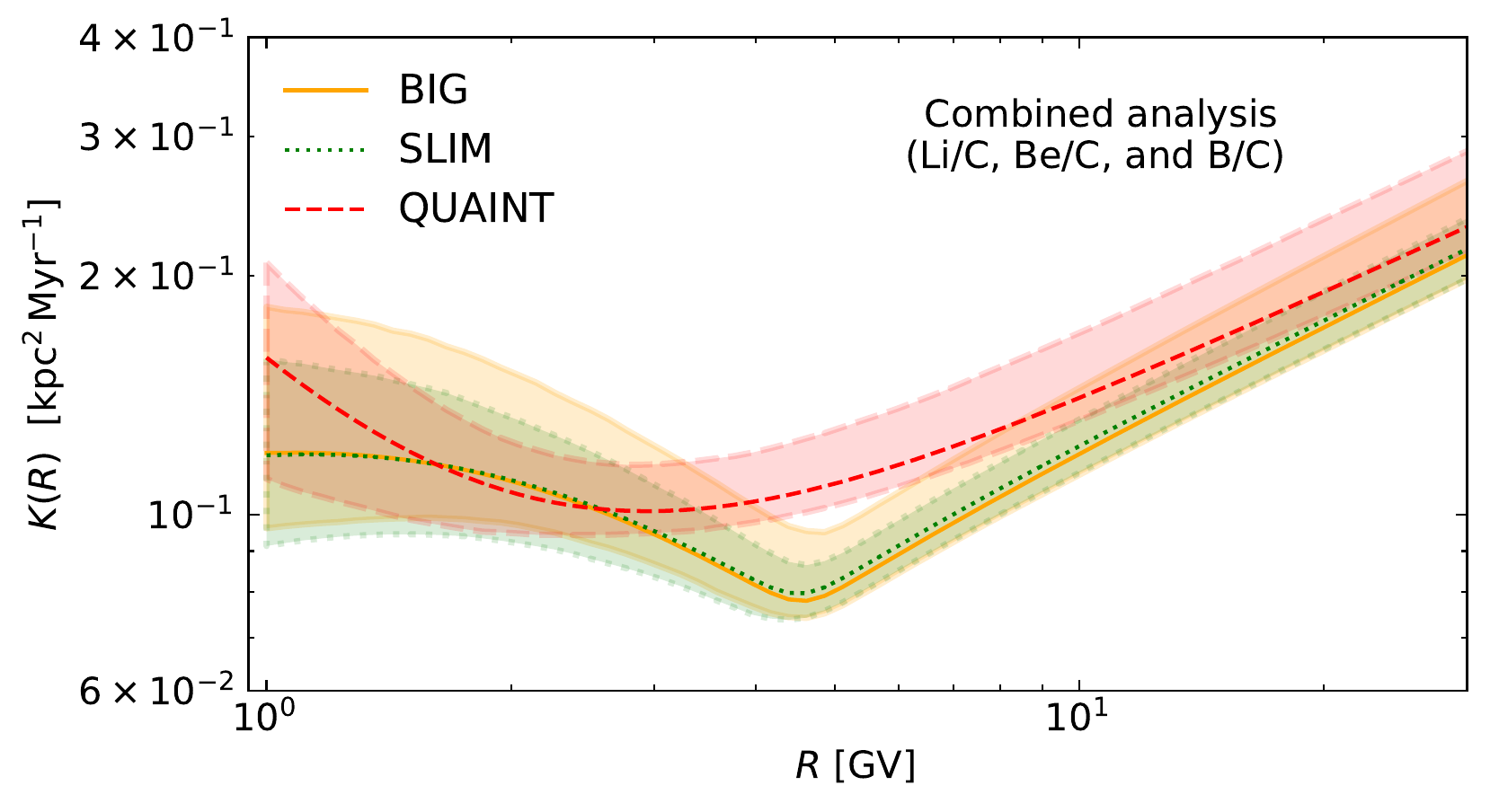}
  \caption{Best-fit and $1\sigma$ contours for $K(R)$, see Eq.~(\ref{eq:def_K}), reconstructed from the best-fit transport parameters (and their covariance matrix) from the combined LiBeB analysis. Three models are shown: BIG{} (orange), \SLIM{} (green), and \QUAINT{} (red). In the low-rigidity range, the factor $\beta$ in Eq.~(\ref{eq:def_K}) makes $K(R)$ dependent on the CR species, and we show here the result for $A/Z=2$. See text for discussion.}
  \label{fig:KR_BSQ}
\end{figure}
\paragraph{Low-rigidity parameters ($V_c,V_a,\eta,\delta_l,R_l$)}
There is also a significant change with respect to the B/C-only analysis in this regime. A low-rigidity break is well-identified at $R_l\approx 4.6\pm 0.3$ with $\delta_l\approx 0.63\pm0.3$ for both \BIG{} and \SLIM{}, whereas \BIG{} was compatible with no break in the B/C-only analysis \citepads{2019PhRvD..99l3028G}. The diffusion coefficient in \QUAINT{} only enables a modification in the sub-relativistic regime---$\beta^\eta$ term in Eq.~(\ref{eq:def_K})---and with $\eta\approx -1.6\pm0.5$, a clear upturn is observed compared to the B/C-only analysis (compatible with $\eta=0$). Concerning reacceleration, whereas \BIG{} could find best-fit regions with large $V_a$ (up to 80~km~s$^{-1}$ in the B/C analysis), the combined analysis shrinks \BIG{} towards \SLIM{} (neither reacceleration, nor convection); even in \QUAINT{} the need for reacceleration is halved.

\paragraph{Low-rigidity break}
Figure \ref{fig:KR_BSQ} shows $1\sigma$ contours of the diffusion coefficient\footnote{See \citetads{2019PhRvD..99l3028G} for the high-rigidity behaviour. At variance with the latter paper, where contours were defined as the overall envelopes obtained from varying the correlated transport parameters within $1\sigma$, we calculate here at each rigidity $R_i$ the $1\sigma$ range from the distribution of $K(R_i)$.}, providing a direct illustration of the low-rigidity break: the preference of a break or upturn is significant in all configurations. We investigate in App.~\ref{app:LE-data} whether low-energy data (ACE-CRIS data \citealtads{2009ApJ...698.1666G}) can provide similar but independent conclusions. They do not, but are nevertheless in broad agreement with models derived from the AMS-02 LiBeB constraints.
The low-rigidity diffusion upturn is thus supported by three observations: (i) the combined analysis of various species points towards a break or upturn at GV rigidities; (ii) low-energy data are consistent with the presence of an upturn; (iii) although \BIG{} has the largest number of free parameters, its parameter space prefers to shrink to that of the minimal configuration \SLIM{}, which favours a low-rigidity break.

\subsection{Propagation uncertainties in benchmark models}

From the previous analysis (combined fit of Li,/C, Be/C, and B/C AMS-02 data), we assess the propagation uncertainties on calculated secondary-to-primary ratios.
From the best-fit transport (and nuisance parameters) and their correlations, we draw $N$ realisations of the parameters, calculate the associated CR fluxes, and extract from their distribution the desired quantiles (in each rigidity bin) for any species.

\begin{figure}[t]
  \includegraphics[width=\columnwidth]{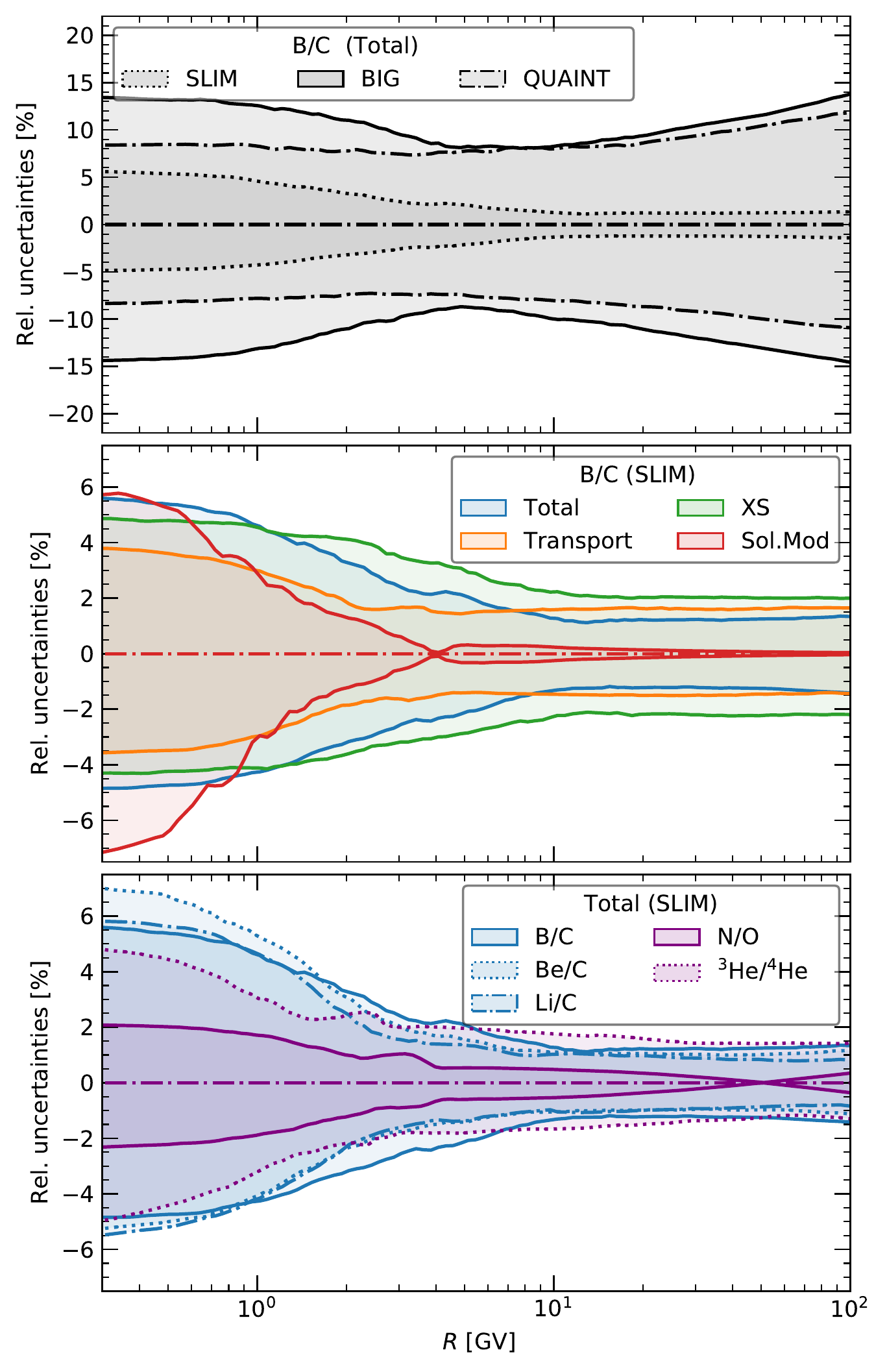}
  \caption{Propagation relative uncertainties ($1\sigma$ contours w.r.t. median) as a function of rigidity based on the constraints set by the combined LiBeB analysis. {\em Top panel}: comparison of total transport uncertainties from \BIG{} (solid), \SLIM{} (dashed), and \QUAINT{} (dash-dotted). {\em Middle panel}: separate uncertainty contributions from nuclear cross sections (green), Solar modulation (red), and transport (range) in \SLIM{}. {\em Bottom panel}: comparison (in \SLIM{}) of various calculated secondary-to-primary ratios within the combined LiBeB constraint.}
  \label{fig:model_uncertainties}
\end{figure}
We show in Fig.~\ref{fig:model_uncertainties} the $68\%$ contours ($1\sigma$) w.r.t. the median for various models (top panel), various parameters (middle panel), and various secondary-to-primary ratios (bottom panel). In the configuration with the fewest number of free parameters (\SLIM{}), the model uncertainties are at the level or even below the data uncertainties shown in Fig.~\ref{fig:sub_error_diags}. At variance, the larger the number of parameters the more degenerate the configuration, so that \QUAINT{} and \BIG{} model uncertainties encompass the data uncertainties. Nevertheless, for all transport configuration, the rigidity at which the calculation is best constrained is at $\sim 10$~GV, corresponding to the region where AMS-02 data uncertainties are minimum\footnote{The uncertainties at 100~GV are underestimated as we do not account here for the high-rigidity break uncertainties \citepads{2019PhRvD..99l3028G}.}.

The total model uncertainties account for model parameter correlations, and include transport, cross sections, and Solar modulation. The middle panel of Fig.~\ref{fig:model_uncertainties} shows how each of these ingredients contribute to the error budget: below 1~GV, solar modulation uncertainties are dominant, then cross-section uncertainties are dominant from $\sim 1$ to $\sim 10$~GV, and transport and cross-section uncertainties are equally important above $\sim10$~GV. Because of correlations, the total uncertainties (blue contours) are smaller than the individual ones.

The bottom panel of Fig.~\ref{fig:model_uncertainties} illustrates the fact that, once the propagation parameters have been constrained by some specific secondary-to-primary ratios, all other similar ratios contribute the same level of modelling uncertainties\footnote{The N/O crossing point at 50~GV is artificial and comes from the enforced normalisation of the total N flux to fix its primary component.}. This is not surprising as the baseline modelling uncertainty is from the transport parameters, which applies to all propagated species.

\subsection{Discussion}

It is interesting to compare our results to those of \citetads{2019PhRvD..99j3023E}. These authors rely on a similar 1D propagation model, but discard reacceleration arguing that in general, it is incompatible with the models based on self-generated waves that they consider. They first take AMS-02 fluxes for B, C, N, and O in combinations differing from ours, from which they obtain $\delta=0.63$, $V_c=7~{\rm km}~{\rm s}^{-1}$, and $K_0=1.1\times 10^{28}~{\rm cm}^2~{\rm s}^{-1}$, the latter corresponding to $K_{10}/L=0.039~{\rm kpc}~{\rm Myr}^{-1}$. This is to compare with the typical values we find $\delta \in [0.45, 0.53]$ and $K_{10}/L \in [0.020,0.031]$kpc.Myr$^{-1}$ from our configurations (reacceleration or pure diffusion). \citetads{2019PhRvD..99j3023E} do not provide an estimate of the uncertainty, but our results seems difficult to reconcile with theirs\footnote{Given that the normalisation $K_0$ is degenerate with $\mu$, the gas surface density, different $K_0$ values from different studies can sometimes be understood as the use of different $\mu$ in the models \citepads{2010A&A...516A..67M}. However, different $\delta$ can hardly be reconciled.}. The origin of the difference may be related to the fact that these authors do not account for a low-rigidity break and only fit the data above 10 GV. It may also be related to the fact that using more primary data than secondary ones in the fit biases the determination of the transport parameters (see comments above). We underline that while their parameters do not lead to good fits to H and He data, ours do, both at low rigidity as shown here, or at higher rigidities as checked in \citetads{Boudaud:2019efq}.

Another interesting comparison can be achieved with the results of \citetads{2020ApJ...889..167B}. In this latter the authors use the \galprop{} and HelMod codes \citepads{BOSCHINI20192459} to fit absolute fluxes of Li, Be, B, C, N and O, as well as the B/C ratio. While their main focus is on the interstellar spectrum and on the hypotheses around the high rigidity break, we can still compare their best fit for the intermediate rigidity parameters. They obtain $\delta=0.415\pm 0.025$ and $K_{10}/L\approx 0.047\pm0.008~{\rm kpc}~{\rm Myr}^{-1}$, which stands many sigma away from ours. This can be explained by the methodology which is quite different: mainly, they do not allow for low-rigidity break in the diffusion coefficient and do not treat the systematic uncertainties with covariance matrices as we do. Interestingly the authors also report an overproduction of Be and a deficit of Li at high energies. Since we have chosen to use nuisance parameters to handle nuclear cross-sections uncertainties, we do not experience the same difficulty, and show that mild variation (within the current uncertainties) of the normalisation can resolve this tension---to be specific,  \citetads{BOSCHINI20192459} add primary Li to correct for a 20-25\% deficit, while we increase the total Li production cross sections by $13\%$ (see Sect.~\ref{sec:aboutxs}).

\section{Accommodating \hettohef{} and N/O data}
\label{sec:NO_3He4He}

There are two other AMS-02 ratios possibly relevant to our study, namely the isotopic ratio \hettohef{} \citepads{2019PhRvL.123r1102A}, and the partly primary N/O ratio \citepads{2018PhRvL.121e1103A}. For reasons discussed below, these ratios have specificities that prevent them to be readily employed in a combined analysis. However, they still provide useful constraints complementary to the ones set by Li, Be, and B.

\subsection{Motivations and complications}

\subsubsection{Fitting N/O}

The N/O ratio evolves from a secondary N fraction of $\sim 70\%$ at a few GV to $\lesssim 30\%$ above 1~TV \citepads{2018PhRvL.121e1103A}. In principle, it is not ideal to study transport parameters, as it is no longer independent from source parameters. Yet, it enables a test of the universality of transport for a heavier species and is worth considering.

As shown in \citetads{2019PhRvD..99l3028G}, primary fluxes and in particular C and O AMS-02 data can be nicely reproduced in our model, by fitting the transport parameters on B/C and then using a simple universal power-law source spectrum (not fitted). We checked that C and O AMS-02 data are also reproduced in our combined LiBeB analysis. The primary content of N is thus expected to be correctly predicted using the same source spectral index as that of C and O. With this caveat, fitting N/O is expected to bring complementary constraints on the transport parameters, all the more because their broken-down data uncertainties are similar to those of Li/C, Be/C and B/C (see App.~\ref{app:cov_mat}). In all analyses involving N/O data below, an extra nuisance parameter for the production cross section of N is added (see Table~\ref{tab:xs_nuis}).

\subsubsection{Fitting \het{} and \hef{}}
Waiting for AMS-02 deuterium data, the pure secondary \het{} is the best option to test the universality of transport towards lighter nuclei. In addition, the \hettohef{} ratio was shown to provide complementary and competitive constraints compared to those obtained from B/C \citepads{2012A&A...539A..88C,2012Ap&SS.342..131T,2019PhLB..789..292W}.
However, with the high precision of AMS-02 data \citepads{2019PhRvL.123r1102A} on an unprecedented energy range, directly fitting \hettohef{} is no longer recommended. As explained below, unbiased conclusions can only be reached by fitting simultaneously \het{} and \hef{}.

All production cross sections are assumed to follow the straight-ahead approximation, in which the fragment carries out the same kinetic energy per nucleon as the parent. On the other hand, both Solar modulation and diffusion have a similar impact on species with the same $R$.  With $R\approx(A/Z)\sqrt{E_{k/n}(E_{k/n}+2m_p)}$, species with similar $A/Z$ at a given $E_{k/n}$ also have similar $R$ (and vice-versa). As a result, ratios of such secondary-to-primary species are independent of the source spectra when taken per $E_{k/n}$ or $R$, and all feel the same transport at a given $E_{k/n}$ or $R$; an example is B/C, with both the dominant $^{10}$B and $^{12}$C contributions having $A/Z=2$. This is no longer the case for \hettohef{}, having respectively $A/Z=1.5$ and 2, so that neither the fit vs $E_{k/n}$ or $R$ is appropriate. Hence, contrarily to the analysis of LiBeB, we cannot directly fit \hettohef{} and are forced to fit both \het{} and \hef{} simultaneously.

To ensure that \hef{} data are reproduced correctly, we allow for a flexible enough parametric formula for the  \hef{} source term. These extra parameters are added in the combined fit---in practice, the best-fit source spectrum is close to a pure power law. Given that the number of data points is small and that the latter spread over a limited energy range (22 and 25 data for \het{} and \hef{} from 3 to 20~GV), the determination of the transport parameters is expected to remain driven by the LiBeB combined data (205 data points from 3~GV to 2~TV). With these provisos, the fit on \het{} provides a complementary constrain on the transport parameters. To account for uncertainties in the production cross section of \het{}, an extra nuisance parameter is added in the fit (see Table~\ref{tab:xs_nuis}) whenever \het{} data are considered.

Owing to the experimental challenges of separating these isotopes, the two high-precision datasets at hand are those of AMS-02 \citepads{2019PhRvL.123r1102A} and PAMELA \citepads{2013ApJ...770....2A,2016ApJ...818...68A}. The BESS-Polar~II data are not considered here, because only preliminary analyses are available \citepads{2015ICRC...34..425P,2017ICRC...35..210P}. For the sake of consistency, only AMS-02 data are fit, but PAMELA data are used for post-fit visual inspection. The choice remains as whether to fit AMS-02 data as a function of kinetic energy per nucleon or rigidity, both being provided in \citetads{2019PhRvL.123r1102A}. We argued in \citetads{2019A&A...627A.158D} that opting for one or the other brought different systematic uncertainties in the B/C case (because of the unknown isotopic content of the elements biasing the conversion). However, here, the conversion to go from different energy types is exact for \het{} and \hef{}. In the AMS-02 analysis \citepads{2019PhRvL.123r1102A}, \het{} and \hef{} fluxes and systematics are provided as a function of kinetic energy per nucleon, from three separate sub-detectors. A covariance matrix of uncertainties can be built either in $E_{k/n}$ or $R$ (see App~\ref{app:covHe}), so that any of them can be used with the same end result.

\subsection{Transport parameters from LiBeB, N/O, $^3$He, and $^4$He}
\begin{figure}[t]
  \includegraphics[width=\columnwidth]{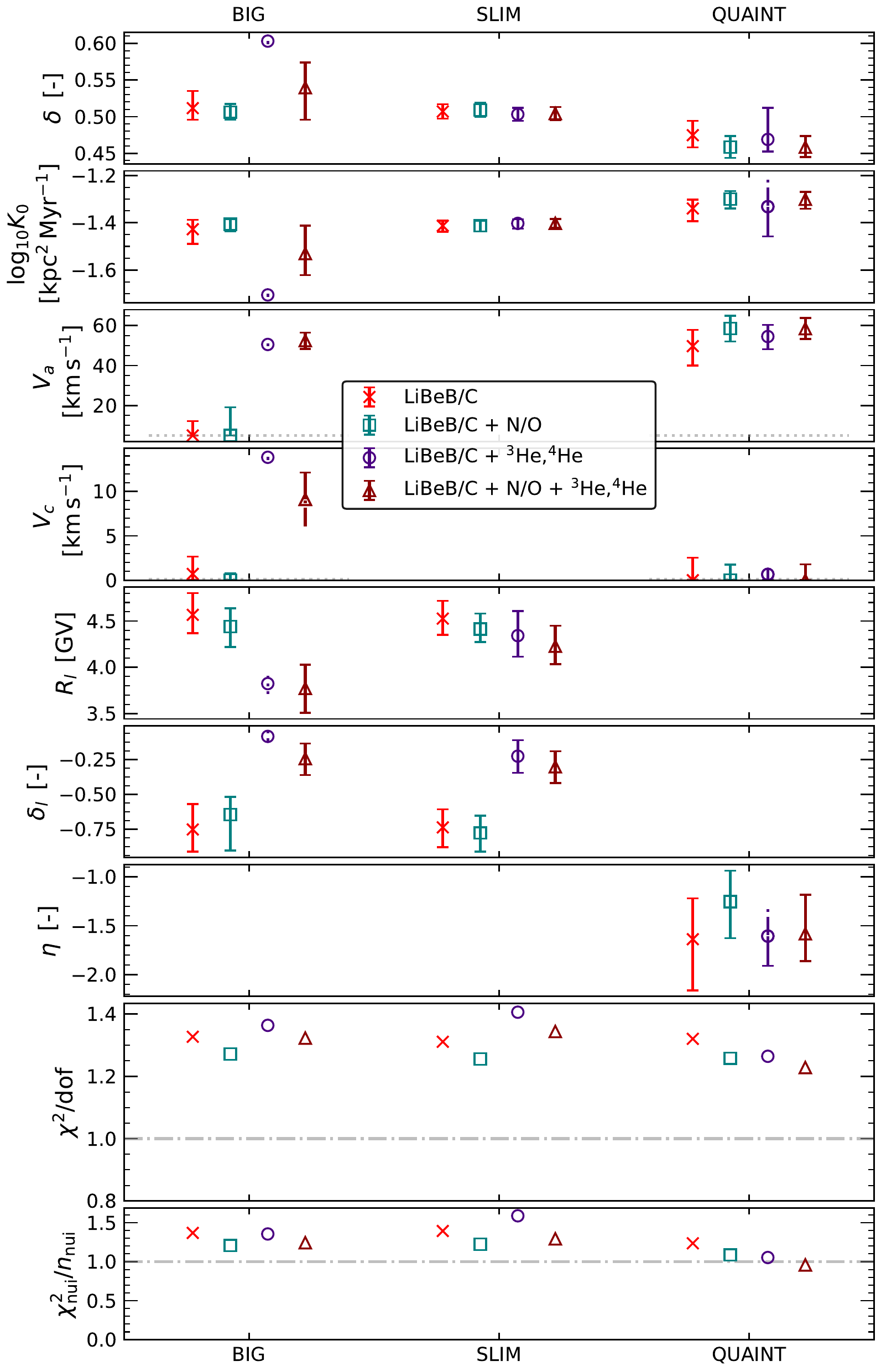}
  \caption{Best-fit transport parameters and uncertainties with the corresponding \chimindof{} and specific contribution from nuisance parameters (next-to-last and last columns) for different combinations of AMS-02 data: combined fit of Li/C, Be/C, and B/C (or LiBeB for short, red crosses); LiBeB and N/O combined (turquoise squares); LiBeB and \hettohef{} combined (purple circles);  LiBeB, \hettohef{}, and N/O combined (brown triangles). From left to right, models \BIG{}, \SLIM{}, and \QUAINT{}.}
  \label{fig:fit_LiBeBHeN}
\end{figure}
Figure~\ref{fig:fit_LiBeBHeN} shows the best-fit parameters from the simultaneous analysis of AMS-02 Li/C, B/C, Be/C (red crosses), further combined with N/O (turquoise squares) or \het{} and \hef{} (violet circles), or with both (brown triangles). 

For all configurations (\BIG{}, \SLIM{}, and \QUAINT{}), adding N/O in the fit (turquoise squares) improves the \chimindof{} and \chipernui{} values and better constrains the transport parameters. This indicates that N/O data were already consistent with the results of the LiBeB combined fit. The situation is less clear-cut when adding \het{} and \hef{}. Whereas the data are easily accommodated for by the model for \QUAINT{} (same \chimindof{} and parameter values), the fit is not so good for \SLIM{}, with an increased \chimindof{} and \chipernui{} and a $\delta_l$ marginally compatible with the LiBeB analysis. Even more significant is the behaviour of \BIG{}: whereas in the case of the LiBeB combined analysis, \BIG{}'s parameter space shrank towards \SLIM{}, the few \het{} data (compared to LiBeB ones) completely shift \BIG{}'s parameter space towards \QUAINT{} (with reacceleration and also convection). This behaviour remains the same for the combined LiBeB, N/O, \het{}, and \hef{} combined fit (brown triangles). This demonstrates the strong sensitivity of \het{} to the low-rigidity transport parameters.

\begin{figure}[t]
  \includegraphics[width=\columnwidth]{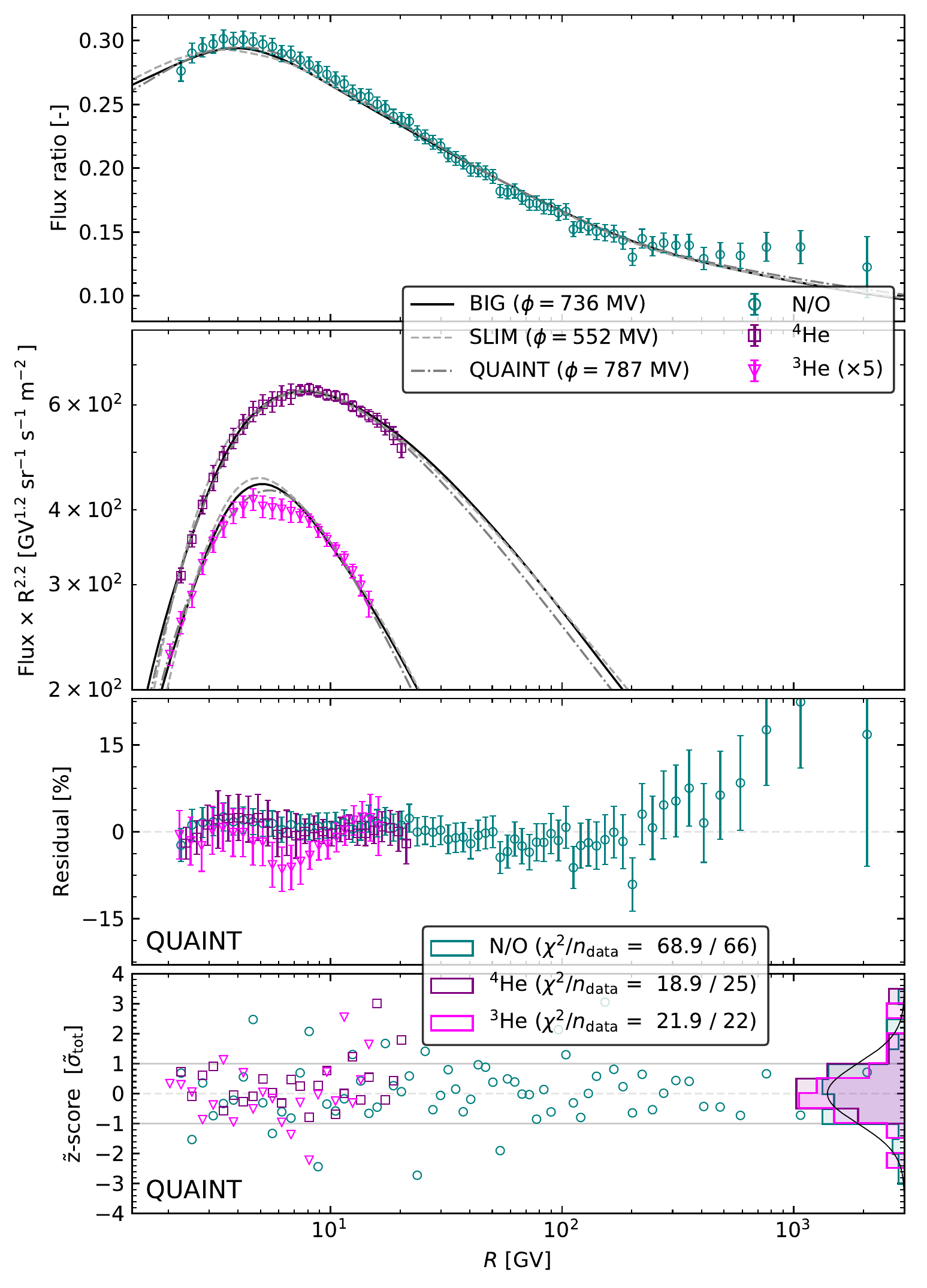}
   \caption{Model calculation and data (first and second panel), residuals (third panel), and $\tilde{z}$-score (fourth panel) for N/O (turquoise circles), \hef{} (red squares), and \het{} (pink triangles)---for readability, He isotopic fluxes are multiplied by $R^{2.2}$. The two top panels show the model calculations for \BIG{}, \SLIM{}, and \QUAINT{} along with their respective post-fit modulation level, whereas the two bottom panels are restricted to the best-fit configuration \QUAINT{}. The insert shows the distance between the model and specific datasets, calculated from $\chi^2=\sum_i \tilde{z_i}$. All models are calculated from the best-fit parameters of the combined analysis of Li/C, Be/C, B/C, N/O, \het{}, and \hef{}.}
  \label{fig:LiBeB_model_vs_data_check}
\end{figure}
Figure~\ref{fig:LiBeB_model_vs_data_check} shows a comparison between the models and data (top panels) and the corresponding residual and $\tilde{z}$-score (\QUAINT{} only, bottom panels) for the combined fit of AMS-02 Li/C, Be/C, B/C, N/O, \het{}, and \hef{} data. The above conclusions and goodness of fit to the data are illustrated in the various panels. All models fit equally well N/O data (top panel) and \hef{} data (second panel). For the latter, the very good fit $\chi^2_{\hef{}}/n_{\rm data}=(22.7,\,23.8,\,18.9)/25$ for (\BIG{}, \SLIM{}, \QUAINT{}) validates the procedure depicted in the previous section, i.e. it ensures the fit of \het{} is based on the correct spectrum of its main progenitor. We obtain $\chi^2_{\het{}}/n_{\rm data}=(25.5,\,41.4,\,21.9)/22$ for (\BIG{}, \SLIM{}, \QUAINT{}). All models (second panel) slightly overshoot the data at a few GV, with \QUAINT{} giving a very good fit, followed by \BIG{}, but with \SLIM{} giving a poor fit.

\subsection{Impact of data correlation length}
\label{sec:corrlength}

\begin{figure}[t]
  \includegraphics[width=\columnwidth]{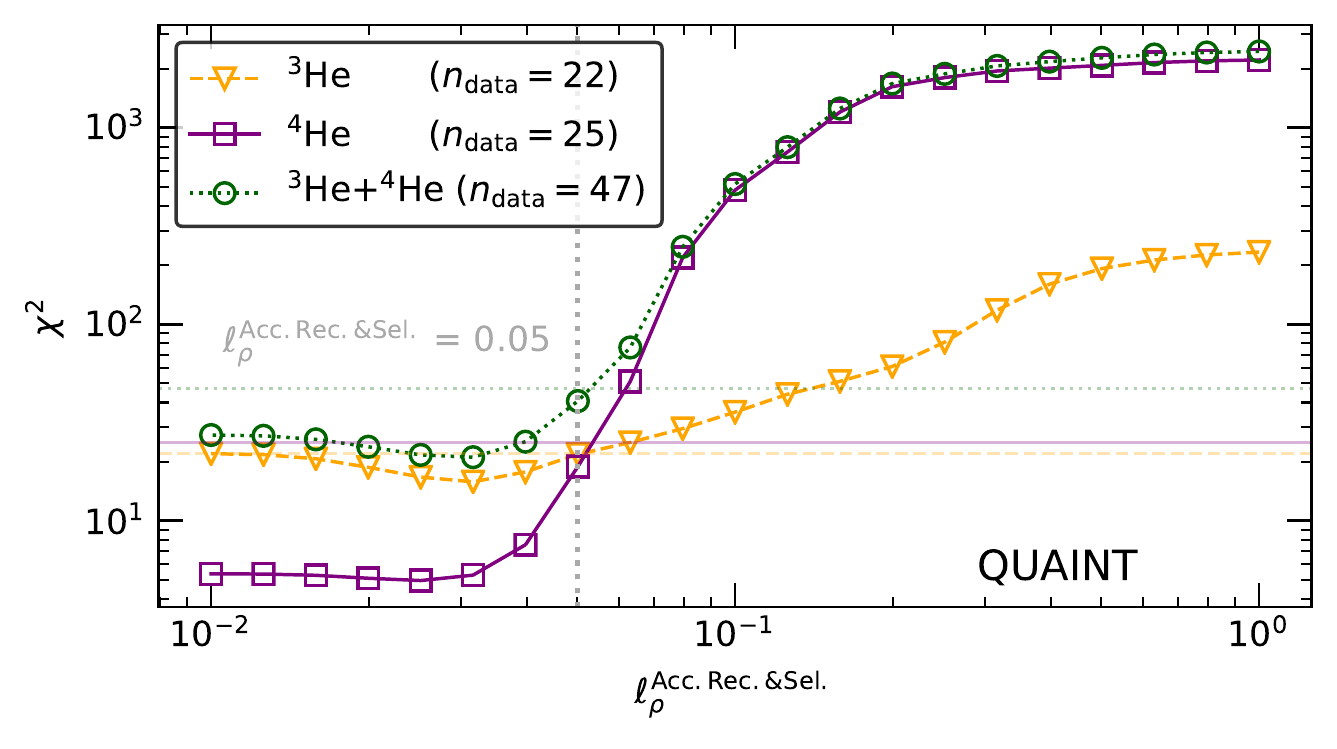}
   \caption{Distance (post-fit $\chi^2$) between data and models varying the correlation length $\ell_\rho^{\rm Acc. Rec.\&Sel.}$ (in decade) of \het{} and \hef{} data for \QUAINT{}. Horizontal dotted line highlight the number of data, indicating roughly the position at which $\chi^2/{\rm d.o.f.}\approx 1$.}
  \label{fig:chi2_vs_lacc}
\end{figure}

The goodness of fit of He isotopic data highlighted in the previous section must however be taken with a grain a salt. The quantitative agreement between the model and the data depends on the correlation length taken for \het{} and \hef{} data, as illustrated in Fig.~\ref{fig:chi2_vs_lacc}. The latter shows that the acceptance systematic uncertainties---which dominates the uncertainty budget (see Fig.~\ref{fig:sub_error_diags_He}) and whose correlation length value is not strongly determined---strongly impact the conclusions.

The calculation in the previous section and in Fig.~\ref{fig:LiBeB_model_vs_data_check} corresponded to $\ell_\rho^{\rm Acc. Rec.\&Sel.}=0.05$. From Fig.~\ref{fig:chi2_vs_lacc}, we conclude that if we were to have chosen a shorter (resp. longer) correlation length, we would have concluded that prediction for the He isotopes were in perfect agreement (resp. in tension) with the data. This would however mostly leave unchanged the values of the best-fit transport parameters. This situation is very similar to that of the B/C case, studied in detail in \citetads{2019A&A...627A.158D}, for which $\ell_\rho^{\rm Acc. res.}$ was set to 0.1---a value also used for LiBeB and N/O data in this analysis, see App.~\ref{app:covLiBeBN}.

\subsection{Discussion}

We can briefly compare our results to the work of \citetads{2019PhLB..789..292W}, which is the only analysis using recent \hettohef{} data. These authors analyse PAMELA data with the \galprop{} code, which relies on the same production cross sections we use here, i.e. \citetads{2012A&A...539A..88C}\footnote{It was implemented in \galprop{} by \citetads{2015ICRC...34..555P}.}. They obtain a good fit to $^2$H/$^4$He, \hettohef{}, H, and He data, but then their model undershoot \pbar{} and B/C data by many $\sigma$. Although our fit is not perfect, we clearly do not face the same issues. Several reasons could explain this difference, like the use of cross-section nuisance parameters and covariance matrices of uncertainties. Another likely reason could be that fitting high-precision primary species (e.g. H and He)---whose flux depend on both the source and transport parameters, and with fewer and less precise data for the secondary species ($^2$H/$^4$He, \hettohef{}, or even $\pbar{}$)---may bias the transport parameter determination \citepads{2012A&A...539A..88C}. The same issue might be present in several recent studies, for instance, \citetads{2016ApJ...824...16J}, \citetads{2016PhRvD..94l3019K}, \citetads{2019PhLB..789..292W}, and \citepads{2019PhRvD..99j3023E}. It could and should be fully assessed with the help of simulated data in future analyses.

\section{Nuisance parameters post-fit values}
\label{sec:aboutxs}

In the previous section, we found (with some caveats) that all the data could be reproduced by all of our model configurations. For this conclusion to hold, we must however check that nuisance parameters behave as expected. Given that the number of data points ($\sim 300$) is much larger than the number of nuisance parameters ($\sim 10$), the latter degrees of freedom could be used at a cheap cost to improve the fit. If so, this would lead to conflicts with the priors and uncertainties expected on these parameters, disfavouring the associated model configuration (here, \BIG{}, \SLIM{}, or \QUAINT{}). We check below that this is not or only mildly the case.

\subsection{Consistency of Solar modulation values}

As described in Sect.~\ref{sec:phi_nui}, AMS-02 data in the analyses are set to $\phi_{\rm prior} = 676\pm 100$~MV. Post-fit values of the AMS-02 modulation level were only shown in the plots of the previous section, and we now comment on them.

If we come back to the combined analysis of the LiBeB ratios, the legend of Fig.~\ref{fig:LiBeB_model_vs_data} shows that the post-fit AMS-02 modulation level for all configurations is within 30~MV of the values taken for the prior. Including low-energy data in the fit (App.~\ref{app:LE-data}) also leads to consistent post-fit modulation levels for all datasets.
\begin{figure}[t]
  \includegraphics[width=\columnwidth]{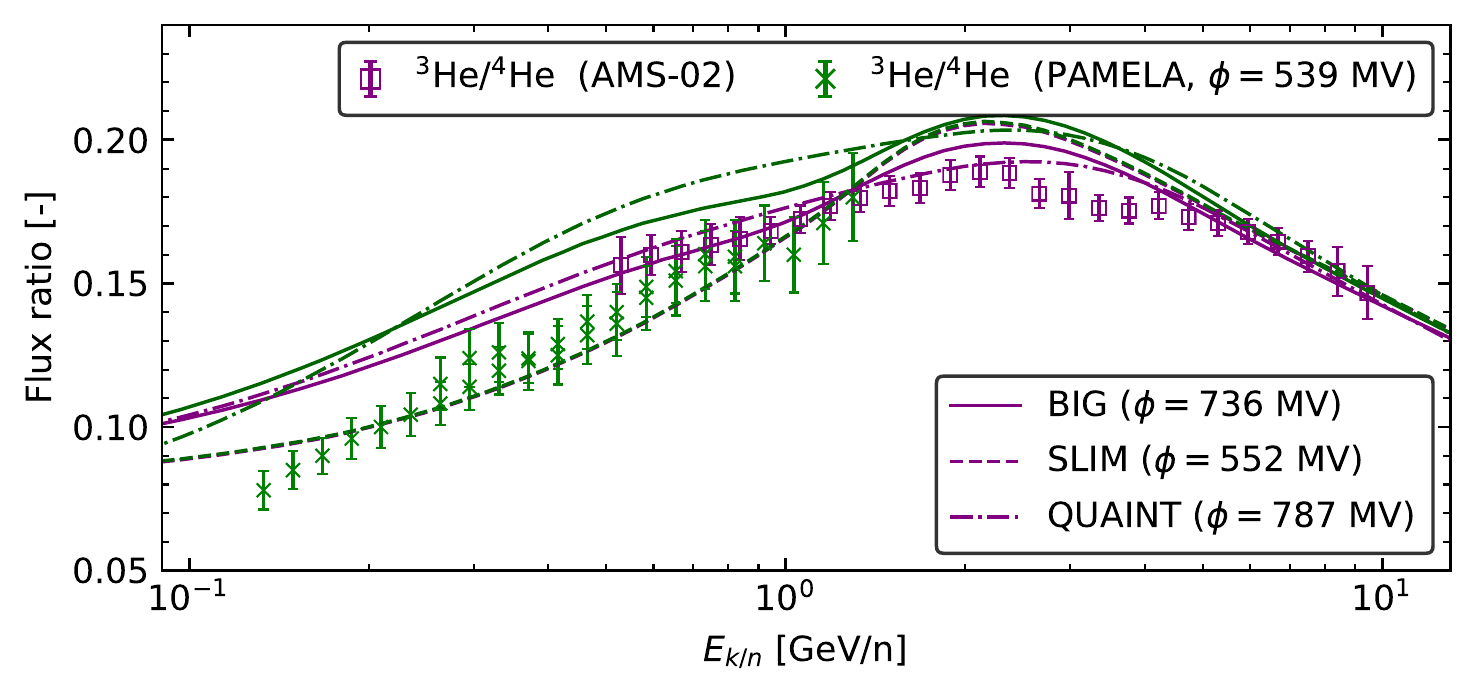}
   \caption{\hettohef{} as a function of $E_{k/n}$ for AMS-02 (purple squares) and PAMELA (green crosses)---the latter combine data from two different analysis, respectively PAMELA-ToF and PAMELA-Calorimeter \citepads{2016ApJ...818...68A}. All models are calculated from the best-fit parameters of the combined analysis of Li/C, Be/C, B/C, N/O, \het{}, and \hef{} AMS-02 data. Different line styles show different model configurations (\BIG{}, \SLIM{}, and \QUAINT{}) and their respective post-fit solar modulation level in purple. Green lines show the same model calculations but modulated at the PAMELA expected value of 539~MV.}
  \label{fig:3he4he_vs_ekn}
\end{figure}
For the combined fit with all species, the post-fit values are read off from Fig.~\ref{fig:3he4he_vs_ekn}, which provides a further illustration of the He isotopes goodness of fit. The modulation levels in the legend correspond to post-fit values of AMS-02 data. Whereas \BIG{} leads to closest value to the prior, \SLIM{} and \QUAINT{} are respectively $1\sigma$ below and above it, starting to be close to their allowed uncertainties.

It is also instructive to compare the model predictions with PAMELA data \citepads{2016ApJ...818...68A}. For the latter the modulation is estimated to be 539~MV at which the model calculations are taken (green lines): the best model in that case is \SLIM{}, because AMS-02 and PAMELA data are then expected to be equally modulated, though this is not very realistic. On the other hand, \BIG{} and \QUAINT{} both overshoot by $\sim 2\sigma$ the PAMELA data points. In the context of a very challenging experimental measurement, it is difficult to conclude on the relevance of this difference. The latter could also be related to Solar modulation features as we need to modulate two isotopes with different $A/Z$ values.

At this stage, we conclude that post-fit values obtained for the modulation level of AMS-02 data are within the expected uncertainties. As highlighted by the comparison to PAMELA data, further data taken at different periods in the modulation cycles could be very useful to conclude on the best transport configuration and on the consistency of solar modulation levels.

\subsection{Consistency of nuclear cross sections}

We recall that for each selected reaction (inelastic or production), the NSS scheme enables several nuisance parameters, see App.~\ref{app:nuis_xs}. The most important one is a normalisation, ${\cal N}_X$, involved in the calculation of the CR quantity $X$. All these nuisance parameters have a prior ${\cal N}_{\rm prior}\approx1$ with a width $\sigma\approx 5\%$ and $\sigma\approx 10\%$ for inelastic and production cross sections respectively.

We look below into the correlation between this parameter and the normalisation of the diffusion coefficient, for several quantities and fit configurations.

\begin{figure*}[t]
{\footnotesize \hspace{3.25cm} Inelastic cross sections
       \hspace{5.75cm} Effective production cross sections \\}
  \includegraphics[width=0.97\columnwidth]{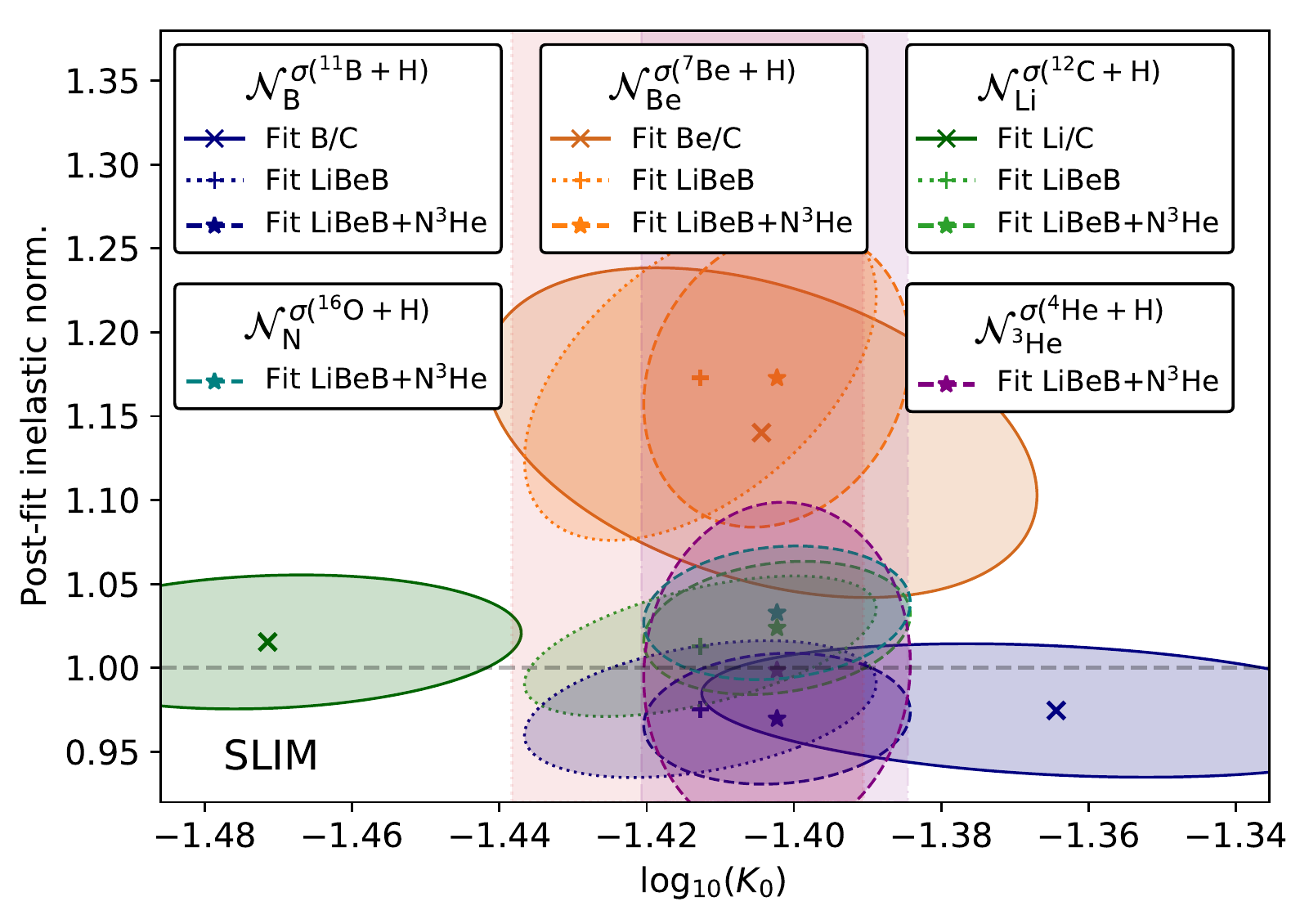}
   \hspace{0.8cm}
  \includegraphics[width=0.97\columnwidth]{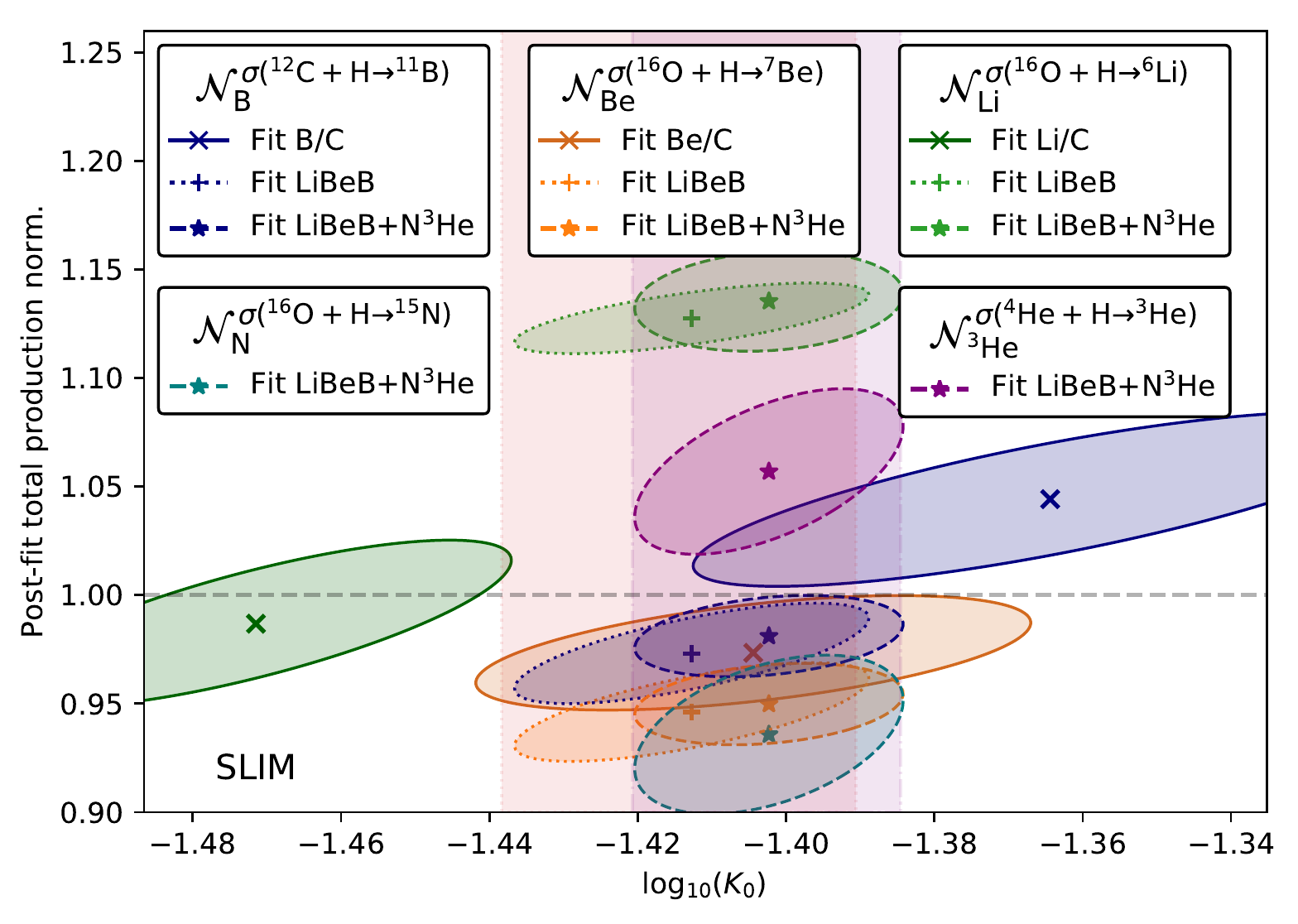}\\
  \includegraphics[width=0.97\columnwidth]{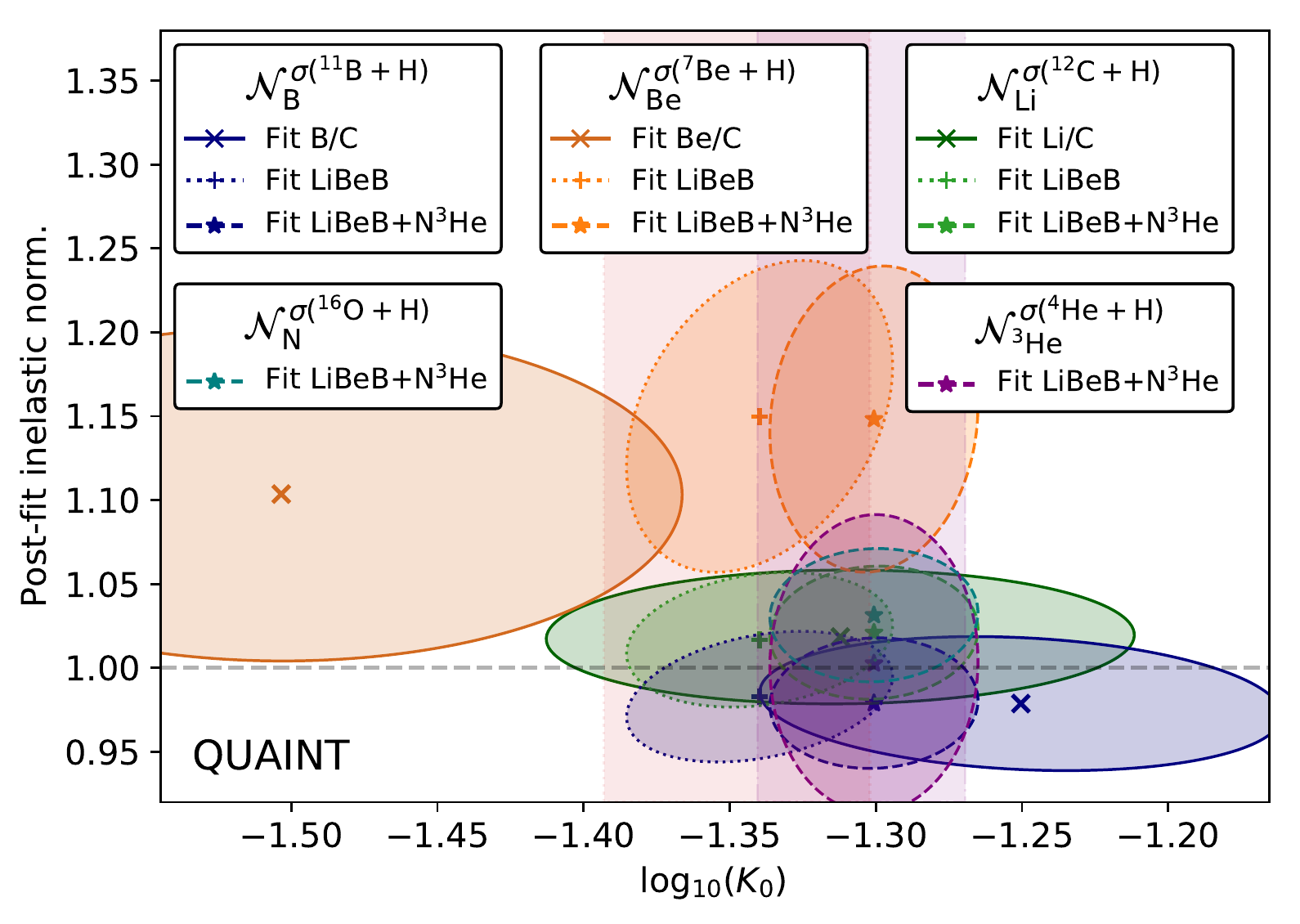}
  \hspace{0.8cm}
  \includegraphics[width=0.97\columnwidth]{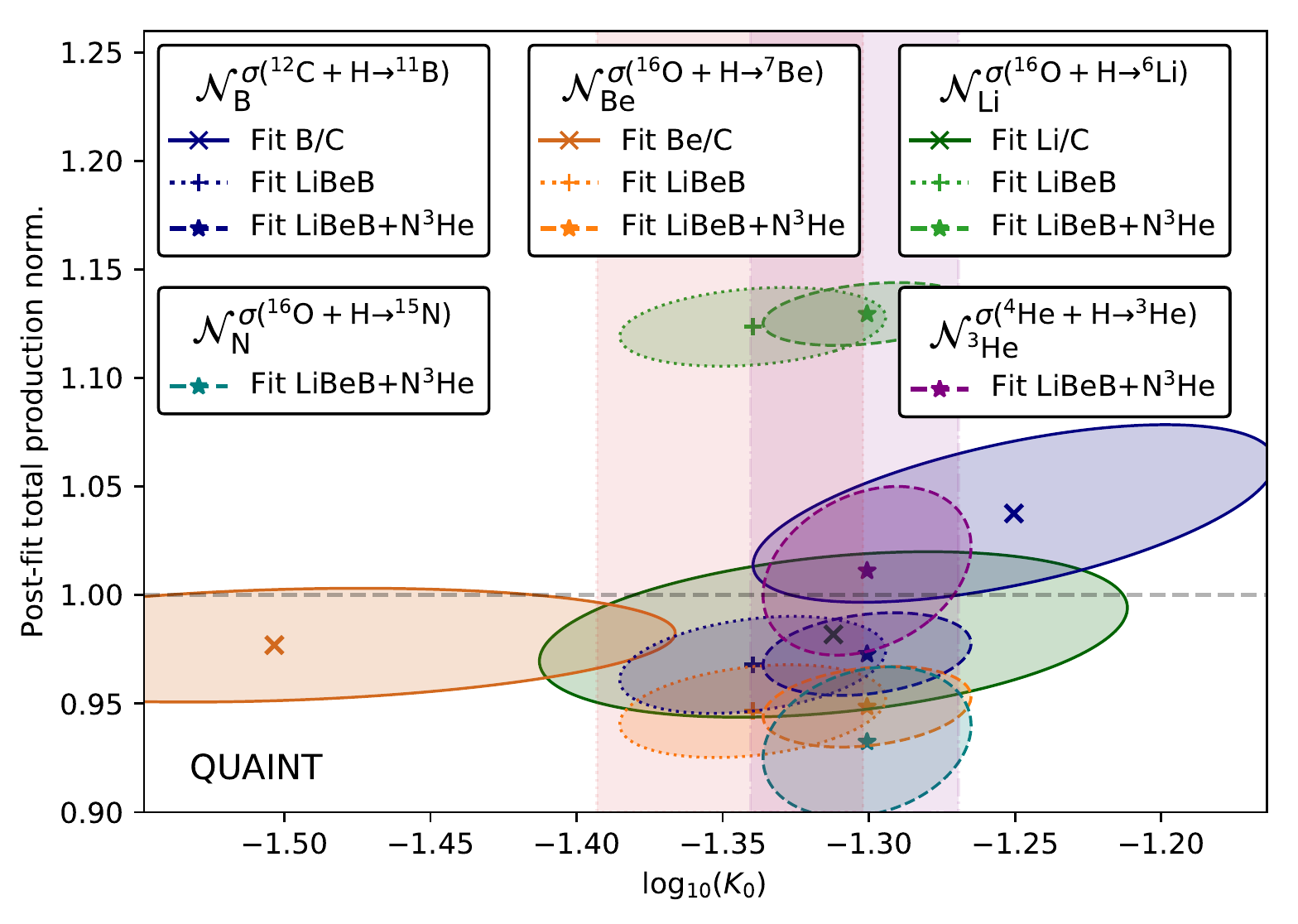}
  \caption{Correlations between the normalisation of the inelastic (left panels) or production (right panels) cross sections and $\log_{10}(K_0)$, for models \SLIM{} (top panels) and \QUAINT{} (bottom panels). The various colours correspond to the normalisations of cross sections involved in different CR species: Li (green), Be (orange), B (blue), N (turquoise), and \het{} (crimson). The various symbols and line styles correspond to the post-fit values and ellipses (for the previously listed species) from various fit configurations: separate Li/C, Be/C, and B/C fits (`$\times$' symbols and solid lines), combined Li/C, Be/C, and B/C fit (LiBeB for short, `$+$' symbols and dotted lines), combined LiBeB+N/O+\het{}+\hef{} fit (`$\star$' symbols and dashed lines). The horizontal grey dashed lines highlights the case of using unmodified cross section datasets (T99 for inelastic and G17 for production, see App.~\ref{app:nuis_xs}), i.e. ${\cal N}^{\rm reac.}_{\rm CR}=1$: the subscript in ${\cal N}$ indicates the CR species associated to this nuisance parameter; the superscript shows the reaction used as a proxy for this CR. For production cross sections (right panels), the normalisation parameters are not directly the value of the nuisance parameter, but are effective parameters, see Eq.~(\ref{eq:xs_prod_eff}). See text for discussion.}
  \label{fig:xs_K0}
\end{figure*}

\subsubsection{Inelastic cross-section normalisation}

The left panels of Fig.~\ref{fig:xs_K0} show correlations for inelastic cross sections in the plan `XS norm'-$\log_{10}(K_0)$, for \SLIM{} (top) and \QUAINT{} (bottom). The ellipses are calculated from the correlation matrix of best-fit parameters returned by \hesse{}; their widths are constructed from a trade-off between values found by \hesse{} and \minos{}. All ellipses give the directions towards which $\log_{10}(K_0)$ would move if the cross section normalisation were to change (and vice-versa).

We first focus on the solid lines, corresponding to the cross-section normalisation parameters involved for the calculation of Li/C (${\cal N}_{\rm Li}$ in green), Be/C (${\cal N}_{\rm Be}$ in orange), or B/C (${\cal N}_{\rm B}$ in blue). When these ratios are fitted separately, different best-fit $\log_{10}(K_0)$ values are obtained. The combined Li/C, Be/C, and B/C analysis forces the latter to move towards the same value. As expected, the new ellipses (dotted lines) and best-fit values (cross symbols) moved from the old ones (plus symbols) mostly along the strongest correlation directions (ellipse principal axis). The normalisation nuisance parameters ${\cal N}_{\rm Li}$ and ${\cal N}_{\rm B}$ are within $\pm 5\%$ of the initial cross-section values (i.e. 1), but ${\cal N}_{\rm Be}\approx 1.15$; the same behaviour is observed for both \SLIM{} and \QUAINT{}. This slightly larger value is possibly related to the possible statistical fluctuation on the AMS-02 low-energy data points discussed in Sect.~\ref{sec:remarks}. The combined analysis with LiBeB, N/O, and He isotopes leads to similar conclusions (compare dash-dotted line ellipses and dotted ones).

We recall that these normalisations are actually proxies for a much larger list of reactions (see App.~\ref{app:nuis_xs}). We also recall that inelastic reactions overall impact the calculation at the level of a few percent and at low rigidity only (see left panels of Fig.~\ref{fig:xs_impact}). We can furthermore state that these `effective' nuisance parameters are well-behaved and that their posterior values are within the expected uncertainties.

\subsubsection{Production cross-section normalisation}
\label{sec:norm_xsprod}
We now turn to the more impacting case of production cross sections, in which model calculations for Li/C, Be/B, etc. can change by $\sim 10\%$ on the whole rigidity range depending on the selected nuclear datasets (see right panels of Fig.~\ref{fig:xs_impact}).

In Fig.~\ref{fig:xs_K0}, the right panels show correlation plots between the post-fit normalisation of the overall production of various species (Li, Be, B, N, and \het{}) and the normalisation of the diffusion coefficient $\log_{10}(K_0)$. To account for the fact that the specific reactions (used as nuisance parameters) are merely proxies and only represent a fraction $x$ of the total production of a CR species under scrutiny, we rescale our normalisation nuisance parameters ${\cal N}^{\rm single~reac~\equiv~proxy}_{\rm CR}$, to obtain\footnote{For the various reactions and CR species, the value of $x$ is reported in square brackets in Table~\ref{tab:xs_nuis}.}
\begin{equation}
  {\cal N}^{\rm eff}_{\rm CR}=(1-x) + x\,{\cal N}_{\rm CR}^{\rm proxy} = 1 + x\,({\cal N}^{\rm proxy}_{\rm CR}-1)\,.
  \label{eq:xs_prod_eff}
\end{equation}
The quantity shown in the $y$-axis of the plot can directly be taken as the global uncertainty on the total production of the species considered. The solid and dashed lines show the ellipses from separate and combined fits of Li/C, Be/C, and B/C. In the combined fits (cross symbols and dotted ellipses), the normalisations again move in the direction of the correlation to reach the best $\log_{10}(K_0)$. If N/O and He isotopes are added to the combined fit (star symbols and dashed ellipses), a further but minor displacement occurs. Similar behaviours are observed for both \SLIM{} (top) and \QUAINT{} (bottom).

From the position of the ellipses in the bottom panel of Fig.~\ref{fig:xs_K0}, we can refine the statement made in the previous section: the model is able to accommodate for all data, although it requires some small but significant modification of the production cross sections w.r.t. to the initial cross section taken. The overall production must be a few percent different for B and \het{}, $\sim 5\%$ for Be and N, but $\sim 12\%$ for Li. As illustrated in the Supplemental Material of \citetads{2018PhRvC..98c4611G}, a ten to fifteen cross section difference is easy to obtain for many individual channels. This translate into a similar (or smaller) `effective' uncertainty for the overall production if all cross-section reactions are weakly correlated (or uncorrelated), see \citealtads{2018PhRvC..98c4611G}).

\begin{figure}[t]
  \includegraphics[width=\columnwidth]{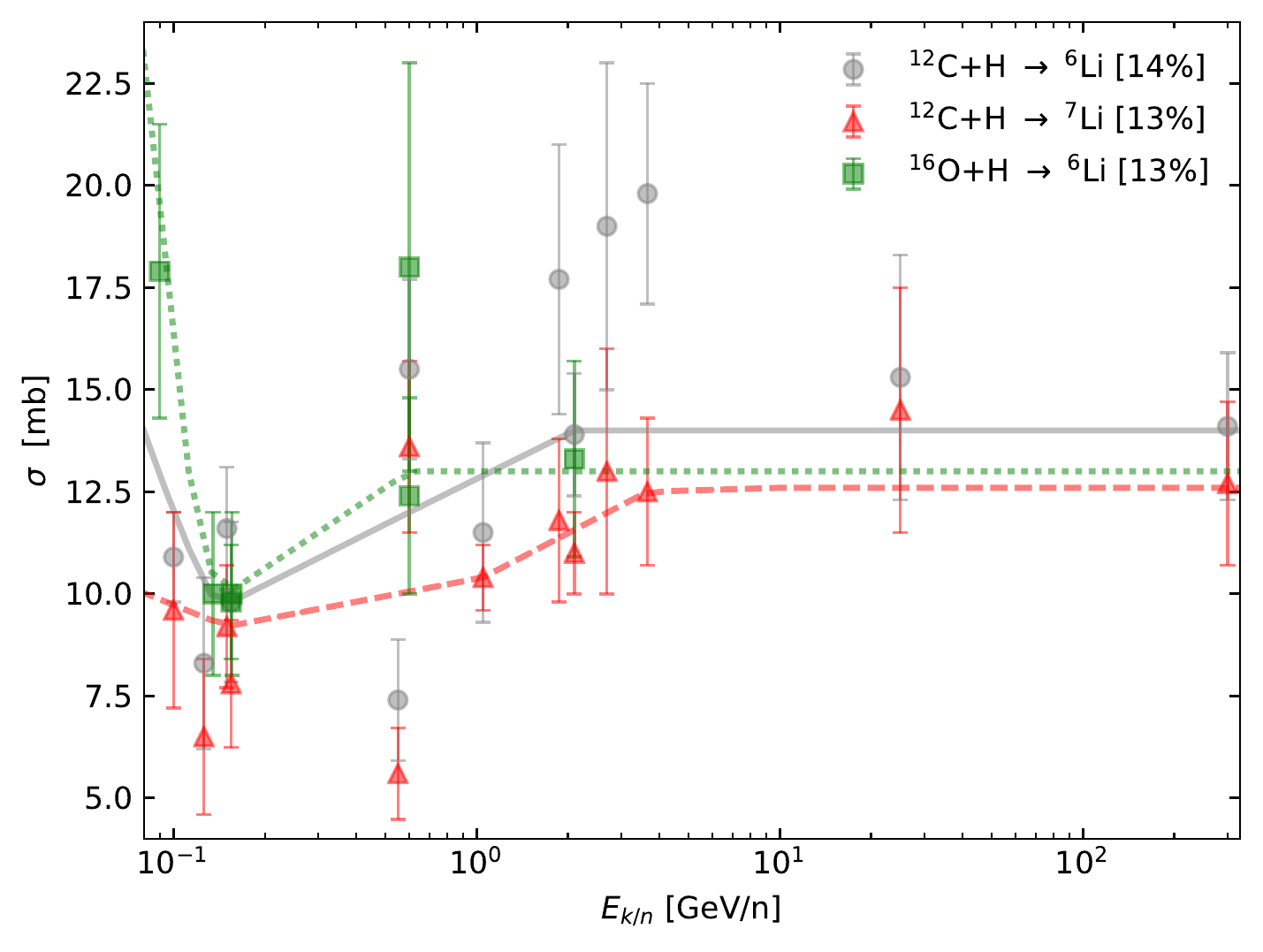}
  \includegraphics[width=\columnwidth]{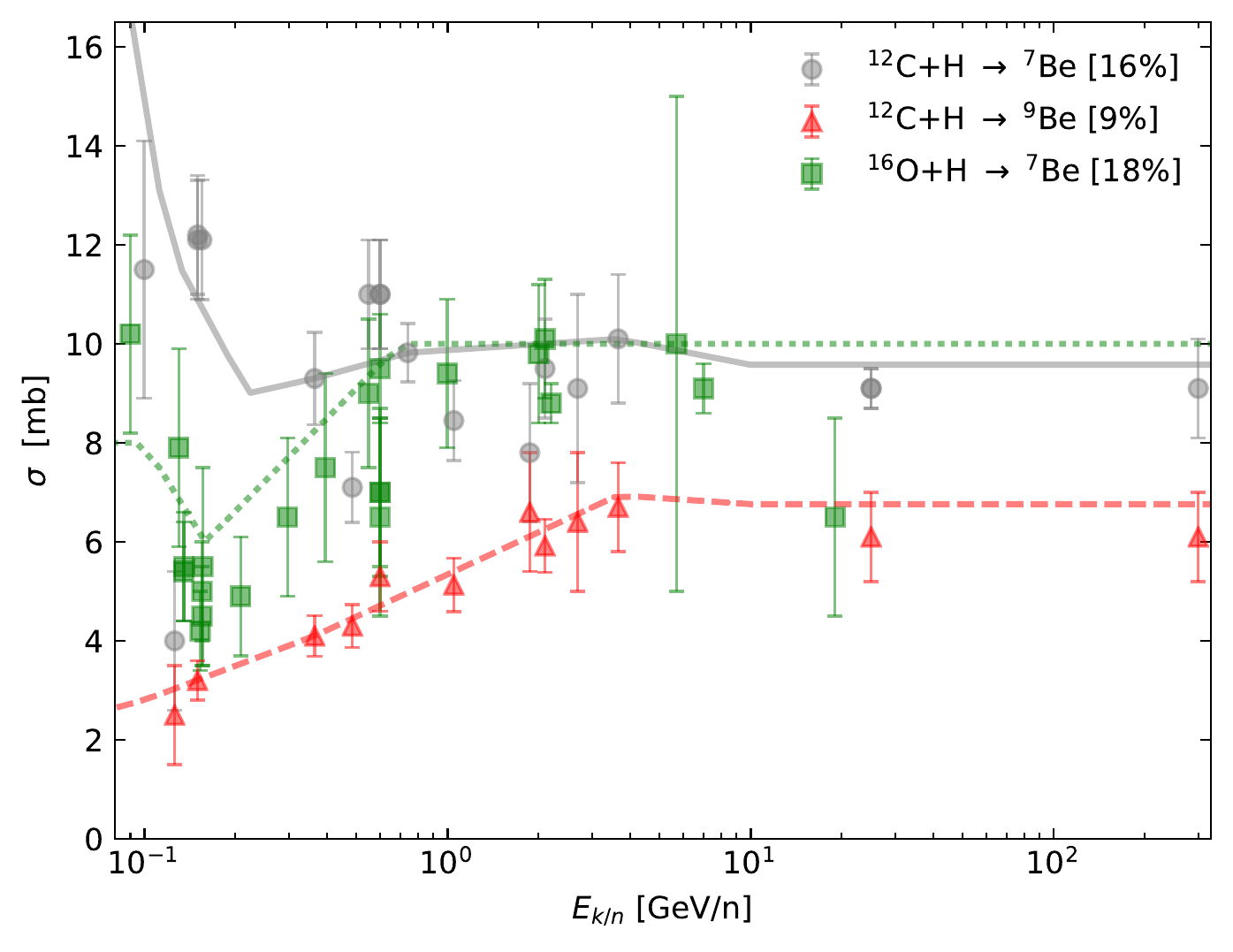}
  \caption{Data (symbols) and model (lines) comparison for the most important production cross sections leading to Li (top panel) and Be (bottom panel) fluxes. Their respective contribution to the total flux production is reminded in brackets. The data are taken from \citetads{2018PhRvC..98c4611G} and the model is the GALPROP parametrisation \citepads{2001ICRC....5.1836M,2003ICRC....4.1969M}. See text for discussion.}
  \label{fig:prod_xs}
\end{figure}
From this analysis, we conclude that all production cross sections end up within their expected values, even Li, which reaches the limit of its allowed uncertainties. The nuclear data for the later are however scarce. For illustrative purpose, we show in Fig.~\ref{fig:prod_xs} a comparison between the data (symbols) and the benchmark G17 parametrisations \citepads{2001ICRC....5.1836M,2003ICRC....4.1969M} used in our analysis, for the most important production cross sections contributing to Li and Be fluxes \citepads{2018PhRvC..98c4611G}. The required normalisation variation of a few percent for Be and $\sim 15\%$ for Li are completely allowed by the present data quality. This whole discussion illustrates that at the level of precision of AMS-02 data, production cross sections matter a lot in the context of combined analyses. Better nuclear data are mandatory to better assess the goodness of fit, and possibly reveal tensions between CR flux calculations and data for different species.

\section{Conclusions}
\label{sec:conclusions}

We have shown that a propagation model is able to successfully reproduce all secondary-to-primary ratios recently published by the AMS-02 collaboration, i.e. Li/C, Be/C, B/C \citepads{2018PhRvL.120b1101A}, N/O \citepads{2018PhRvL.121e1103A}, and He isotopes \citepads{2019PhRvL.123r1102A}. These model's configurations are based on recent benchmark transport scenarios introduced in \citetads{2019PhRvD..99l3028G} and updated here. In the context of a high-rigidity break at $\sim200$~GV, they confirm a diffusion slope in the intermediate regime of $\delta$ in the range $[0.45, 0.53]$.

The combined analysis of different AMS-02 ratios (Li/C, Be/C, and B/C) or the separate analysis of these ratios combining AMS-02 and lower-energy data (ACE-CRIS) show an equal preference either for a low-rigidity break in the diffusion coefficient (at $\sim 4.6$~GV with a slope change of $\sim0.7$) or an upturn below a few GV. This effective change in the diffusion behaviour could reveal a decrease of the CR pressure as CRs reach the non-relativistic regime ($\beta$ dependence), or be related to some dissipation of the turbulence power spectrum (rigidity dependence). As in \citetads{2019PhRvD..99l3028G} where only B/C data where considered, two disjoint regions of the transport parameter space provide viable solutions to match the LiBeB data: a purely diffusive regime with a low-rigidity break (configuration dubbed \SLIM{}), or a convection/reacceleration solution with either a diffusion break or an upturn of the diffusion slope at the non-relativistic transition (configurations dubbed \BIG{} and \QUAINT{}); based on the data analysed till now, there is no strong quantitative argument for choosing one or the other. These two configurations are also able to reproduce N/O data and \het{} and \hef{} fluxes. However, we find that \het{} data are extremely sensitive to reacceleration, and considering them or not in the combined analysis moves the best-fit parameters between the two preferred regions of the parameter space. At variance with the B/C analysis only, the \BIG{} configuration requires both reacceleration ($\sim50$~km~s$^{-1}$) and convection ($\sim10$~km~s$^{-1}$) in the combined analysis of all species.  It is possible that either/both the parametrisation of the low-energy cross-sections or/and diffusion coefficient is too simple. Combining other secondary species---when released by the AMS-02 collaboration (e.g., \deut{}, F, Na\dots up to subFe)---or understanding the low-energy interstellar Voyager data \citepads{2016ApJ...831...18C} should help deciphering the low-energy transport of CRs.

Compared to other similar efforts in the literature \citepads{2016ApJ...824...16J,2016ApJ...831...18C,2016PhRvD..94l3019K,2019PhLB..789..292W,2019PhRvD..99j3023E,2020ApJ...889..167B}, the decisive factor in our approach is to account for nuisance parameters for nuclear cross sections and for rigidity correlations in the systematics of AMS-02 data---as in previous studies, we also use nuisance parameters for Solar modulation levels. Firstly, as already demonstrated on B/C in \citetads{2019A&A...627A.158D}, the value of the correlation length in specific data systematics is crucial not to bias the determination of the transport parameters. These correlations are difficult to evaluate and not provided in the AMS publications, but the latter contain sufficient information to build them from educated guesses. We further show here, that these correlations strongly impact the quantitative estimate of the goodness of fit to the data, especially for \het{}, hence hampering our ability to draw statistically sound conclusions on the universality of our effective model for light species.
Secondly, nuisance parameters provide extra degrees of freedom which allow one to directly propagate various uncertainties in the sought transport parameters. For instance, we find that below a few GV, Solar modulation and production cross sections are the dominant sources of uncertainties; above a few GV, transport and production cross sections are the dominant ones. In any analysis, post-fit values of the nuisance parameters should not wander too far away from their allowed ranges, and we checked they all stay within their $1\sigma$ values in our analyses. These values are especially important and interesting for the case of production cross sections. Whereas in the case of a single secondary-to-primary fit, the production cross-section normalisation is partly degenerate with $K_0$ (normalisation of the diffusion coefficient), this degeneracy is lifted in combined secondary-to-primary ratio analyses (because the same $K_0$ value is enforced). Inspecting the post-fit values for the production cross sections, we find deviations going from a few percent (for \het{}, Be, B, and N) up to $15\%$ for Li with respect to the nuclear model values. This is in the ballpark of the estimated uncertainties for the production of these species \citepads{2019PhRvD..99l3028G}, but this strengthens the need for better nuclear data in order to fully benefit from AMS-02 data precision; better nuclear data are also needed to draw stronger conclusions on the consistency of transport for all CR species.

\begin{acknowledgements}
We thank our CR colleagues at Annecy and Montpellier, and in particular J.~Lavalle, A.~Marcowith, P.~Salati, P.~Serpico for their very helpful feedback and discussions. We also thank A.~Oliva for very useful comments.
This work has been supported by the `Investissements d'avenir, Labex ENIGMASS', by the French ANR, Project DMAstro-LHC, ANR-12-BS05-0006, and by Villum Fonden under project no. 18994.
\end{acknowledgements}

\appendix

\section{Covariance matrices of systematic errors}
\label{app:cov_mat}

The correlation between the AMS-02 data points can be taken into account by using covariance matrices for the fit. However, the latter are not available and must be estimated from the information found in the AMS-02 publications. Following \citetads{2019A&A...627A.158D}, the relative covariance $(C_{\rm rel}^\alpha)_{ij}$ between rigidity bin $R_i$ and $R_j$ is taken to be
\begin{equation}
(C_{\rm rel}^\alpha)_{ij} = \sigma^\alpha_i \sigma^\alpha_j \exp\left(-\frac{1}{2}
\frac{(\log (R_i/R_j)^2} {(\ell_\rho^\alpha)^2} \right)\,,
\label{eq:correl}
\end{equation}
with $\sigma_i^\alpha$ the relative uncertainty of error type $\alpha$ at bin $i$ and $\ell_\rho^\alpha$ the correlation lengths for error type $\alpha$ (in unit of rigidity decade). The correlation length $\ell_\rho$ of each systematic was carefully chosen to best reflect the physics process behind the associated systematics \citepads{2019A&A...627A.158D}.
From the correlation length we can also form correlation matrices
\begin{equation}
{\rm c}_{ij}^{\alpha} = \frac{{\cal C}_{ij}^{\alpha}}{\sqrt{{\cal C}_{ii}^{\alpha} \times {\cal C}_{jj}^{\alpha}}}\,.
\end{equation}

\subsection{Li/C, Be/C, B/C (or ratios with O), and N/O}
\label{app:covLiBeBN}
\begin{figure}[t]
	\includegraphics[width=\columnwidth]{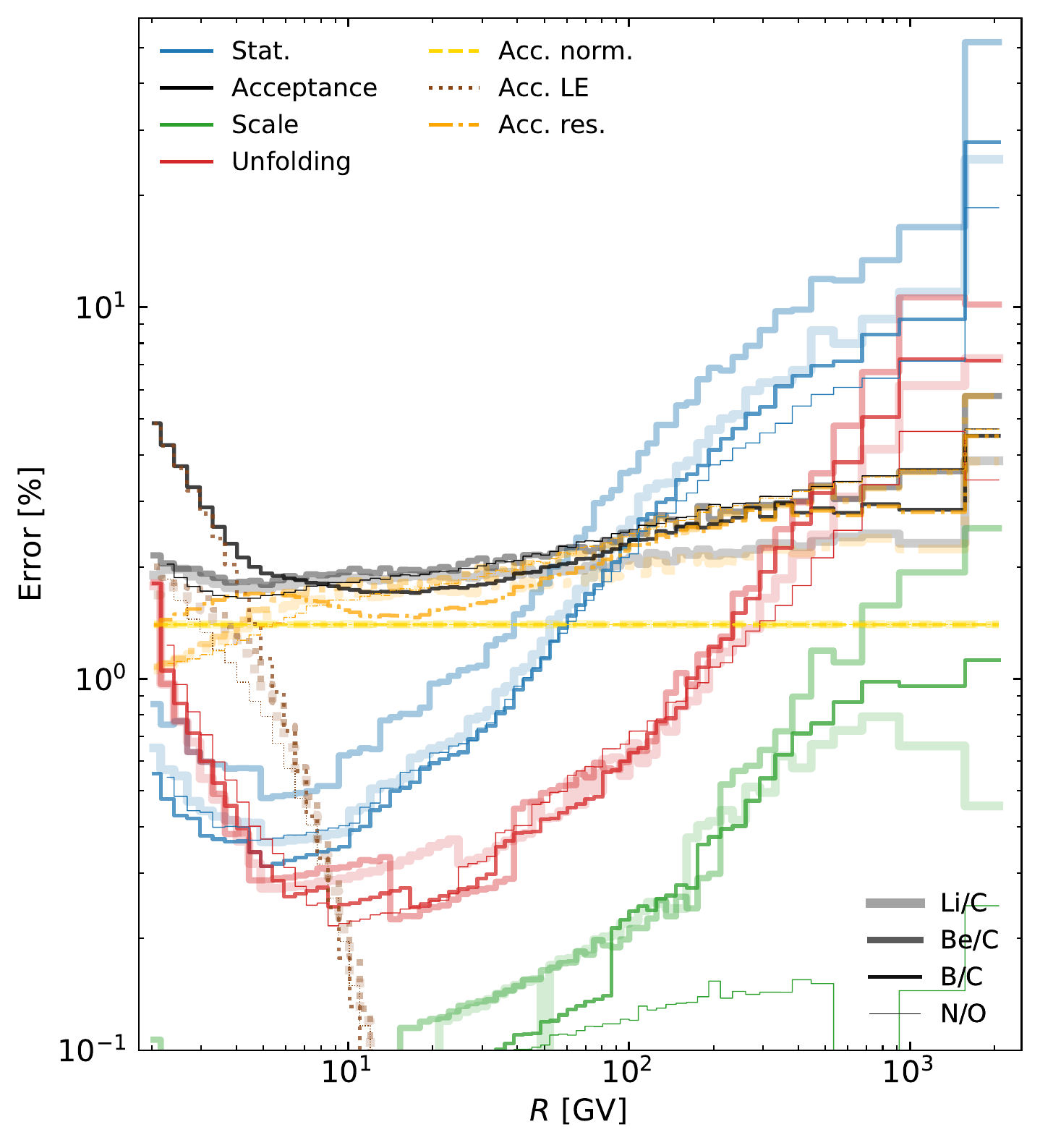}
	\caption{Comparison of statistical and systematic uncertainties for N/O, B/C, Be/C, and Li/C (from thin to thick shaded lines). Values are taken from \citetads{2018PhRvL.120b1101A,2018PhRvL.121e1103A}, except for `Acc. norm.', `Acc. LE', and `Acc. res.', which are broken-down from {\em Acc.} uncertainties. See text for details.}
	\label{fig:sub_error_diags}
\end{figure}
Figure~\ref{fig:sub_error_diags} shows statistical (Stat.) and systematic uncertainties provided by the AMS-02 collaboration \citepads{2018PhRvL.120b1101A}, i.e. acceptance (Acc.), scale (Scale), and unfolding (Unf.); the three line thickness's (from thin to thick) and shades (from dark to light) correspond to N/O, B/C, Be/C, and Li/C ratios respectively. As motivated and detailed in \citetads{2019A&A...627A.158D}, `Acc.' is further broken down in three more systematics, `Acc. norm.', `Acc. LE', and `Acc. res.'.

For all species, we observe that (i) at high rigidity, statistical uncertainties are dominant, especially for Be (medium-thick solid blue line), the less abundant of all the secondary species, see Fig.~\ref{fig:LiBeB_model_vs_data}; (ii) at intermediate rigidities, `Acc. res.', the most difficult systematics to derive a correlation length for (see below), is dominant; (iii) at low-rigidity, `Acc. LE', which has a short correlation length (see below), is dominant, especially for B/C (thin-dashed black line). Despite some small differences in uncertainties between the species, we can conclude they all have the same status in the context of fits and conclusions that can be drawn from their fits (see Sect.~\ref{sec:LiBeB}).

\begin{figure}[t]
	\includegraphics[width=\columnwidth]{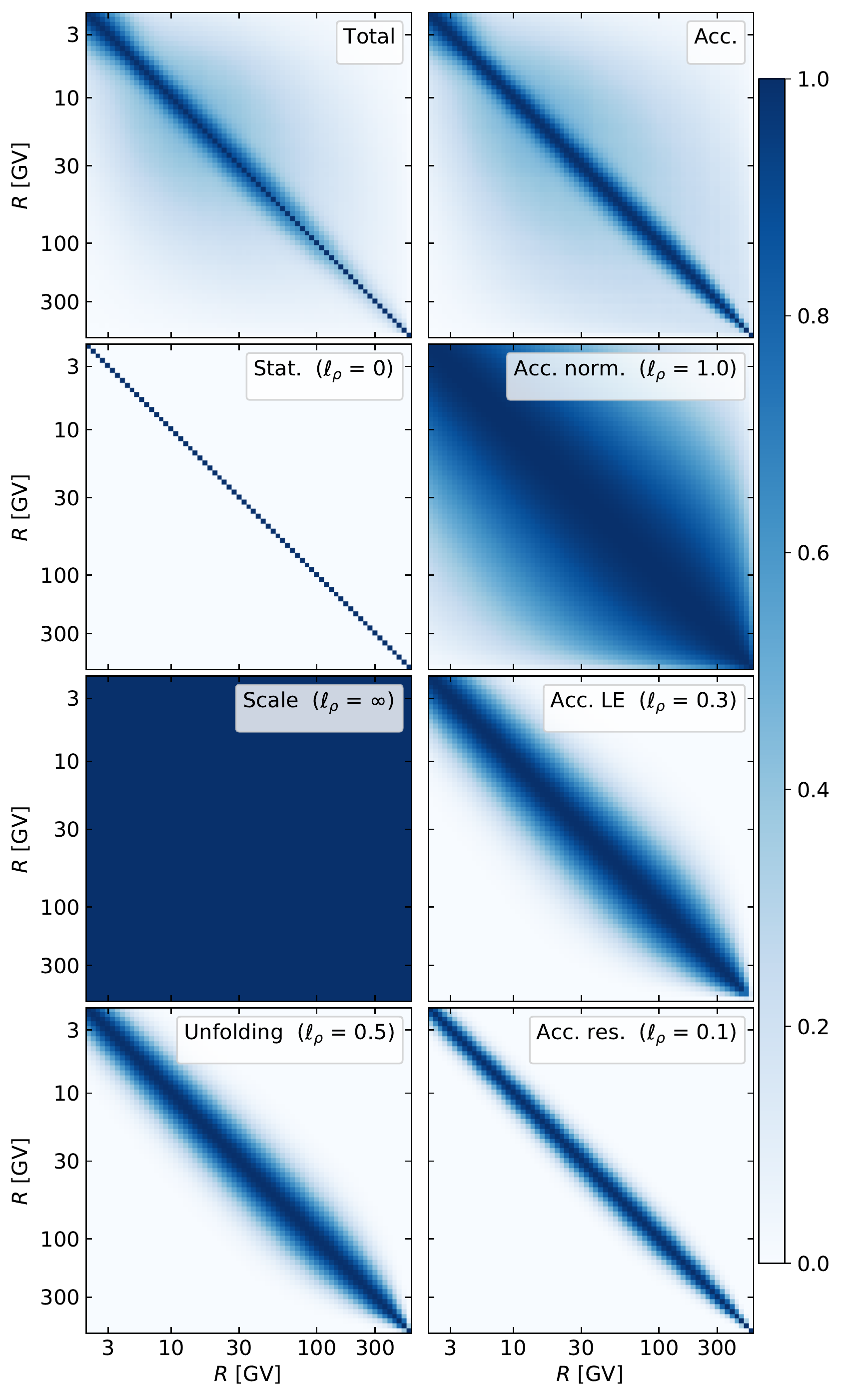}
	\caption{Correlation matrices colour-coded from zero (no correlation, white) to one (full correlations, blue). The three bottom panels correspond to the systematics whose amplitude are shown in Fig.~\ref{fig:sub_error_diags}. The top right panel shows the correlation matrix from all acceptance uncertainties, and the top left panel that from all uncertainties combined.}
	\label{fig:sub_error_matrices}
\end{figure}
The correlation matrices ${\rm c}_{ij}^{\alpha}$ are shown in Fig.~\ref{fig:sub_error_matrices}. They are taken to be the same for the three secondary-to-primary ratios Li/C, Be/C, and B/C (and also ratios to O). Indeed, these neighbour species have similar interactions in the detector, hence the same correlation lengths for their systematics. As illustrated in the various panels, statistical uncertainties (`Stat.', $\ell_\rho=0$) are fully uncorrelated by definition, whereas the scale systematics (`Scale', $\ell_\rho=\infty$), is taken to be an overall normalisation. The `Acc.' systematics has several components, going from a quite correlated (`Acc. norm.', $\ell_\rho=1$ decade) to less correlated (`Acc. res.', $\ell_\rho=0.1$ decade) component. The overall covariance matrix for `Acc.' (top right panel) is a non trivial combination of the correlation matrices (three bottom right panels) and relative uncertainties shown in Fig.~\ref{fig:sub_error_diags}: it is dominated by `Acc. LE' at low rigidity and by `Acc. res.' above, with large wings from `Acc. norm.'. The correlation length of all combined uncertainties (top left panel) reflects the dominance of statistical uncertainties at high rigidities, and is again a non-trivial combination of all shown systematics. We recall that the exact choice of the correlation lengths for most systematics is not critical when fitting the data, except for `Acc. res.' \citepads{2019A&A...627A.158D}: for the latter, consistency arguments from the B/C analysis provided a preferred range, and for definiteness $\ell_\rho^{\rm Acc. res.}=0.1$ was chosen in order to have $\chimindof{}\approx 1$ in the B/C analysis.

\subsection{$^3$He and $^4$He}
\label{app:covHe}

\begin{figure}[t]
  \includegraphics[width=\columnwidth]{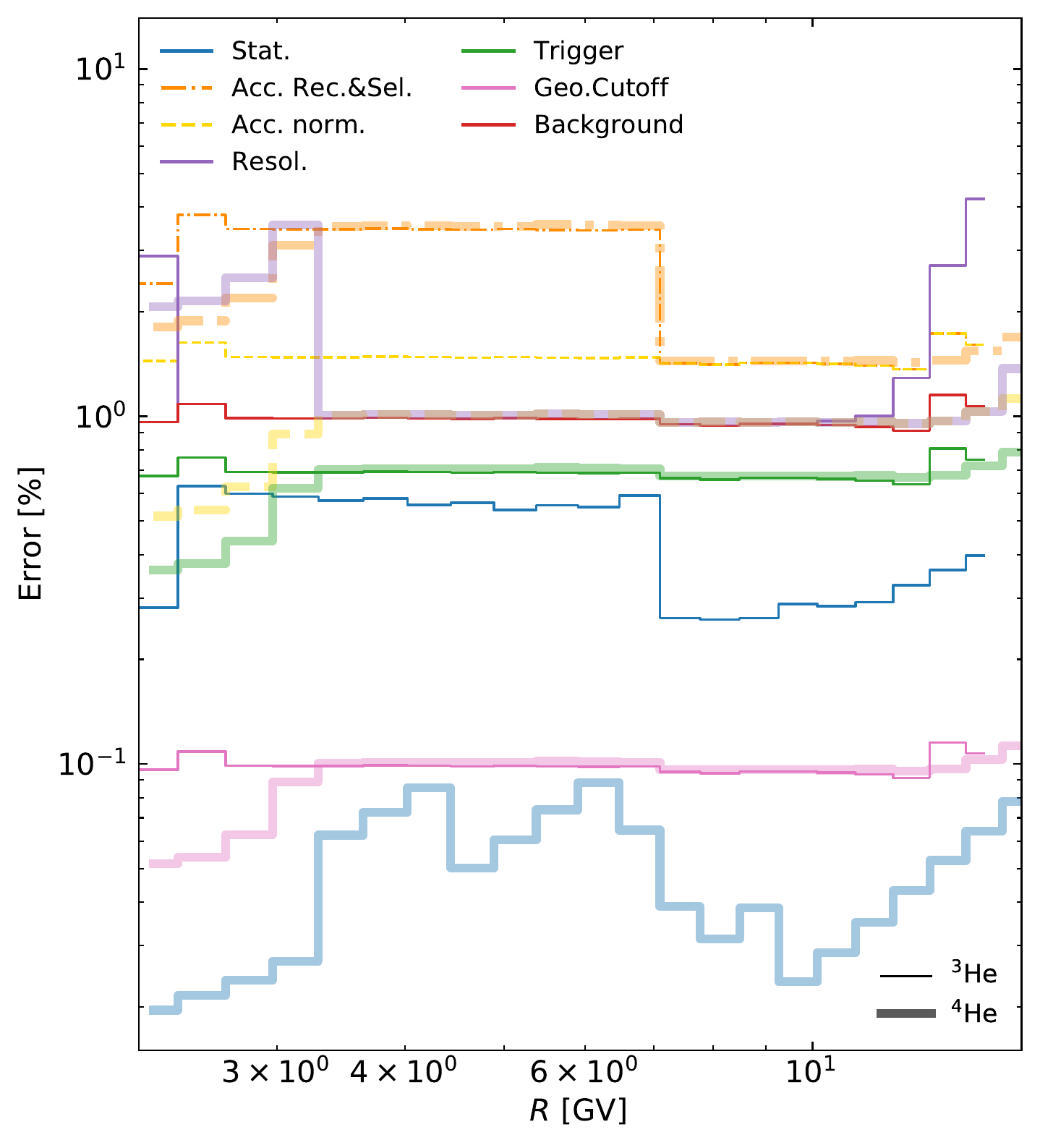}
  \includegraphics[width=\columnwidth]{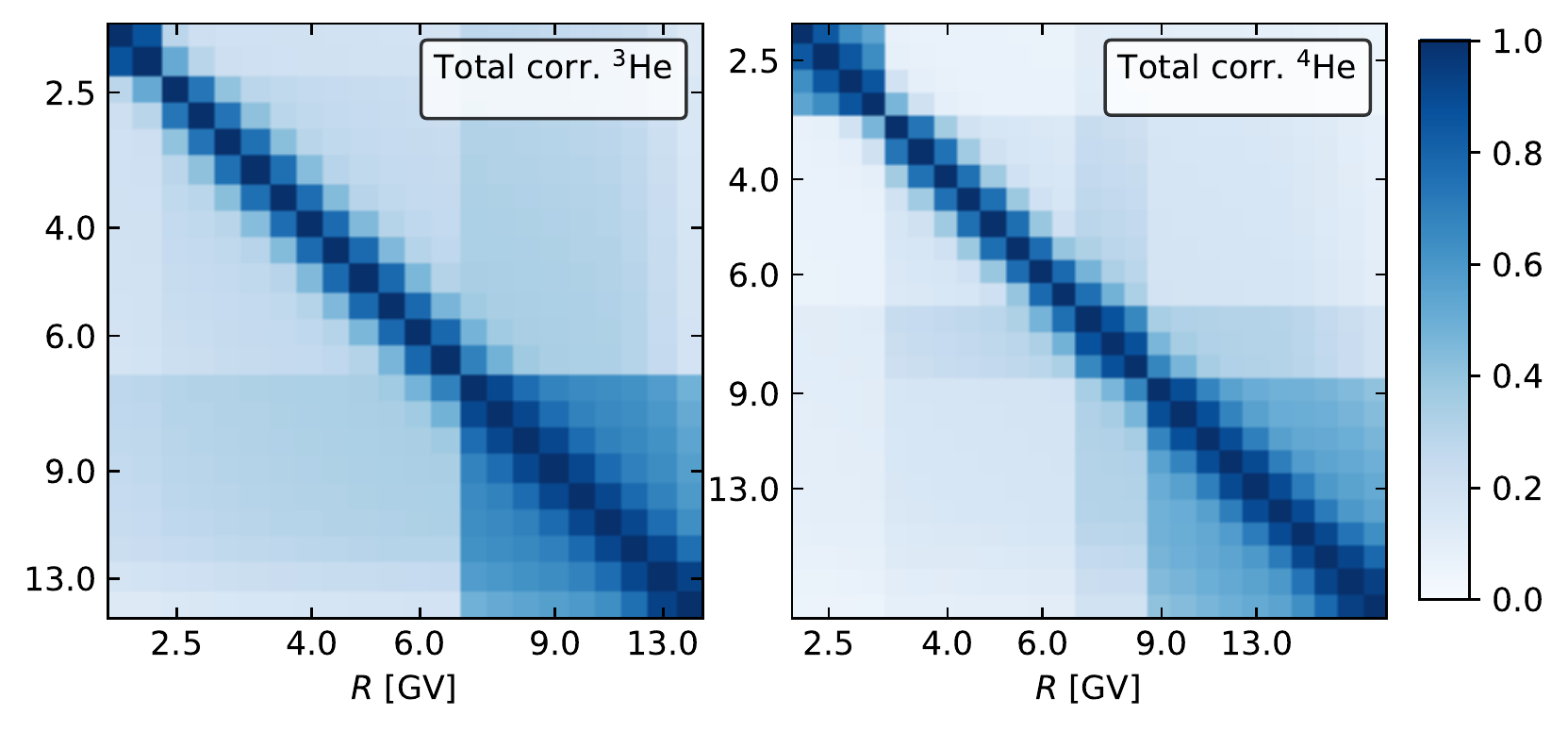}
  \caption{{\em Top panel}: comparison of statistical and systematic uncertainties for \het{} (thin lines) and \hef{} (thick lines) AMS-02 data. {\em Bottom panel}: correlation matrix for \het{} (left) and \hef{} (right) as a function of $R$. All quantities are reconstructed from \citetads{2019PhRvL.123r1102A}, see text for details.
  \label{fig:sub_error_diags_He}}
\end{figure}

The various contributions to the AMS-02  \het{} and \hef{} systematics are broadly described in \citetads{2019PhRvL.123r1102A}. They originate from:
\begin{itemize}
    \item the rigidity and $\beta$ resolution function, `Resol.';
    \item the trigger efficiency, `Trigger';
    \item the geomagnetic cutoff, `Geo. cutoff';
    \item reconstruction and selection efficiencies for acceptance, `Acc. Rec.\& Sel.';
    \item inelastic cross sections for acceptance, `Acc. norm';
    \item background contamination, `Background'.
\end{itemize}
We stress that the break-down by the AMS-02 collaboration of systematics in various categories is different for He isotopes and Li, Be, B, and N elements. This is related to the different analyses required for elemental or isotopic flux reconstructions. Nevertheless, some similarities exist. In particular, `Geo.~cutoff' and `Acc.~Rec.\& Sel.' for the He isotopes correspond to `Acc.~LE' and `Acc.~res' for the LiBeBN elements.

For each systematics, we interpret the contributions at different rigidity values
as a piece-wise power-law function. The latter are then rescaled
at each rigidity bin so that the sum of all systematic errors matches the total
systematic errors provided in \citetads{2019PhRvL.123r1102A}.
This leads to our model for the AMS-02  \het{} and \hef{} systematics,
shown on the top panel of Fig.~\ref{fig:sub_error_diags_He}.

The covariance matrix associated with these systematics is then built from
Eq.~(\ref{eq:correl}) based on an educated guess for their correlation length,
$\ell_\rho$ (in unit of energy decade).
In addition, as the fluxes are reconstructed using different detectors for the $\beta$
measurements (TOF, RICH-NaF, RICH-AGL) with different selections and acceptances, one can assume that
some contributions in the systematic are uncorrelated between the different associated rigidity regions.
Here we assume that only the `Resol.' contribution corresponds to this case. For the later, all covariance
matrix elements corresponding to two different regions are set to 0 and the covariance matrix is then a block diagonal matrix.

For the correlation lengths, we take:
\begin{itemize}
    \item $\ell_\rho^{\rm Resol.}=0.3$, as uncertainties from rigidity response function affect intermediate scales;
    \item $\ell_\rho^{\rm Trigger}=1.0$, because the uncertainty on the detector response affecting the trigger efficiency should produce a systematics strongly correlated for different rigidities;
    \item $\ell_\rho^{\rm Geo.\,cutoff}=0.3$, and this contribution is similar to $\ell_\rho^{\rm Acc.\,LE}$ for the Li, Be, B, and N elements (same correlation length assumed);
    \item $\ell_\rho^{\rm Acc.\,norm}=\ell_\rho^{\rm Background}=1.0$, since the uncertainty on cross sections is mainly on their normalisations and then produce a strongly correlated systematic;
    \item $\ell_\rho^{\rm Acc.\,Rec. \& Sel.}=0.05$, but similarly to $\ell_\rho^{\rm Acc.\,res}$ for elements, this number cannot be easily defined. As for $\ell_\rho^{\rm Acc.\,res}$, this systematics dominates the total error budget of the flux. The dependence
        of \chimindof{} with this correlation length was discussed in \citetads{2019A&A...627A.158D} for the B/C case, and it is discussed here in Sect.~\ref{sec:corrlength} for \het{}- and \hef{}-related fits.
\end{itemize}

The bottom panel of Fig.~\ref{fig:sub_error_diags_He} shows the total correlation matrix for \het{} and \hef{}.
The contributions of the block diagonal matrices from the `Resol.' systematics are visible.

\section{Nuisance parameters for Li, Be, B cross sections}
\label{app:nuis_xs}

\newcommand{\cwa}{0.48\columnwidth}
\newcommand{\cwb}{0.495\columnwidth}
\newcommand{\cwc}{0.49\columnwidth}
\newcommand{\cwd}{0.49\columnwidth}
\begin{figure*}[t]
{\footnotesize \hspace{2.5cm} Impact of inelastic cross sections
       \hspace{5.cm} Impact of production cross sections \\}
   \subfigure{\includegraphics[width = \cwa]
      {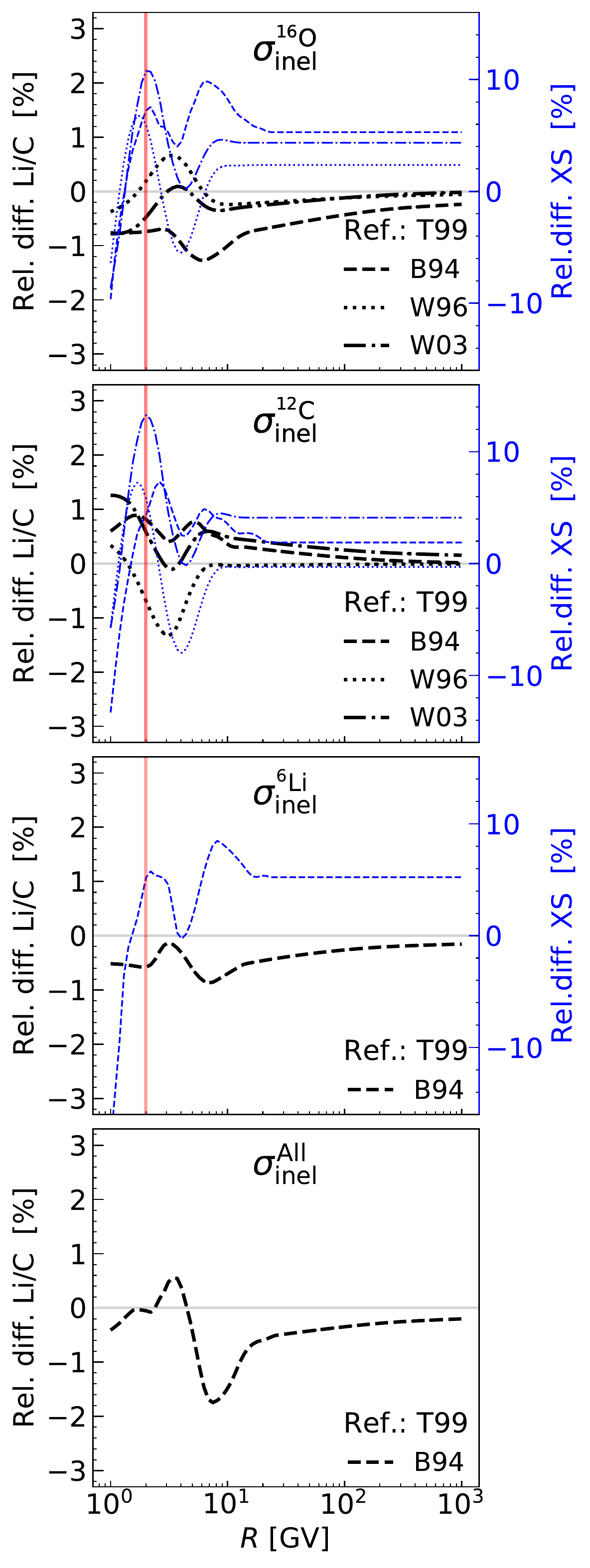}}
   \subfigure{\includegraphics[width = \cwa]
   {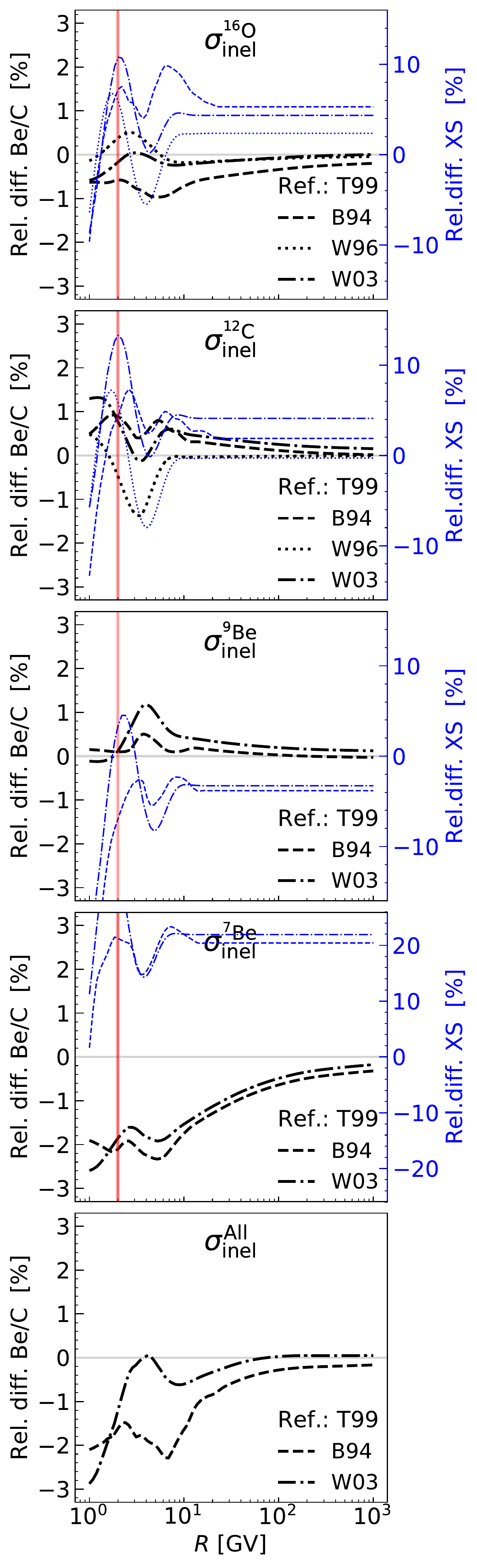}}
   \hspace{0.5cm}
   \subfigure{\includegraphics[width = \cwb]
   {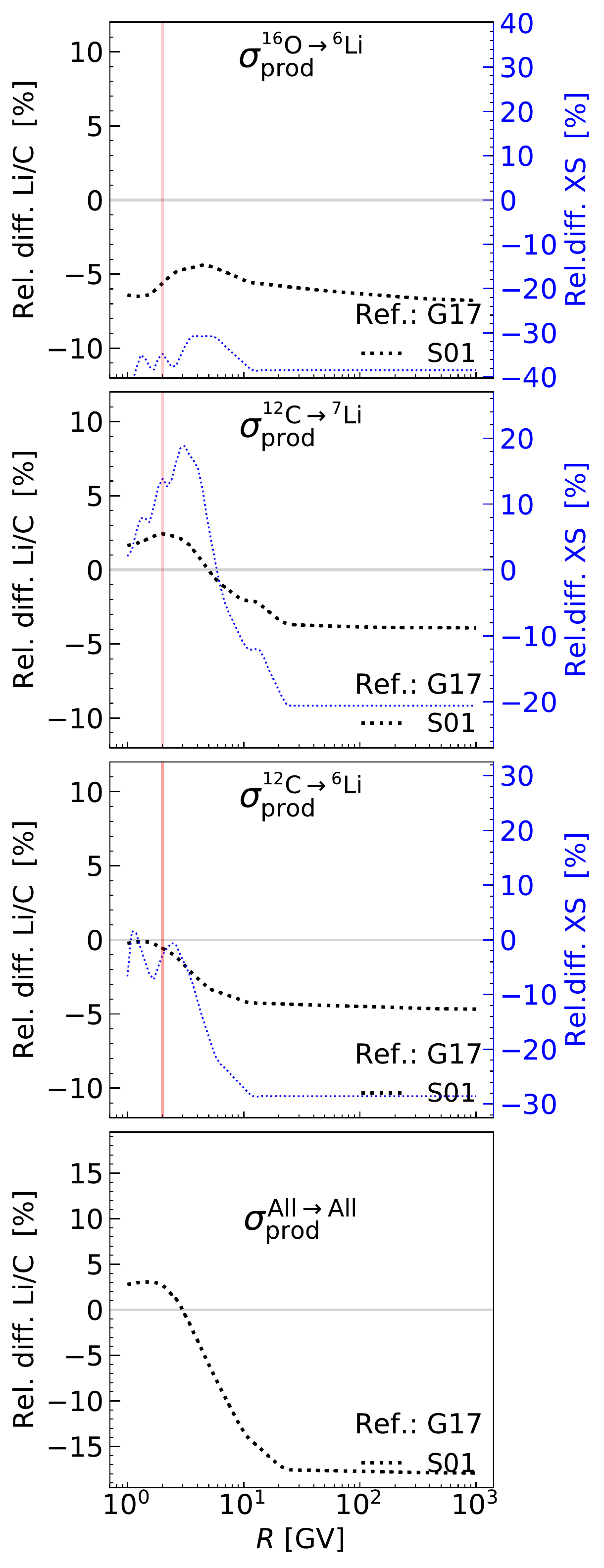}}
   \subfigure{\includegraphics[width = \cwb]
   {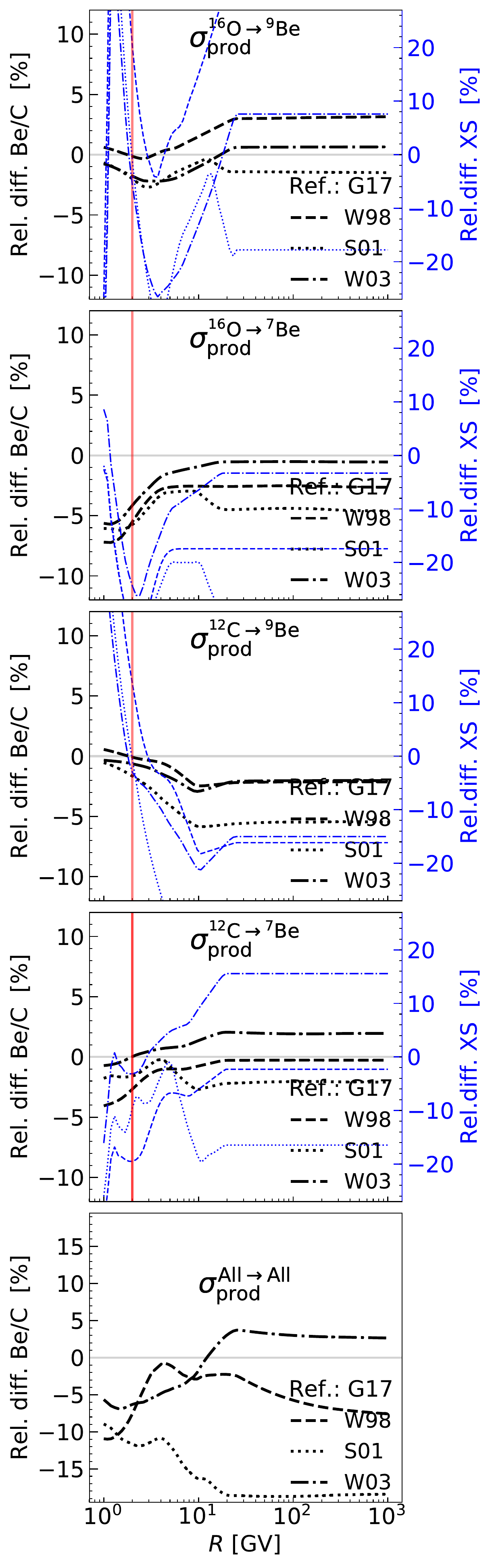}}
   \caption{Impact of inelastic (left) and production (right) cross-section uncertainties on Li/C (first and third column) and Be/C (second and fourth column) ratio for specific reactions as a function of rigidity. In each panel, the relative difference between (inelastic or production) cross section parametrisations w.r.t. a reference one is shown in blue (associated to the right-hand side $y$-axis), whereas the impact on the calculated ratio (in model \SLIM{}) of varying the cross section is shown in black (associated to the left-hand $y$-axis). To guide the eye, the vertical red line indicates the rigidity of the first AMS-02 data point. The bottom panels show the overall impact when all reactions (i.e. all nuclei in the network) are replaced. Inelastic cross sections are B94~\citepads{BarPol1994}, W96~\citepads{1996PhRvC..54.1329W}, T99~\citepads{1996NIMPB.117..347T,1999NIMPB.155..349T}, and W03~\citepads{2003ApJS..144..153W}. Production cross sections are W98 \citepads{1998ApJ...508..940W,1998ApJ...508..949W,1998PhRvC..58.3539W}, S01 (A. Soutoul, private communication), W03~\citepads{2003ApJS..144..153W}, and G17~\citepads{2001ICRC....5.1836M,2003ICRC....4.1969M}.}
   \label{fig:xs_impact}
\end{figure*}

Following \citetads{2019A&A...627A.158D}, the uncertainties in the inelastic and production cross sections for CR analyses are dealt with nuisance parameters in the $\chi^2$. As shown and validated on mock data, the presence of cross section nuisance parameters ensures a minimally biased determination of the transport parameter. It also naturally propagates cross-section uncertainties to all derived propagation parameters \citepads{2019A&A...627A.158D}. This approach was successfully used in \citetads{2019PhRvD..99l3028G} for the B/C analysis, and it is repeated here for all our minimisation studies; other approaches have been used in the literature to assess and propagate the cross section uncertainties to CR fluxes \citepads{2017PhRvD..96j3005T,2018JCAP...01..055R,2019PhRvD..99j3023E}.

\paragraph{Impact of cross section uncertainties}
The calculation of any CR quantity involves a large network of nuclear reactions, and it is not possible in practice to include uncertainties for all the reactions. Instead, we focus on the most impacting ones and use them as a proxy to capture the overall effect of the whole network \citepads{2019A&A...627A.158D}. Lists of reactions ranked by decreasing contribution for the production of Li, Be, B, and N can be found in \citetads{2018PhRvC..98c4611G}. For illustration, we show in Fig.~\ref{fig:xs_impact} a selection of these reactions in the context of the calculation of the Li/C (first and third column) and Be/C (second and fourth column) ratios\footnote{We do not show plots for B/C as they are similar and were already presented in \citetads{2019A&A...627A.158D}.}; the two leftmost panels are related to inelastic cross sections, the two rightmost to production cross sections. The blue curves---associated to the right-hand side $y$-axis ticks and labels---illustrate the relative differences between several cross section parametrisations available (from $\sim10\%$ to $\sim 25\%$). The black curves---associated to the left-hand side $y$-axis ticks and labels---show the impact of these differences on the calculated secondary-to-primary ratio: the $\sim10\%$ difference between inelastic cross sections translate in a $\sim1-2\%$ difference only on Li/C and Be/C, whereas the $\sim25\%$ differences between production cross sections translate in $\sim5-10\%$ differences on the CR ratios; see \citetads{2019A&A...627A.158D} for the origin of these differences. The bottom row shows the impact of changing all cross sections of the network at once (and not just one at a time): the full variation is typically captured by the combination of the few above-selected cross sections, validating the choice of using the latter as proxies for the whole network.

\begin{figure}[t]
   \subfigure{\includegraphics[scale=0.37]
      {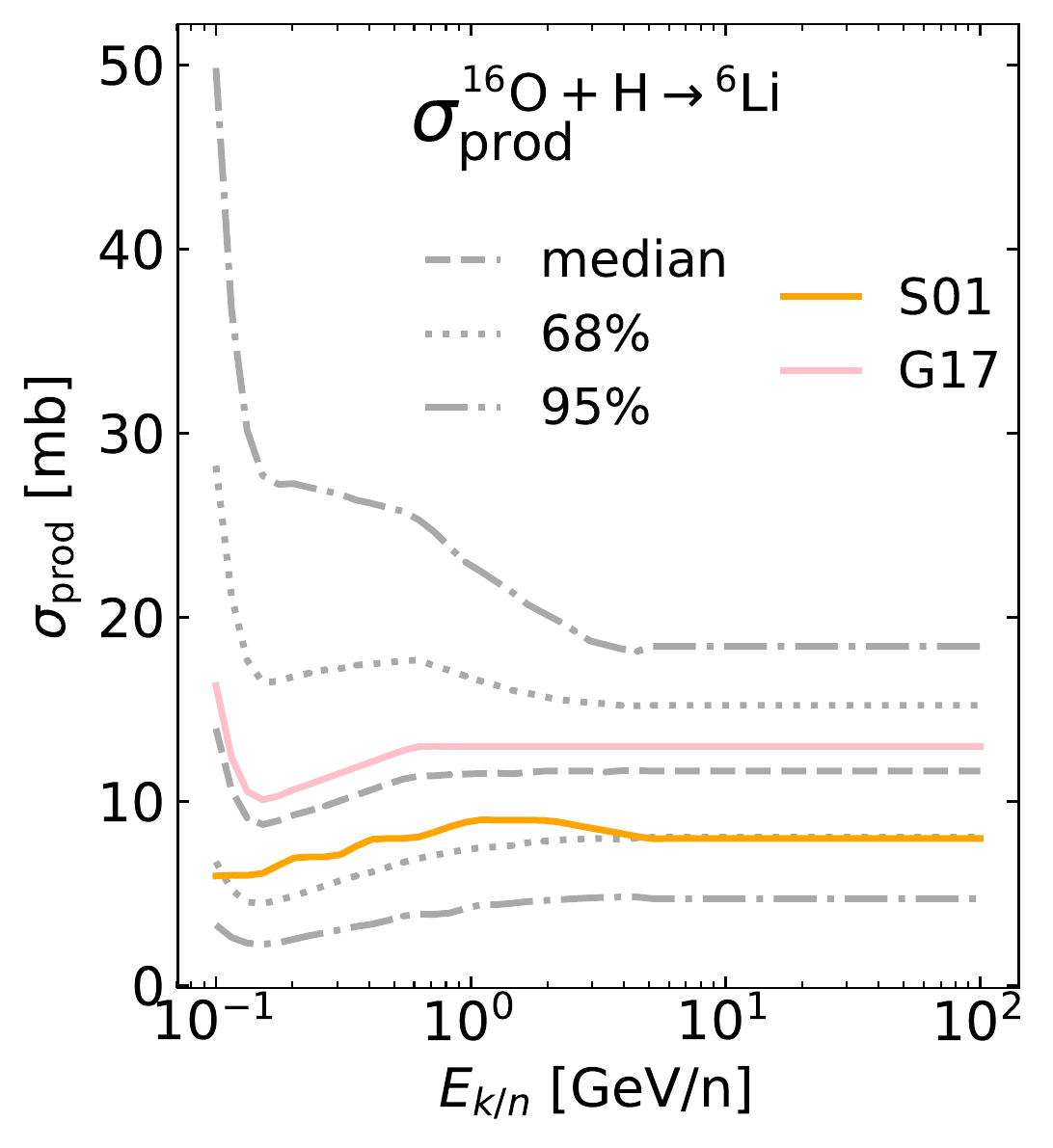}}
   \subfigure{\includegraphics[scale=0.37]
      {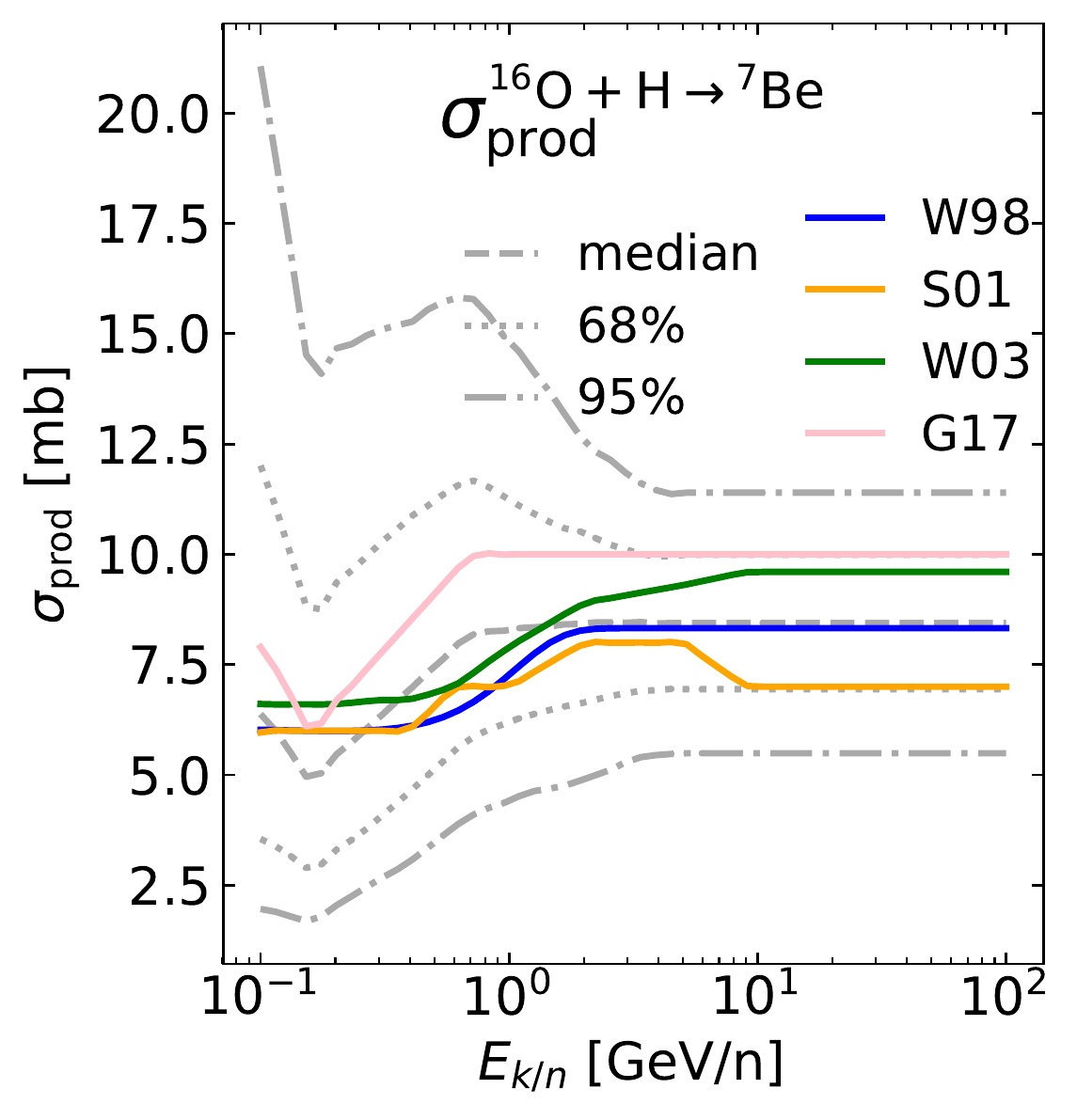}}
   \caption{Illustration of the NSS scheme used for cross section nuisance parameters. Colour-coded solid lines correspond to existing cross section parametrisations,  W98 \citepads{1998ApJ...508..940W,1998ApJ...508..949W,1998PhRvC..58.3539W}, S01 (A. Soutoul, private communication), W03~\citepads{2003ApJS..144..153W}, and G17~\citepads{2001ICRC....5.1836M,2003ICRC....4.1969M}. Grey lines correspond to the median, $68\%$, and $95\%$ CLs resulting from the use of NSS nuisance parameters, see Eqs.~(\ref{eq:NSS_Norm}) to (\ref{eq:NSS_Slope}), taken from Table~\ref{tab:xs_nuis}.}
   \label{fig:xs_nuis}
\end{figure}

\begin{table}[t]
\caption{List of nuclear reactions for which the mean ($\mu$) and width ($\sigma$) of the NSS nuisance parameters are picked (last three columns). For the production cumulative cross sections, see Eq.~(\ref{eq:xs_cumul}), the number in brackets report which fraction the specific reaction contributes to the overall production of the CR species---taken from \citetads{2012A&A...539A..88C} for \het{} and \citetads{2018PhRvC..98c4611G} for all others. The second column reports (in parenthesis) the maximal impact the associated cross section uncertainty has on the calculated secondary species (read off Fig.~\ref{fig:xs_impact}): numbers in ({\bf boldface}) highlight the most impacting reactions used as nuisance parameters in our analysis.}
\label{tab:xs_nuis}
\centering
{
\footnotesize
\begin{tabular}{l c c c c c c}
\hline\hline
Reaction &\hspace{-0.75cm}Impact $(\frac{\Delta\sigma}{\sigma})^{\rm XS}$\hspace{-0.75cm}  & Norm. & Scale & Slope \\
         & on flux & $\; \mu \,|\, \sigma$ & $\mu \,|\, \sigma$ & $\mu \,|\, \sigma$\\
\hline
&{\em $^3$He}&&&\\[1mm]
\!\!$^{3}$He+H     & (1.8\%)          & 1.00\,|\,0.15 & 1.2\,|\,0.5 & n/a \\
\!\!$^{4}$He+H   & (\textbf{5.0\%}) & 1.00\,|\,0.10 & 1.0\,|\,0.25\!\!\! & n/a \\[1mm]
\!\!\!$^{16}$O+H$\veryshortarrow$$^{3}$He \,\,\,[5\%]\!\!\!\! & ({2.1\%})        & 1.10\,|\,0.30 &  n/a  & 0.10\,|\,0.10  \\
\!\!\!$^{12}$C+H$\veryshortarrow$$^{3}$He \,\,\,[5\%]\!\!\!\! & (1.5\%)          & 1.10\,|\,0.30 &  n/a  & 0.05\,|\,0.15 \\
\!\!\!$^{4}$He+H$\veryshortarrow$$^{3}$He [80\%]\!\!\!\!\!\! & (\textbf{7.3\%}) & 1.00\,|\,0.10 &  n/a  & \!\!0.00\,|\,0.025\!\!\!\!\! \\[4mm]
&{\em Li}&&&\\[1mm]
\!\!\!$^{16}$O+H     & (1.2\%) & 1.03\,|\,0.04 & 0.7\,|\,0.5 & n/a \\
\!\!\!$^{12}$C+H     & (\textbf{1.3\%}) & 1.01\,|\,0.04 & 0.8\,|\,0.5 & n/a \\
\!\!\!$^{6}$Li+H     & (0.8\%)         & 1.02\,|\,0.04 & 0.7\,|\,0.4 & n/a \\[1mm]
\!\!\!$^{12}$C+H$\veryshortarrow$$^{7}$Li [12\%]\!\!\!\!\! & (3.9\%)   & 0.90\,|\,0.12 & n/a &\!\!0.03\,|\,0.15\!\! \\
\!\!\!$^{12}$C+H$\veryshortarrow$$^{6}$Li [14\%]\!\!\!\!\! & (4.7\%)   & 0.87\,|\,0.15 & n/a &\!\!0.00\,|\,0.15\!\! \\
\!\!\!$^{16}$O+H$\veryshortarrow$$^{6}$Li [14\%]\!\!\!\!\! & (\textbf{6.8\%})  & 0.89\,|\,0.28 & n/a &\!\!0.00\,|\,0.15\!\! \\[4mm]
&{\em Be}&&&\\[1mm]
\!\!\!$^{16}$O+H     & (0.9\%)            & 1.03\,|\,0.04 & 0.7\,|\,0.5 & n/a \\
\!\!\!$^{12}$C+H     & (1.4\%)            & 1.01\,|\,0.04 & 0.8\,|\,0.5 & n/a \\
\!\!\!$^{9}$Be+H     & (1.1\%)            & 0.95\,|\,0.06 & 0.7\,|\,0.4 & n/a \\
\!\!\!$^{7}$Be+H     & (\textbf{2.7\%})   & 1.10\,|\,0.10 & 0.7\,|\,0.4 & n/a \\[1mm]
\!\!\!$^{16}$O+H$\veryshortarrow$$^{9}$Be \,\,\,[5\%]\!\!\!\!\! & (3.2\%)           & 1.00\,|\,0.30 & n/a &\!\!0.00\,|\,0.15\!\! \\
\!\!\!$^{12}$C+H$\veryshortarrow$$^{9}$Be  \,\,\,[9\%]\!\!\!\!\! & (5.9\%)           & 0.87\,|\,0.20 & n/a &\!\!0.03\,|\,0.15\!\! \\
\!\!\!$^{12}$C+H$\veryshortarrow$$^{7}$Be [16\%]\!\!\!\!\! & (4.0\%)           & 1.00\,|\,0.25 & n/a &\!\!0.00\,|\,0.15\!\! \\
\!\!\!$^{16}$O+H$\veryshortarrow$$^{7}$Be [18\%]\!\!\!\!\! & (\textbf{7.2\%})  & 0.85\,|\,0.15 & n/a &\!\!0.00\,|\,0.15\!\! \\[4mm]
&{\em B}&&&\\[1mm]
\!\!\!$^{16}$O+H     & (0.8\%)          & 1.03\,|\,0.04 & 0.7\,|\,0.5 & n/a \\
\!\!\!$^{12}$C+H     & (1.0\%)          & 1.01\,|\,0.04 & 0.8\,|\,0.5 & n/a \\
\!\!\!$^{11}$B+H     & (\textbf{1.7\%}) & 0.98\,|\,0.04 & 0.7\,|\,0.4 & n/a \\[1mm]
\!\!\!$^{12}$C+H$\veryshortarrow$$^{10}$B \,\,\,[7\%]\!\!\!\!\! & (2.5\%)           & 1.07\,|\,0.15 & n/a &\!\!0.00\,|\,0.15\!\! \\
\!\!\!$^{16}$O+H$\veryshortarrow$$^{11}$B [18\%]\!\!\!\!\! & (4.0\%)           & 0.96\,|\,0.18 & n/a &\!\!0.00\,|\,0.15\!\! \\
\!\!\!$^{12}$C+H$\veryshortarrow$$^{11}$B [34\%]\!\!\!\!\! & (\textbf{7.1\%})  & 1.10\,|\,0.12 & n/a &\!\!0.03\,|\,0.15\!\! \\[4mm]
&{\em N }&&&\\[1mm]
\!\!\!$^{16}$O+H  & (\textbf{1.8\%}) & 1.03\,|\,0.04 & 0.70\,|\,0.50 &  n/a \\
\!\!\!$^{15}$N+H  & (1.0\%) & 1.00\,|\,0.05 & 0.70\,|\,0.50 &  n/a \\
\!\!\!$^{14}$N+H  & (1.6\%) & 1.02\,|\,0.07 & 0.70\,|\,0.50 &  n/a \\[1mm]
\!\!\!$^{16}$O+H$\veryshortarrow$$^{14}$N [20\%]\!\!\!\!\!& (1.7\%) & 1.00\,|\,0.15 &   n/a   & 0.00\,|\,0.05 \\
\!\!\!$^{16}$O+H$\veryshortarrow$$^{15}$N [50\%]\!\!\!\!\!& (\textbf{5.9\%}) & 0.90\,|\,0.15 &   n/a   & 0.05\,|\,0.10 \\
\hline
\end{tabular} 
}
\end{table}

\paragraph{Normalisation, Scale, and Slope (NSS) and nuisance parameters}
To properly propagate cross-section uncertainties on the calculated secondary-to-primary ratios, the trick is to find some parametrisations or transformation laws that allow to encompass the sets of possible cross section values. As stressed in \citetads{2019A&A...627A.158D}, this is not a trivial task given the variety of production cross section shapes. The transformation laws used here are the ones introduced in \citetads{2019A&A...627A.158D} and used in \citetads{2019PhRvD..99l3028G}, namely a combination of normalisation, energy scale, and low-energy slope applied to some reference cross-section dataset:
\begin{flalign}
 \sigma^{\rm Norm.}_{\rm new}(E_{k/n}) &= \mathrm{Norm} \times \sigma_{\rm ref}(E_{k/n}) \label{eq:NSS_Norm}\\
 \sigma^{\rm Scale}_{\rm new}(E_{k/n}) &= \sigma_{\rm ref}\left(E_{k/n} \times \mathrm{Scale}\right) \label{eq:NSS_Scale}\\
 \sigma^{\rm Slope}_{\rm new}(E_{k/n}) &= \left\lbrace
   \begin{matrix}
    \displaystyle
    \sigma_{\rm ref}(E_{k/n})\quad\text{if}\; E_{k/n} \geq E_{k/n}^{\text{thresh.}}; \label{eq:NSS_Slope}\\
    \displaystyle
    \sigma_{\rm ref} (E_{k/n}) \times \left(\frac{E_{k/n}}{E_{k/n}^{\text{thresh.}}}\right)^\text{ Slope} \text{otherwise.}
  \end{matrix}\right.
\end{flalign}
These transformations are illustrated in Fig.~\ref{fig:xs_nuis}, where we show the available cross-sections parametrisations (coloured lines) along with envelopes generated from the NSS transformation laws (grey lines). For a given reaction, the nuisance parameters to used in a secondary-to-primary ratio fit are the mean $\mu$ and scatter $\sigma$ of each transformation law. Actually, to keep the number of free parameters as low as possible, only two out of the three transformations are used, namely a normalisation and energy scale for inelastic cross sections, and a normalisation and low-energy slope for production cross sections. In practice, these parameters are chosen so that the median and $1\sigma$ contours of the NSS-generated cross sections visually encompass the range of existing cross-section parametrisations (see Fig.~\ref{fig:xs_nuis}). We refer the reader to \citetads{2019A&A...627A.158D} for further justification.

Table~\ref{tab:xs_nuis} gathers the NSS $\mu$ and $\sigma$ values used in the analyses described in the main text. In this analysis, the production cross-section reactions used correspond to cumulative cross sections, i.e. the production of a given species $Y$ accounting for all short-live fragments $Y_i$ (decaying into $Y$):
\begin{equation}
\label{eq:xs_cumul}
  \sigma^{\rm c}_{X+{\rm H}\rightarrow Y} =
    \sigma_{X+{\rm H}\rightarrow Y} + \sum_i \sigma_{X+{\rm H}\rightarrow Y_i^\star} \times {\cal B}r(Y_i^\star\rightarrow Y)\,.
\end{equation}
This is important to calculate correctly how much a reaction contributes to the total, as these number are used in Sect.~\ref{sec:norm_xsprod} to interpret the post-fit values of the nuisance parameters. To be explicit, in Table~\ref{tab:xs_nuis}, the fraction given in bracket for the production of $^{11}$B and $^{15}$N account respectively for the significant contributions of the short-lived $^{11}$C and $^{15}$O \citepads{2018PhRvC..98c4611G}.

\section{Minimisation convergence}
\label{app:min_conv}

Many results in this work are obtained by fitting \usine{} models to data, based on the \minuit{} algorithm \citepads{1975CoPhC..10..343J}. Because of the large number of free parameters in some configurations, it is important to check the reliability of the algorithm for finding the minimum of the $\chi^2$ valley.

To do so, \minuit{} was compared to two different minimisation algorithms. Technically, a python interface was implemented for $\chi^2$ calls in \usinebis{} (this will be part of the next release), allowing one to use alternative algorithms from various python packages: (i) \minuit{} (\texttt{iMinuit} package) is the state-of-the-are minimisation algorithm based on Migrad, which uses gradients to find the minimum of a scalar function, by iterative estimations of the Hessian matrix; (ii) Powell (\texttt{SciPy} package), does not rely on gradients but uses iterative line searches instead, less prone to numerical instabilities that could arise in gradient calculations; (iii) the conjugate gradient (CG) method (\texttt{SciPy} package), is another gradient-based method which solves iteratively the linear algebraic problem of finding a minimum of a scalar function.

\begin{figure}[t]
	\label{fig:min_conv_plot}
	\includegraphics[width=\columnwidth]{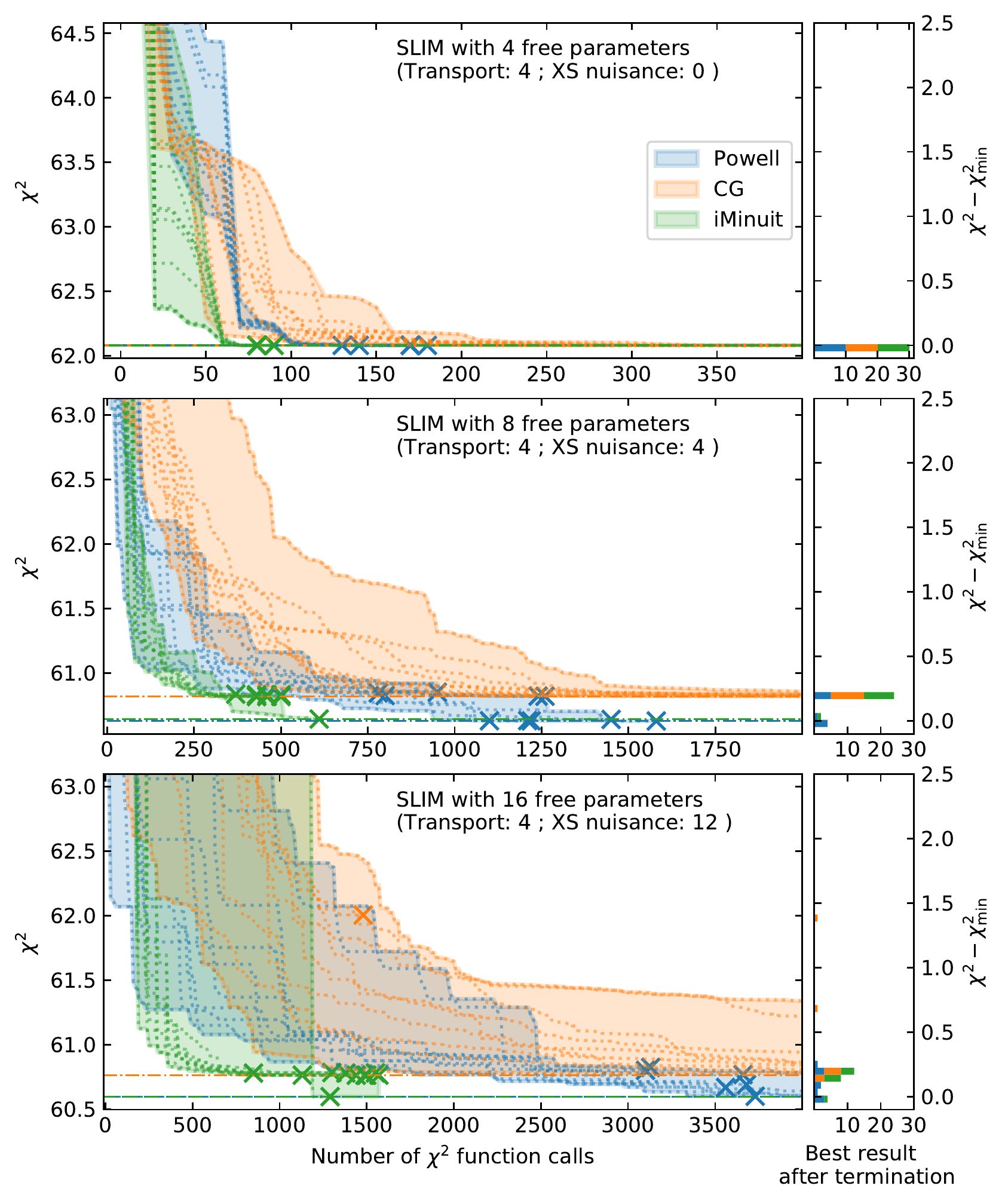}
	\caption{Comparison of minimiser convergence speed and robustness. The dotted lines mark the $\chi^2$ progress towards its minimum as a function of the number of calls, until the algorithm stops (cross symbol). Several algorithms are shown, namely Powell (blue), GC (red), and \minuit{} (green), and the horizontal dash-dotted lines highlight the best-fit obtained for each algorithm. Ten different starting points are run, and the distribution of the best $\chi^2_\mathrm{min}$ found are projected on the right-hand panel to show that sometimes the algorithm is stuck in a local minimum. From top to bottom, the number of nuisance parameters in the model increases. See text for discussion.}
\end{figure}
These three minimisers where tested on the same tasks and conditions, bookkeeping their performance and convergence over time, as measured by the number of $\chi^2$ calls---the typical calculation time for one configuration is $\sim \mathcal{O}(1\,\mathrm{s})$, so that the number of calls is roughly the number of seconds. These performances are shown in Fig.~\ref{fig:min_conv_plot} for a growing number of nuisance parameters in the analysis (from top to bottom). Two main conclusions can be taken from these plots. First, compared to the other algorithms (in blue and red), the \minuit{} algorithm (in green) always provides the fastest route to the minimum. Second, with an larger number of parameters, i.e. a more complex parameter space, the chances to end up in a local minimum increase (see bottom panel). No minimiser is immune to this, and the only option to overcome this difficulty is to use many trial starting points (represented by the many crosses).

From these tests, we conclude that \minuit{} remains the best algorithm and that in order to find the `true' minimum, the minimisation must be repeated for $\mathcal{O}(100)$ different starting points; keeping the best $\chi^2$ among all those obtained. This is the procedure we used for all the results presented in the main text.

\section{Combining AMS-02 with lower-energy data}
\label{app:LE-data}

\begin{figure}[t]
  \includegraphics[width=\columnwidth]{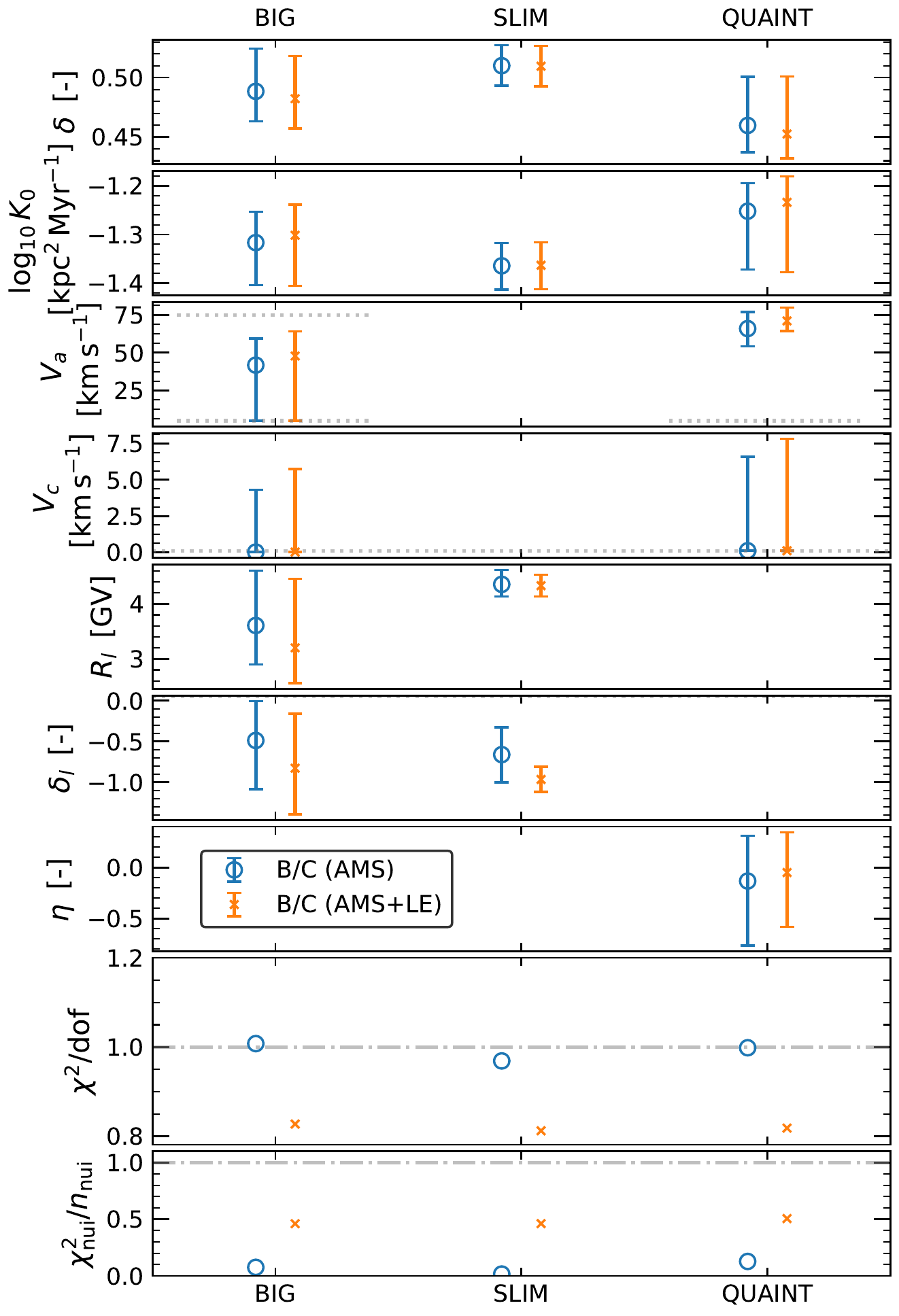}
  \caption{Best-fit transport parameters and uncertainties for \BIG{}, \SLIM{}, and \QUAINT{} (first, second, and third column) from B/C analysis. Large circles correspond to fits on AMS02 data only, whereas small crosses to the combined fits on AMS-02 and LE-data. The two bottom panels show \chimindof{} and \chipernui{}, defined in Eqs.~(\ref{eq:chi2}) and (\ref{eq:chi2nuis}) respectively.}
  \label{fig:BCwwoLEdata}
\end{figure}

In this appendix, we explore the consistency of low-energy data (LE-data for short) with AMS-02 data, and we focus below on B/C data; the datasets used are detailed in Table~\ref{tab:LiBeBC_data}. These LE-data cover top-of-atmosphere (TOA) energies of ten to one hundred GeV/n. ACE-CRIS data are the most precise, with uncertainties at the level of $\sim 6\%$, i.e. only twice that of AMS-02. We do not consider in the following the recent Voyager~1 data covering a similar range in interstellar (IS) energies because our models do not match these data. This discrepancy was also observed in the analysis of \citetads{2016ApJ...831...18C}, and whether this is an issue with the models, the cross sections, specifics of the local interstellar medium, or even the data---which should be explored soon, as Voyager~2 recently crossed into interstellar space  \citepads{2019NatAs...3.1013S}, 6 years after Voyager~1--- calls for a dedicated study that goes beyond the scope of this paper.

\begin{table}[t]
\caption{List of experiments with their data-taking periods and associated expected Solar modulation level for the low-energy B/C dataset considered in the analysis.}
{
\footnotesize
\label{tab:LiBeBC_data}
\begin{tabular}{rcl}
\hline\hline
Experiment (period)                 & $\phi_{\rm prior}$ & Reference\\
\hline \\[-1em]
   ACE-CRIS ('97/08-'98/04)              & \!\!\!528~MV          \!\!\! & \citetads{2013ApJ...770..117L} \\
   ACE-CRIS ('01/05-'03/09)              & \!\!\!872~MV          \!\!\! & \citetads{2013ApJ...770..117L} \\
   ACE-CRIS ('09/03-'10/01)              & \!\!\!445~MV          \!\!\! & \citetads{2013ApJ...770..117L} \\
       IMP8 ('74/01-'78/10)              & \!\!\!540~MV          \!\!\! & \citetads{1987ApJS...64..269G} \\
\hspace{-5mm}ISEE3-HKH ('78/08-'81/04)   & \!\!\!742~MV          \!\!\! & \citetads{1988ApJ...328..940K}\!\!\!\!\!\!\!\!  \\
\hspace{-9mm}Ulysses-HET ('90/10-'95/07) & \!\!\!732~MV          \!\!\! & \citetads{1996AaA...316..555D} \\
\hspace{-5mm}Voyager1\&2 ('77/01-'98/12) & \!\!\!700~MV$^\dagger$\!\!\!\!\! & \citetads{1999ICRC....3...41L} \\
  \hline
\end{tabular}
\tablefoottext{$\dagger$}{From the publication, the prior should have been 450~MV (weighted average modulation at different positions in the Solar cavity), but all the analyses were done when we spotted our mistake. We believe it to have a marginal impact only on the results.}
}
\end{table}

Figure~\ref{fig:BCwwoLEdata} shows the best-fit transport parameters and associated $\chi^2$ for fits to B/C AMS-02 data only (large circles), or to B/C AMS-02 and LE-data together (small crosses). First, there is no impact for  \BIG{} because the model has too many low-rigidity competing transport parameters (break, $\eta$, and to some extent reacceleration and convection). Second, the impact is maximal for the simplest model, \SLIM{}, where the low-rigidity break value is better constrained with  $\delta_l\simeq 1$. Lastly, the situation is in-between for \QUAINT{}, with a slightly more constrained `low-rigidity' parameters $\eta$ and $V_a$; moreover, because the latter parameter couples to high rigidities, constraints on high-rigidity parameters are also slightly improved. We find similar trends for Li/C and Be/C (not shown).

\begin{figure}[t]
  \includegraphics[width=\columnwidth]{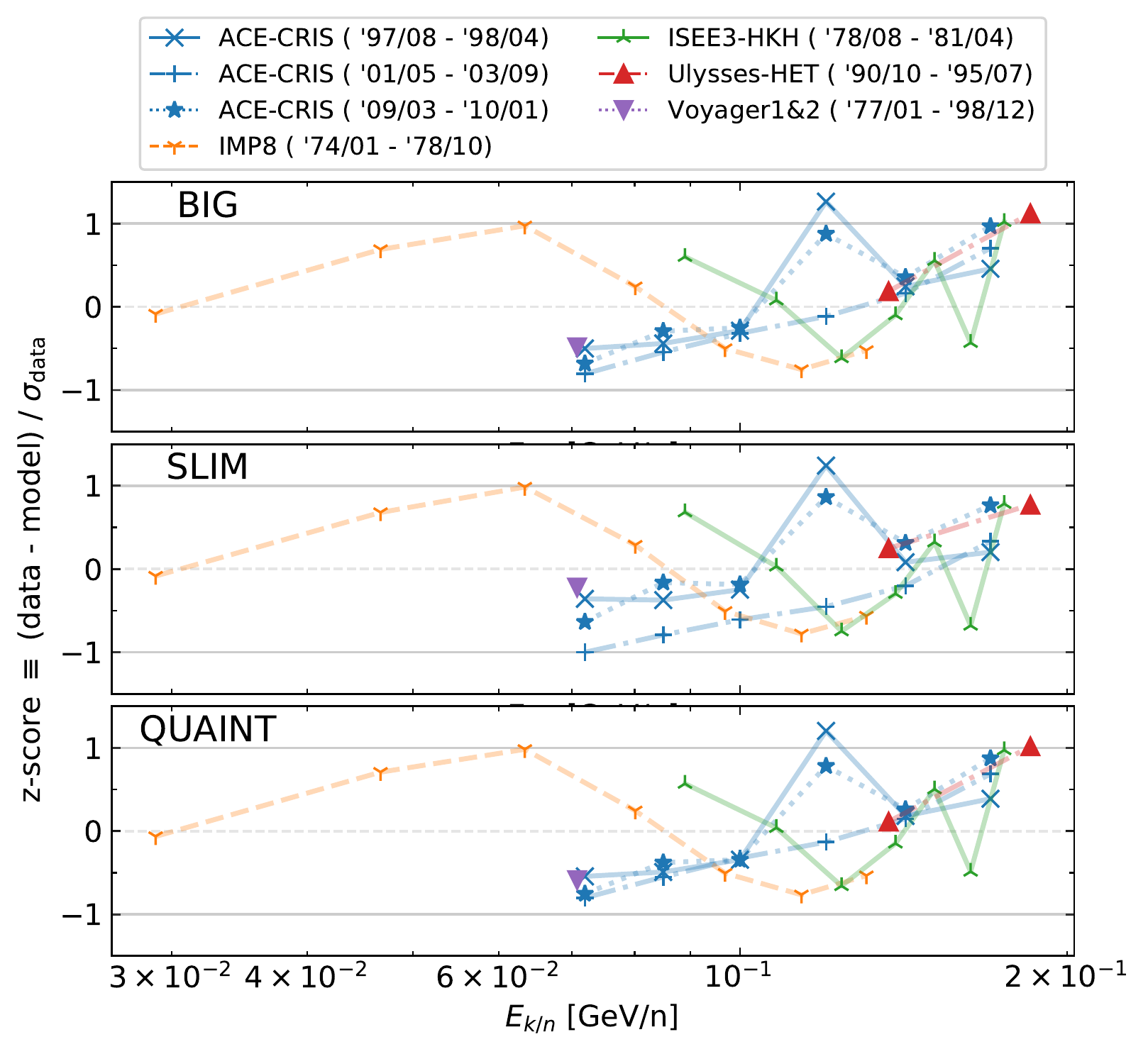}
  \includegraphics[width=\columnwidth]{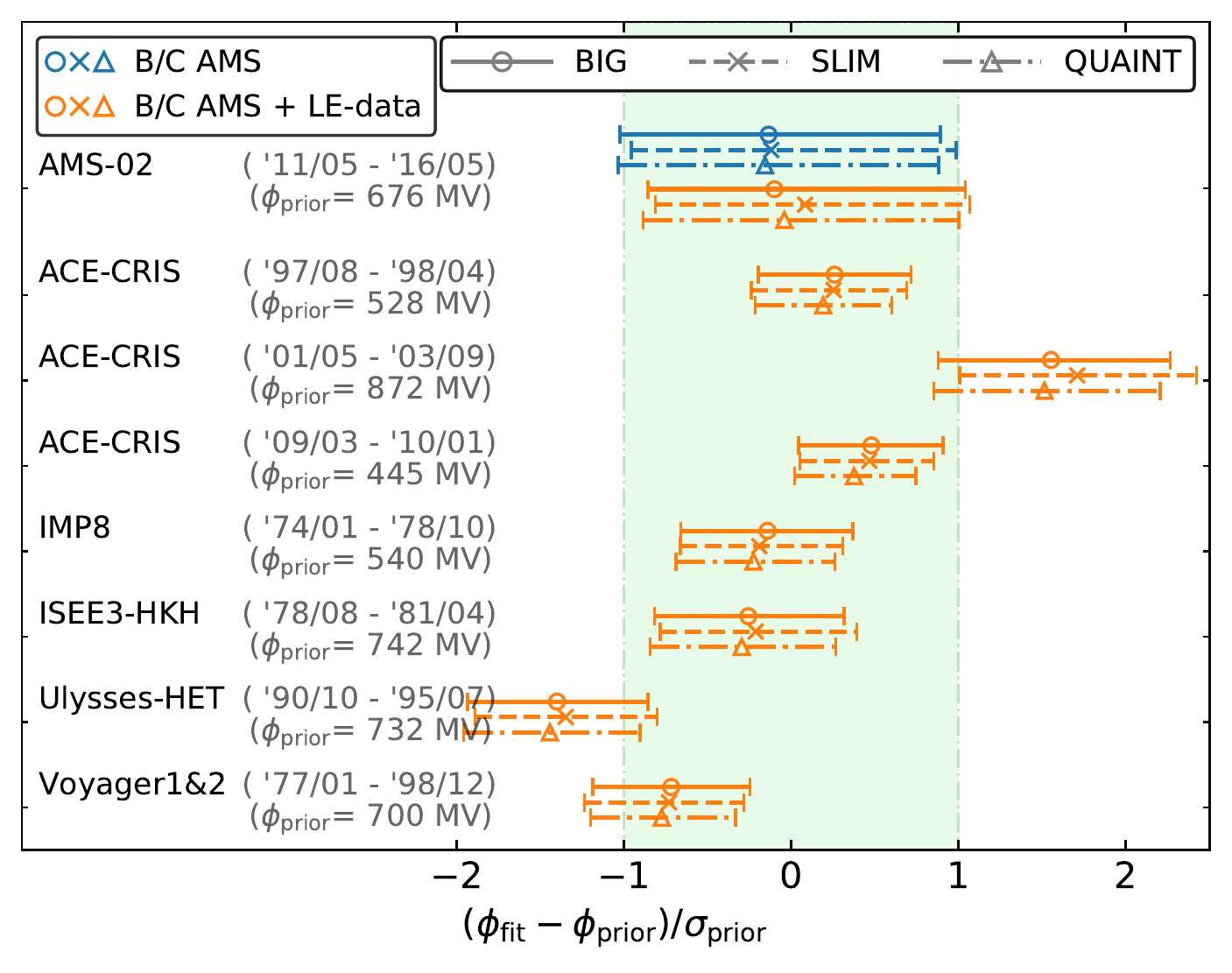}
  \caption{{\em Top panel:} $z$-score=(model-data)/$\sigma_{\rm data}$ of B/C LE-data used in the AMS and LE-data combined analysis for \BIG{} (top), \SLIM{} (middle), and \QUAINT{} (bottom). To guide the eye, data points from the same experiments are linked. {\em Bottom panel:} Post-fit residuals and $1\sigma$ uncertainty of the Solar modulation nuisance parameters for \BIG{} (solid lines), \SLIM{} (dashed lines), and \QUAINT{} (dash-dotted lines). The blue lines show the residuals for the fit to AMS-02 data only, and orange lines for AMS-02 and LE-data---one nuisance per dataset (per data taking period). For comparison, the greenish vertical band shows the $1\sigma$ nuisance prior range.}
  \label{fig:fit_BCwwoLE}
  \vspace{-0.25cm}
\end{figure}
Figure~\ref{fig:fit_BCwwoLE} shows in the top panel the well-behaved residuals (within $1\sigma$) of LE-data from the combined fit, while the bottom panel illustrates that the compatibility between LE-data and the model is not tapped in the extra nuisance parameters: the post-fit modulation levels, which cover Solar minimum to Solar maximum periods, all fall within $1\sigma$ of their input values, for all models (\BIG{}, \SLIM{}, and \QUAINT{}). The same conclusions could have been directly read off the bottom panels of Fig.~\ref{fig:BCwwoLEdata}: \chimindof{} decreases with LE-data (very good compatibility of LE-data with AMS-02 data) and \chipernui{} only mildly increases.

\begin{figure}[t]
  \includegraphics[width=\columnwidth]{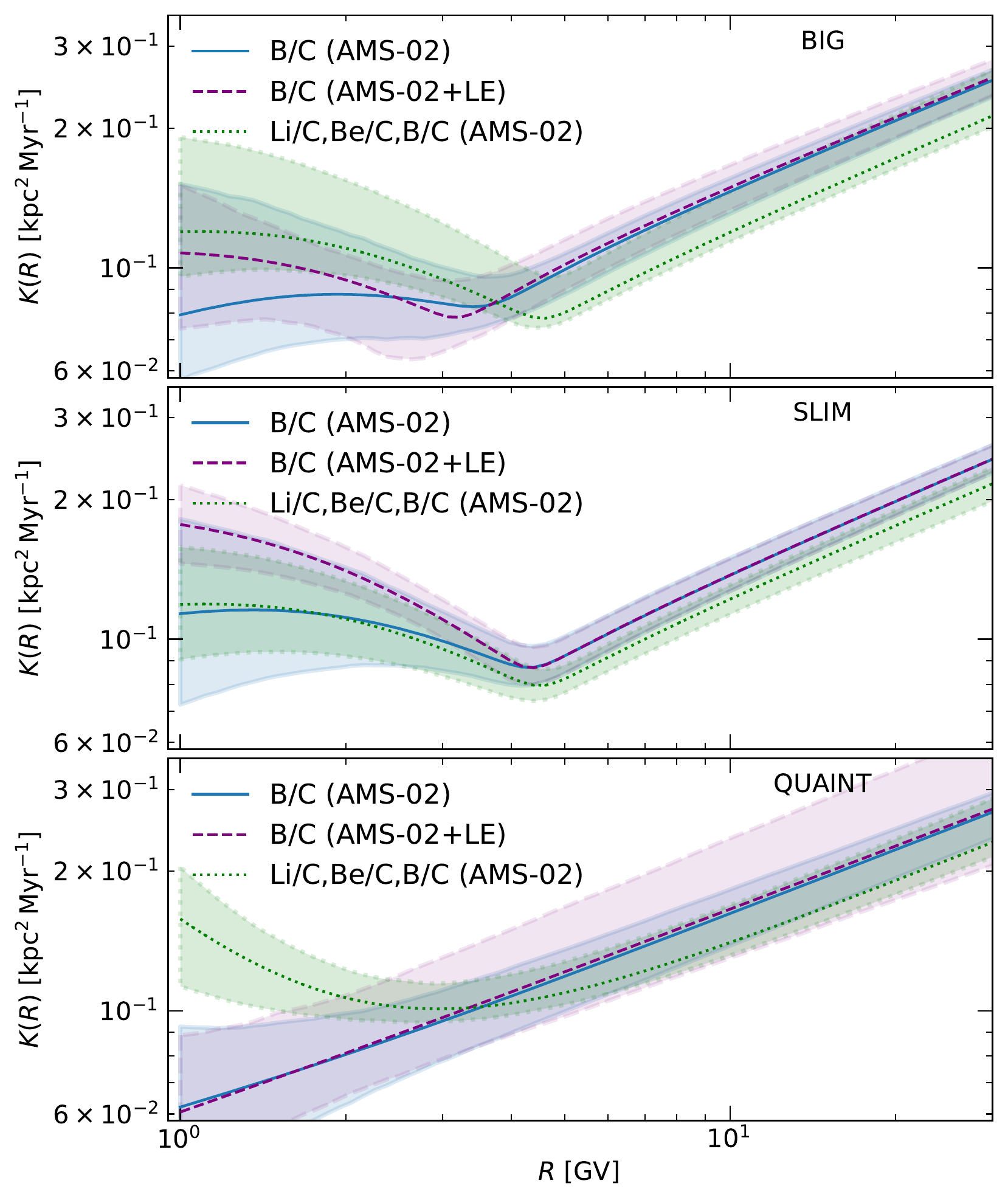}
  \caption{Best-fit and $1\sigma$ contours for $K(R)$ in \BIG{} (top), \SLIM{} (middle), and \QUAINT{}  (bottom) for three different analyses: fit to AMS-02 B/C data (blue contours), fit to AMS-02 B/C and LE-data (purple contours), and combined fit of AMS-02 Li/C, Be/C, and B/C data (green contours).}
  \label{fig:KR_LE}
\end{figure}
Figure \ref{fig:KR_LE} shows $1\sigma$ contours of the diffusion coefficient for B/C data in model \BIG{} (top), \QUAINT{} (middle), and \SLIM{} (bottom). Compared to the analysis based on AMS-02 data only (blue solid lines, adding LE-data in the fit (purple dashed lines) strengthens the presence of a break for \BIG{} and \SLIM{}, but not for \QUAINT{}---similar results are also obtained if relying on Li/C (or Be/C) instead of B/C data (not shown). For comparison purpose, we also show the contours obtained from combining Li/C, Be/C, and B/C (green dotted lines), as discussed in Sect.~\ref{sec:LiBeB}: the latter analysis now shows without ambiguity a break for all configurations.

\bibliographystyle{aa} 
\bibliography{libeb}
\end{document}